\documentclass[12pt,english]{article}
\usepackage{amsmath}
\usepackage{amsthm}

\usepackage{helvet}
\usepackage[lf]{FiraMono}
\usepackage{newtxmath}
\usepackage[T1]{fontenc}
\usepackage[utf8]{inputenc}
\usepackage[table]{xcolor}
\usepackage{colortbl}
\usepackage{verbatim}
\usepackage{xr}
\usepackage{float}
\usepackage{graphicx}
\usepackage[letterpaper]{geometry}
\usepackage[authoryear]{natbib}

\makeatletter

\providecommand{\tabularnewline}{\\}
\floatstyle{ruled}
\newfloat{algorithm}{tbp}{loa}
\providecommand{\algorithmname}{Algorithm}
\floatname{algorithm}{\protect\algorithmname}

\theoremstyle{plain}
\newtheorem{theorem}{Theorem}
\newtheorem{corollary}{Corollary}
\newtheorem{lemma}{Lemma}
\newtheorem{proposition}{Proposition}
\theoremstyle{definition}
\newtheorem{assumption}{Assumption}
\usepackage{hyperref}
\usepackage{pdfpages}

\ifdefined\showcaptionsetup
 \PassOptionsToPackage{caption=false}{subfig}
\fi
\usepackage{subfig}
\makeatother

\usepackage{babel}
\begin{document}
\title{Efficient Bootstrap Calibration of Generalized Posteriors with Importance
Reweighting}
\author{Masahiro Tanaka \thanks{Faculty of Economics, Fukuoka University, Fukuoka, Japan. Address:
8-19-1, Nanakuma, Jonan, Fukuoka, Japan 814-0180. E-mail: m.tanaka.tt@fukuoka-u.ac.jp.}}
\maketitle
\begin{abstract}
The learning rate of a generalized posterior governs posterior uncertainty
but typically has no canonical value. Generalized posterior calibration
(GPC) selects this rate by matching bootstrap-estimated credible-interval
coverage to its nominal level, often requiring repeated posterior
simulation. We propose GPC via importance reweighting (GPC-IRW), which
reuses draws from a reference posterior fitted to the original data
across bootstrap resamples and candidate learning rates. For quantile-based
intervals of a scalar functional, we express the coverage criterion
in terms of bootstrap posterior cumulative distribution functions,
avoiding computation of individual bootstrap credible intervals. Safeguards
based on importance-weight effective sample size and proximity to
coverage-decision boundaries invoke reweighting from cached bootstrap-specific
draws or direct posterior simulation when necessary. Under the stated
regularity conditions, with numerical error tending to zero, we show
that median-centered calibration preserves the first-order learning-rate
target and achieves asymptotically nominal coverage. The approach
extends to independent clusters through cluster-level resampling.
Simulations yield coverage and interval lengths comparable to those
of existing GPC algorithms at substantially lower computational cost.
In an exoplanet mass--radius analysis, GPC-IRW achieves 6.7- and
9.6-fold speedups over MCMC-based calibration in the single-planet
and full-sample analyses, respectively, while illustrating the effects
of within-host dependence.
\end{abstract}

\section{Introduction}

Likelihood-based Bayesian inference provides a convenient framework
for uncertainty quantification, but its conclusions can be sensitive
to assumptions about the data-generating distribution. Generalized
Bayesian inference, including Gibbs posteriors, offers an alternative
in which a chosen loss is used in place of the negative log likelihood
\citep{Zhang2006entropy,Zhang2006information,Jiang2008,Bissiri2016,Syring2023}.
Given observations $Y_{1},\ldots,Y_{n}$, a loss $\ell(\theta;y)$,
and a prior $p(\theta)$, the generalized posterior is proportional
to
\[
\exp\left\{ -\omega\sum^{n}_{i=1}\ell(\theta;Y_{i})\right\} p(\theta),
\]
where $\omega>0$ is the learning rate. This formulation does not
require a fully specified sampling distribution and accommodates losses
chosen for a particular inferential objective or for robustness to
departures from a working model; see \citet{Martin2022} and \citet{Nott2024}
for overviews. Unlike a correctly specified likelihood, however, a
general loss does not determine a canonical scale for posterior uncertainty,
making learning-rate selection essential.

An inappropriate learning rate can yield credible sets with frequentist
coverage far from their nominal level, even when the posterior concentrates
around the intended loss minimizer. In regular problems, expected-loss
curvature governs generalized posterior spread, whereas the sampling
variability of the empirical loss minimizer also depends on score
variability; the resulting uncertainty scales need not agree under
misspecification \citep{Kleijn2012,Miller2021}. Approaches to selecting
the learning rate include predictive criteria, information matching,
and methods based on the loss-likelihood bootstrap \citep{Gruenwald2017,Holmes2017,Lyddon2019};
see \citet{Wu2023} for a comparison. \citet{Onizuka2026} proposes
a closed-form rate using weighted Fisher divergence. Another approach
adjusts posterior draws after simulation to match a suitable sampling
covariance \citep{Shaby2014,Tamano2025}.

Generalized posterior calibration (GPC), introduced by \citet{Syring2019},
instead selects the learning rate by matching bootstrap-estimated
credible-set coverage to the nominal level. A straightforward implementation
requires posterior simulation for each bootstrap resample at each
candidate learning rate, potentially involving many Markov chain Monte
Carlo (MCMC) runs. Computational improvements include particle reweighting
across learning rates \citep{Tanaka2024} and sequential Monte Carlo
(SMC) sampling \citep{Tanaka2025,DelMoral2006}; \citet{Syring2019}
also suggested importance sampling for small learning-rate changes.
Nevertheless, maintaining bootstrap-specific posterior approximations
can be costly, and importance reweighting requires adequate overlap
between the reference and target posteriors. In closely related work,
\citet{Sugasawa2026} develops a Dirichlet reweighting (DRW) approach
to GPC. Both DRW and the present approach reuse draws from an original-data
reference posterior to approximate weighted-data posteriors and evaluate
coverage decisions using posterior cumulative distribution functions.
Unlike DRW, the present approach retains the conventional multinomial
bootstrap and differs in its calibration center and diagnostic safeguards.
We also provide theoretical guarantees and a cluster-level resampling
extension, as detailed below.

We develop \emph{generalized posterior calibration via importance
reweighting} (GPC-IRW). Nonparametric bootstrap resamples are represented
by multinomial counts multiplying observation-level losses. Draws
from a reference posterior fitted to the original data are then reweighted
to approximate the bootstrap posteriors across resamples and candidate
learning rates whenever overlap is adequate. For quantile-based credible
intervals of a scalar functional, the coverage decision can be obtained
from each bootstrap posterior cumulative distribution function evaluated
at the original-data posterior median, without computing bootstrap-specific
credible intervals. Retaining bootstrap counts and precomputing weighted
losses allow vectorized evaluation of the importance ratios. To assess
the reliability of each bootstrap coverage decision, we use replicate-specific
effective sample sizes and a diagnostic for uncertainty near the coverage
cutoffs. If either diagnostic fails, the procedure first attempts
reweighting from cached bootstrap-specific posterior draws and resorts
to direct posterior simulation if necessary.

The proposed approach differs from DRW in several respects. GPC-IRW
preserves the multinomial-bootstrap criterion of conventional GPC,
whereas DRW uses Dirichlet weights; although their finite-sample calibration
criteria differ, we establish that the two weighting schemes have
the same first-order asymptotic calibration target. GPC-IRW also uses
the original-data posterior median as its calibration center and assesses
bootstrap-replicate-specific reweighting reliability, beyond monitoring
the effective sample size for the original-data reference posterior.
On the theoretical side, we establish first-order validity of median-centered
calibration and consistency of the computed criterion on fixed finite
evaluation sets, together with asymptotically nominal coverage for
computed solutions satisfying vanishing-error conditions. Finally,
our cluster-level bootstrap accounts for within-cluster dependence
when calibrating a common learning rate. This serves a different purpose
from \citet{Sugasawa2026} hierarchical calibration of group-specific
learning rates with shrinkage.

Related approaches to approximating the effects of data perturbations
include the Bayesian infinitesimal jackknife \citep{Giordano2024},
posterior-covariance bootstrap approximations \citep{Ohkubo2026},
infinitesimal-jackknife standard errors for posterior functionals
\citep{Luo2026}, and influence-function approximations to weighted
refitting \citep{Gibson2026}. Whereas these approaches focus on posterior
summaries or their sampling variability, GPC-IRW estimates bootstrap
posterior distribution functions to evaluate a coverage criterion.

The method and its first-order analysis extend to independent clusters
by resampling at the cluster level. Simulations compare GPC implementations
under independent sampling, assess performance under clustered sampling,
and examine a hinge-loss classification example beyond the smooth-loss
theory. An application to exoplanet mass--radius data illustrates
the computational gains of GPC-IRW and the effects of within-host
dependence on calibration.

Section~\ref{sec:gpc} introduces the setup, Section~\ref{sec:numimpl}
presents the method, Section~\ref{sec:theory} develops the theory,
Sections~\ref{sec:simulation} and~\ref{sec:application} present
the simulation results and application, respectively, and Section~\ref{sec:conclusion}
concludes.

\section{Generalized Posterior Calibration}

\label{sec:gpc}

Let $Y_{1},\ldots,Y_{n}$ be i.i.d.~draws from an unknown distribution
$P$, let $\mathcal{D}_{n}=(Y_{1},\ldots,Y_{n})$, and denote its
observed realization by $(y_{1},\ldots,y_{n})$.  Let $\theta\in\Theta$
denote the parameter, with prior density $p(\theta)$. Let $\mathbb{P}$
denote the repeated-sampling law of $\mathcal{D}_{n}$. For a loss
function $\ell(\theta;y)$, define 
\[
R_{n}(\theta)=\frac{1}{n}\sum^{n}_{i=1}\ell(\theta;Y_{i}).
\]
For a learning rate $\omega>0$, the generalized posterior \citep[e.g.,][]{Bissiri2016}
is
\begin{equation}
\pi_{\omega}(\theta\mid\mathcal{D}_{n})\propto\exp\{-n\omega R_{n}(\theta)\}p(\theta).\label{eq:general-pos}
\end{equation}
Unless otherwise stated, ``posterior'' refers to the generalized
posterior. Let $R(\theta)=\mathbb{E}_{P}[\ell(\theta;Y)]$ and $\theta^{\dagger}=\arg\min_{\theta}R(\theta)$,
assumed unique. For a scalar functional $g(\theta)$, write $g^{\dagger}=g(\theta^{\dagger})$,
and fix $\alpha\in(0,1)$ and quantile levels $0<\beta_{L}<1/2<\beta_{U}<1$
satisfying $\beta_{U}-\beta_{L}=1-\alpha$. Writing $q_{\beta,n}(\omega)$
for the lower posterior $\beta$-quantile of $g(\theta)$, define
the $(1-\alpha)$ quantile-based credible interval as 
\[
C_{\alpha,\omega}(\mathcal{D}_{n})=\bigl[q_{\beta_{L},n}(\omega),q_{\beta_{U},n}(\omega)\bigr].
\]
The quantile levels are fixed across datasets and learning rates,
allowing potentially unequal tail probabilities. The equal-tailed
interval is recovered as the special case $\beta_{L}=\alpha/2$ and
$\beta_{U}=1-\alpha/2$. The population coverage probability is 
\[
c_{n,\alpha}(\omega)=\mathbb{P}\{g^{\dagger}\in C_{\alpha,\omega}(\mathcal{D}_{n})\}.
\]
Posterior calibration seeks a learning rate at which $c_{n,\alpha}(\omega)=1-\alpha$,
or one that comes as close as possible to this nominal level.

Because $P$ and $g^{\dagger}$ are unknown, we approximate the repeated-sampling
distribution using the multinomial (nonparametric) bootstrap, following
the calibration principle of \citet{Syring2019}. We use the median
of the original-data posterior as the bootstrap center, regardless
of the chosen quantile levels. Define the original-data posterior
distribution function 
\[
F^{0}_{n,\omega}(x)=\pi_{\omega}\{g(\theta)\le x\mid\mathcal{D}_{n}\}
\]
and its median 
\[
\check{g}_{n}(\omega)=\inf\{x:F^{0}_{n,\omega}(x)\ge1/2\}.
\]
Thus the center varies with the learning rate.

Let 
\[
M=(M_{1},\ldots,M_{n})\sim\operatorname{Multinomial}\left(n;\frac{1}{n},\ldots,\frac{1}{n}\right)
\]
be the observation counts in an ordinary nonparametric bootstrap resample.\footnote{Scaled Dirichlet weights, as used in the related method of \citet{Sugasawa2026},
are also examined in Section~\ref{sec:dw} of the Supplementary Material.
The resulting calibration criterion has the same first-order target
as the multinomial-bootstrap criterion, and the two weighting schemes
yielded comparable numerical performance within GPC-IRW in the settings
considered.} Conditional on $\mathcal{D}_{n}$, let $\mathbb{P}^{\mathrm{MB}}$
and $\mathbb{E}^{\mathrm{MB}}$ denote probability and expectation
with respect to $M$. Throughout, a superscript $\mathrm{MB}$ denotes
a quantity associated with the multinomial bootstrap. For a realization
$m=(m_{1},\ldots,m_{n})$, define 
\[
R^{\mathrm{MB}}_{n}(\theta;m)=\frac{1}{n}\sum^{n}_{i=1}m_{i}\ell(\theta;y_{i})
\]
and the bootstrap posterior 
\[
\pi^{\mathrm{MB}}_{\omega}(\theta\mid m,\mathcal{D}_{n})\propto\exp\left\{ -\omega\sum^{n}_{i=1}m_{i}\ell(\theta;y_{i})\right\} p(\theta).
\]

For a nonnegative weight vector $v=(v_{1},\ldots,v_{n})$, define
the normalizing constant 
\[
\mathcal{Z}_{n}(\omega;v)=\int_{\Theta}\exp\left\{ -\omega\sum^{n}_{i=1}v_{i}\ell(\theta;y_{i})\right\} p(\theta)\,d\theta.
\]
The corresponding weighted posterior is well defined whenever $0<\mathcal{Z}_{n}(\omega;v)<\infty$.
Let $\Omega=[\underline{\omega},\overline{\omega}]\subset(0,\infty)$
denote the learning-rate search interval, and assume that the original-data
posterior is proper for $\omega\in\Omega$. This need not imply propriety
of every bootstrap posterior, because resampling may delete observations
needed for identification or tail integrability, especially under
an improper prior.

Let $C^{\mathrm{MB}}_{\alpha,\omega}(m)$ be the corresponding quantile
credible interval at the same levels $\ensuremath{\beta_{L},\beta_{U}}$.
We calibrate the posterior using 
\begin{equation}
c^{\mathrm{MB}}_{n,\alpha}(\omega)=\mathbb{E}^{\mathrm{MB}}\left[\mathbb{I}\{\check{g}_{n}(\omega)\in C^{\mathrm{MB}}_{\alpha,\omega}(M)\}\right],\label{eq:multinomial-coverage}
\end{equation}
where $\mathbb{I}\{\cdot\}$ denotes the indicator function. Unlike
the plug-in empirical target considered by \citet{Tanaka2026}, we
use the original-data posterior median; Section~\ref{sec:theory}
shows that the two yield the same first-order calibration target.

If a bootstrap realization produces an improper posterior, we define
its coverage indicator to be zero. Accordingly, the bootstrap posterior
CDFs below are defined only when the corresponding posterior is proper.
Any bounded assignment on improper realizations is asymptotically
equivalent provided that

\[
\mathbb{P}^{\mathrm{MB}}\left\{ 0<\mathcal{Z}_{n}(\omega;M)<\infty\text{ for every }\omega\in\Omega\right\} \xrightarrow{\mathbb{P}}1.
\]
In applications, posterior propriety must either be guaranteed by
the prior--loss specification or verified model-specifically; the
zero-indicator convention is not a numerical diagnostic. A sufficient
condition for propriety for every finite nonnegative weight vector
and every $\omega>0$ is that $p$ is proper and each $\ell(\cdot;y_{i})$
is finite and bounded below on the support of $p$. 

\section{Proposed Procedure}

\label{sec:numimpl}

\subsection{Reweighting}

\label{subsec:reweighting}

Direct evaluation of \eqref{eq:multinomial-coverage} at a candidate
learning rate requires posterior draws for each of the $B$ bootstrap
replicates. For quantile-based intervals, however, calibration requires
only the bootstrap posterior CDF at the current center, which can
be estimated by reweighting a reference posterior when overlap is
adequate. We assume that every proper target induces a continuous
CDF of $g(\theta)$ that is strictly increasing on its support.

Let $\omega_{\mathrm{ref}}\in\Omega$ be a reference learning rate
and let $\theta^{(1)},\ldots,\theta^{(S)}$ be a posterior sample
targeting $\pi_{\omega_{\mathrm{ref}}}(\cdot\mid\mathcal{D}_{n})$.
For a bootstrap count vector $m$ and candidate $\omega$, define
\[
r(\theta;\omega,m)=\exp\left\{ -\sum^{n}_{i=1}(\omega m_{i}-\omega_{\mathrm{ref}})\ell(\theta;y_{i})\right\} .
\]
To reweight the original-data posterior, define 
\[
a(\theta;\omega)=\exp\left\{ -(\omega-\omega_{\mathrm{ref}})\sum^{n}_{i=1}\ell(\theta;y_{i})\right\} .
\]
Thus $r(\theta;\omega,m)$ changes both the bootstrap weights and
the learning rate, whereas $a(\theta;\omega)$ changes only the learning
rate and is used to update the original-data center. Both are exact
change-of-measure ratios up to normalizing constants. Let $M^{(1)},\ldots,M^{(B)}$
be independent copies of $M$, retained throughout calibration. For
brevity, define $a_{s}(\omega)=a(\theta^{(s)};\omega)$ and $r_{sb}(\omega)=r(\theta^{(s)};\omega,M^{(b)})$.

If the target posterior is proper and absolutely continuous with respect
to the reference posterior, the change-of-measure identity is exact:
\[
\mathbb{E}_{\pi_{\omega_{\mathrm{ref}}}(\cdot\mid\mathcal{D}_{n})}[r(\theta;\omega,m)]=\frac{\mathcal{Z}_{n}(\omega;m)}{\mathcal{Z}_{n}(\omega_{\mathrm{ref}};\mathbf{1}_{n})},
\]
where $\mathbf{1}_{n}$ denotes the $n$-dimensional vector of ones.
Absolute continuity is automatic, for example, when all observed-data
losses are finite $p$-almost everywhere. Posterior propriety and
importance-sampling overlap are distinct issues. The direct-sampling
safeguard below is therefore used only for proper targets.

For a proper target, let 
\[
F_{n,\omega,m}(x)=\pi^{\mathrm{MB}}_{\omega}\{g(\theta)\le x\mid m,\mathcal{D}_{n}\}.
\]
Then 
\begin{equation}
F_{n,\omega,m}(x)=\frac{\mathbb{E}_{\pi_{\omega_{\mathrm{ref}}}(\cdot\mid\mathcal{D}_{n})}[\mathbb{I}\{g(\theta)\le x\}r(\theta;\omega,m)]}{\mathbb{E}_{\pi_{\omega_{\mathrm{ref}}}(\cdot\mid\mathcal{D}_{n})}[r(\theta;\omega,m)]},\label{eq:reweighted-cdf}
\end{equation}
and the analogous identity with $a(\theta;\omega)$ gives $F^{0}_{n,\omega}(x)$.
With the coverage indicator set to zero for an improper target,
\[
c^{\mathrm{MB}}_{n,\alpha}(\omega)=\mathbb{E}^{\mathrm{MB}}\left[\mathbb{I}\left\{ \beta_{L}\le F_{n,\omega,M}(\check{g}_{n}(\omega))\le\beta_{U}\right\} \right].
\]
Thus the bootstrap posterior quantiles need not be computed. The Monte
Carlo approximations are 
\begin{equation}
\widehat{F}_{n,\omega,m;S}(x)=\frac{\sum^{S}_{s=1}r(\theta^{(s)};\omega,m)\mathbb{I}\{g(\theta^{(s)})\le x\}}{\sum^{S}_{s=1}r(\theta^{(s)};\omega,m)}\label{eq:estimated-reweighted-cdf}
\end{equation}
for the bootstrap CDF and 
\begin{equation}
\widehat{F}^{0}_{n,\omega;S}(x)=\frac{\sum^{S}_{s=1}a(\theta^{(s)};\omega)\mathbb{I}\{g(\theta^{(s)})\le x\}}{\sum^{S}_{s=1}a(\theta^{(s)};\omega)}\label{eq:estimated-original-cdf}
\end{equation}
for the original-data CDF, yielding the weighted median 
\begin{equation}
\check{g}_{n,S}(\omega)=\inf\{x:\widehat{F}^{0}_{n,\omega;S}(x)\ge1/2\}.\label{eq:estimated-posterior-center}
\end{equation}

For a global reference sample $\theta^{(1)},\ldots,\theta^{(S)}$,
define 
\[
L_{si}=\ell(\theta^{(s)};y_{i}),\qquad L_{s}=\sum^{n}_{i=1}L_{si},\qquad L^{\mathrm{MB}}_{sb}=\sum^{n}_{i=1}M^{(b)}_{i}L_{si}.
\]
Writing $L=(L_{si})\in\mathbb{R}^{S\times n}$ and $M_{B}=(M^{(b)}_{i})\in\mathbb{R}^{n\times B}$,
all bootstrap losses are obtained simultaneously as $L^{\mathrm{MB}}=LM_{B}\in\mathbb{R}^{S\times B}$.
This product is computed when the global reference sample is first
generated and whenever it is refreshed. At any subsequent candidate
learning rate, 
\[
\log r_{sb}(\omega)=-\omega L^{\mathrm{MB}}_{sb}+\omega_{\mathrm{ref}}L_{s},\qquad\log a_{s}(\omega)=-(\omega-\omega_{\mathrm{ref}})L_{s},
\]
so that the importance weights for all $B$ perturbations can be evaluated
by vectorized operations without further loss evaluation. We stabilize
the weights by subtracting the largest log weight before exponentiation.

\subsection{Safeguard}

\label{subsec:safeguard}

For each bootstrap replicate, we first try the global reference, then
a cached bootstrap-specific reference, and resort to direct simulation
only if neither passes the diagnostics. For the original-data posterior,
define the normalized weights and corresponding importance-weight
effective sample size (ESS)  as 
\[
\varpi_{s0}(\omega)=\frac{a_{s}(\omega)}{\sum^{S}_{j=1}a_{j}(\omega)},\qquad\mathrm{ESS}_{0}(\omega)=\left\{ \sum^{S}_{s=1}\varpi_{s0}(\omega)^{2}\right\} ^{-1}.
\]
Because the original-data posterior center enters every bootstrap
calculation, we refresh the global reference sample whenever $\mathrm{ESS}_{0}(\omega)/S<\zeta_{0}$,
for a prescribed $\zeta_{0}\in(0,1)$.

For a bootstrap replicate $b$, the safeguard has a different role.
Calibration requires only the binary decision 
\[
I_{b}(\omega)=\mathbb{I}\left\{ \beta_{L}\le F_{n,\omega,M^{(b)}}(\check{g}_{n}(\omega))\le\beta_{U}\right\} .
\]
Accordingly, we assess whether importance reweighting is sufficiently
accurate to determine this decision rather than requiring a fixed
fraction of the reference sample to remain effective. 

Let 
\[
\widehat{F}_{b}(\omega)=\widehat{F}_{n,\omega,M^{(b)};S}(\check{g}_{n,S}(\omega))
\]
denote the reweighted CDF estimate, and define the normalized weights
and the corresponding importance-weight ESS by 
\[
\varpi_{sb}(\omega)=\frac{r_{sb}(\omega)}{\sum^{S}_{j=1}r_{jb}(\omega)},\qquad\mathrm{ESS}_{b}(\omega)=\left\{ \sum^{S}_{s=1}\varpi_{sb}(\omega)^{2}\right\} ^{-1}.
\]
We then define the variance diagnostic 
\[
\widehat{V}_{b}(\omega)=\sum^{S}_{s=1}\varpi_{sb}(\omega)^{2}\left[\mathbb{I}\{g(\theta^{(s)})\le\check{g}_{n,S}(\omega)\}-\widehat{F}_{b}(\omega)\right]^{2}.
\]
We accept the decision based on reweighting when $\mathrm{ESS}_{b}(\omega)\ge E_{\min}$
and 
\begin{equation}
\min\left\{ |\widehat{F}_{b}(\omega)-\beta_{L}|,|\widehat{F}_{b}(\omega)-\beta_{U}|\right\} >\kappa_{\mathrm{sg}}\sqrt{\widehat{V}_{b}(\omega)},\label{eq:decision-safeguard}
\end{equation}
where $E_{\min}\in(1,S]$ prevents severe weight degeneracy and $\kappa_{\mathrm{sg}}>0$
controls the conservativeness of the decision safeguard. Condition
\eqref{eq:decision-safeguard} keeps the diagnostic band around the
estimated CDF from crossing either coverage cutoff. Like importance-weight
ESS, the operational variance diagnostic does not account for autocorrelation
in the reference MCMC output.

If global-reference reweighting fails either criterion, we avoid repeatedly
sampling the same bootstrap posterior across learning-rate iterations.
Let $K(\leq S)$ denote the number of draws in a bootstrap-specific
reference sample. Suppose that a previous safeguard evaluation for
perturbation $b$, at reference learning rate $\omega_{b,\mathrm{ref}}$,
produced a sample $\vartheta^{(1)}_{b},\ldots,\vartheta^{(K)}_{b}$
targeting $\pi^{\mathrm{MB}}_{\omega_{b,\mathrm{ref}}}(\cdot\mid M^{(b)},\mathcal{D}_{n})$.
Define and retain
\[
Q_{bk}=\sum^{n}_{i=1}M^{(b)}_{i}\ell(\vartheta^{(k)}_{b};y_{i}),\qquad g_{bk}=g(\vartheta^{(k)}_{b}).
\]
At another learning rate $\omega$, this bootstrap-specific reference
sample can be reweighted using 
\[
u_{bk}(\omega)=\exp\{-(\omega-\omega_{b,\mathrm{ref}})Q_{bk}\}.
\]
This is again an exact change-of-measure ratio up to normalization,
now between two learning rates for the same bootstrap perturbation.
We apply the ESS and decision criteria first to the global reference
and then, if needed, to the cached bootstrap-specific reference. Direct
simulation is required only when neither passes; the resulting draws
replace the cache for reuse at subsequent calibration iterations.

\subsection{Learning rate search}

\label{subsec:search}

For a proper target, let $\widehat{F}^{\mathrm{comp}}_{n,\omega,m;S,K}$
denote the CDF estimate obtained from the global reference, a cached
bootstrap-specific reference, or direct simulation according to this
rule; for an improper target, define the corresponding coverage indicator
to be zero. The computed coverage estimate is 
\[
\widehat{c}^{\mathrm{MB}}_{n,\alpha;S,K,B}(\omega)=\frac{1}{B}\sum^{B}_{b=1}\mathbb{I}\left\{ \beta_{L}\le\widehat{F}^{\mathrm{comp}}_{n,\omega,M^{(b)};S,K}(\check{g}_{n,S}(\omega))\le\beta_{U}\right\} .
\]

Increasing $\omega$ typically concentrates the posterior, so that
coverage above the target suggests increasing the learning rate. We
therefore use a stochastic-approximation-style search with Kesten's
adaptive step-size rule \citep{Robbins1951,Kesten1958}. Define 
\[
e_{t}=\widehat{c}^{\mathrm{MB}}_{n,\alpha;S,K,B}(\omega^{(t)})-(1-\alpha),\qquad d_{t}=|e_{t}|.
\]
Let $\Pi_{\Omega}(\cdot)$ denote projection onto $\Omega$, let $\{\gamma_{l}\}$
be a positive non-increasing step-size sequence, and set $l_{0}=1$.
For $t\ge1$, update 
\[
\omega^{(t+1)}=\Pi_{\Omega}(\omega^{(t)}+\gamma_{l_{t}}e_{t}),\qquad l_{t}=l_{t-1}+\mathbb{I}\{e_{t}e_{t-1}<0\}.
\]
Thus the step size is reduced when successive coverage discrepancies
change sign. Although the limiting coverage curve is decreasing, the
finite-sample criterion need not be monotone; we therefore use the
recursion only as a search heuristic and retain the best evaluated
value.

Algorithm~\ref{alg:gpcirw} summarizes the proposed algorithm, GPC-IRW.

\begin{algorithm}
\caption{Generalized posterior calibration via importance reweighting (GPC-IRW)}

\label{alg:gpcirw} Choose an initial learning rate $\omega^{(0)}\in\Omega$,
a tolerance $\epsilon_{\textrm{tol}}>0$, and a maximum number $T$
of coverage evaluations. Obtain a global reference posterior sample
at $\omega_{\mathrm{ref}}=\omega^{(0)}$. Generate and retain $M^{(1)},\ldots,M^{(B)}$,
precompute $L^{\mathrm{MB}}=LM_{B}$, and initialize the bootstrap-specific
reference caches as empty. Set $l_{0}=1$. Evaluate
\[
e_{0}=\widehat{c}^{\mathrm{MB}}_{n,\alpha;S,K,B}(\omega^{(0)})-(1-\alpha),
\]
and set $\omega^{(1)}=\Pi_{\Omega}\{\omega^{(0)}+\gamma_{l_{0}}e_{0}\}$.

For $t=1,\ldots,T-1$: 
\begin{enumerate}
\item If $\mathrm{ESS}_{0}(\omega^{(t)})/S<\zeta_{0}$, refresh the global
reference at $\omega^{(t)}$, set $\omega_{\mathrm{ref}}=\omega^{(t)}$,
and recompute $L$, $L_{s}$, and $L^{\mathrm{MB}}$. Compute $\check{g}_{n,S}(\omega^{(t)})$. 
\item For each $b$, first attempt importance reweighting from the global
reference. If either $\mathrm{ESS}_{b}(\omega^{(t)})<E_{\min}$ or
the decision safeguard \eqref{eq:decision-safeguard} fails, attempt
reweighting from the cached bootstrap-specific reference and apply
the analogous importance-weight ESS and decision criteria. If the
cached reference also fails or no cache exists, obtain $K$ direct
draws from $\pi^{\mathrm{MB}}_{\omega^{(t)}}(\cdot\mid M^{(b)},\mathcal{D}_{n})$
and replace the cache for $b$. 
\item Compute 
\[
e_{t}=\widehat{c}^{\mathrm{MB}}_{n,\alpha;S,K,B}(\omega^{(t)})-(1-\alpha),\qquad d_{t}=|e_{t}|.
\]
If $d_{t}\le\epsilon_{\mathrm{tol}}$, set $\widehat{\omega}^{\mathrm{MB}}=\omega^{(t)}$
and terminate. Otherwise update $\omega^{(t+1)}$ using Kesten's rule. 
\end{enumerate}
If the tolerance is not attained after $T$ evaluations, set $\widehat{\omega}^{\mathrm{MB}}$
to the evaluated learning rate with the smallest $d_{t}$. At the
selected learning rate, compute the original-data $[\beta_{L},\beta_{U}]$
quantile interval by reweighting an adequate global reference sample
or by one final posterior run. 
\end{algorithm}

The initial learning rate affects computational efficiency but not
the calibration criterion. Reweighting is less stable when the target
posterior is substantially more diffuse than the reference, so a high
initial rate may trigger more direct simulations as calibration moves
downward, whereas an unnecessarily low rate may increase the number
of coverage evaluations. A moderately small initial learning rate
may therefore improve computational efficiency.

Cached references affect only the Monte Carlo implementation, not
the exact criterion: a bootstrap perturbation may overlap poorly with
the original-data posterior but strongly with its own posterior at
a nearby learning rate.

Reusing the bootstrap weights provides common random numbers across
iterations, while importance reweighting avoids repeated posterior
simulation whenever overlap is adequate. Because the coverage estimate
is a multiple of $1/B$, the tolerance $\epsilon_{\mathrm{tol}}$
should not be substantially smaller than either the Monte Carlo error
or the finite-$B$ resolution.

\subsection{Extension to independent-cluster sampling}

\label{subsec:clustered}

Suppose instead that the sampling units are i.i.d.~clusters $\mathcal{Y}_{1},\ldots,\mathcal{Y}_{G}\sim P^{\mathrm{CL}}$,
where $\mathcal{Y}_{c}$ contains $N_{c}$ lower-level observations
and $n_{G}=\sum^{G}_{c=1}N_{c}$ denotes the total number of lower-level
observations. Let $\mathcal{D}^{\mathrm{CL}}_{G}=(\mathcal{Y}_{1},\ldots,\mathcal{Y}_{G})$,
and define the cluster-level aggregate loss
\[
\ell^{\mathrm{CL}}(\theta;\mathcal{Y}_{c})=\sum^{N_{c}}_{j=1}\ell(\theta;Y_{cj}),\qquad c=1,\ldots,G.
\]
Then the original-data generalized posterior is
\begin{equation}
\pi_{\omega}(\theta\mid\mathcal{D}^{\mathrm{CL}}_{G})\propto\exp\left\{ -\omega\sum^{G}_{c=1}\ell^{\mathrm{CL}}(\theta;\mathcal{Y}_{c})\right\} p(\theta).\label{eq:cluster-posterior}
\end{equation}
Thus cluster-level resampling changes the bootstrap perturbation but
not the original-data posterior. The corresponding population target
is 
\[
\theta^{\mathrm{CL},\dagger}=\arg\min_{\theta}\mathbb{E}_{P^{\mathrm{CL}}}\{\ell^{\mathrm{CL}}(\theta;\mathcal{Y}_{c})\}.
\]
Because the cluster loss sums rather than averages the observation-level
losses, larger clusters receive greater weight. If cluster size is
informative, this observation-weighted population target may differ
from the equal-cluster-weighted target based on $\mathbb{E}_{P^{\mathrm{CL}}}\{N^{-1}_{c}\ell^{\mathrm{CL}}(\theta;\mathcal{Y}_{c})\}$
\citep{Seaman2014}. We consider superpopulation inference with i.i.d.~clusters,
not general design-based cluster sampling with unequal inclusion probabilities
or other complex-survey features.

For the cluster multinomial bootstrap (CMB), let 
\[
M^{\mathrm{CMB}}=(M^{\mathrm{CMB}}_{1},\ldots,M^{\mathrm{CMB}}_{G})\sim\operatorname{Multinomial}\left(G;\frac{1}{G},\ldots,\frac{1}{G}\right),
\]
and, for a realization $m^{\mathrm{CMB}}$, define
\[
\pi^{\mathrm{CMB}}_{\omega}(\theta\mid m^{\mathrm{CMB}},\mathcal{D}^{\mathrm{CL}}_{G})\propto\exp\left\{ -\omega\sum^{G}_{c=1}m^{\mathrm{CMB}}_{c}\ell^{\mathrm{CL}}(\theta;\mathcal{Y}_{c})\right\} p(\theta).
\]
Equivalently, every observation belonging to cluster $c$ receives
the common bootstrap weight $m^{\mathrm{CMB}}_{c}$.  The corresponding
calibration criterion is defined as in \eqref{eq:multinomial-coverage},
with expectation taken over the CMB counts; the median of the original-data
posterior remains the bootstrap center.

The importance-reweighting construction requires no further modification.
For a global reference posterior at $\omega_{\mathrm{ref}}$, the
change-of-measure ratio for a CMB perturbation is 
\[
r^{\mathrm{CMB}}(\theta;\omega,m^{\mathrm{CMB}})=\exp\left\{ -\sum^{G}_{c=1}(\omega m^{\mathrm{CMB}}_{c}-\omega_{\mathrm{ref}})\ell^{\mathrm{CL}}(\theta;\mathcal{Y}_{c})\right\} .
\]
Hence the CDF representation in \eqref{eq:reweighted-cdf}, the importance-weight
diagnostics, and the bootstrap-specific reference updates in Algorithm~\ref{alg:gpcirw}
apply unchanged.

\section{Theoretical Properties}

\label{sec:theory}

Building on the exact bootstrap calibration theory of \citet{Tanaka2026},
we show that centering on the original-data posterior median preserves
the empirical-minimizer criterion's first-order learning-rate target.
We then establish conditions for consistency of the computed criterion
and validity of a sufficiently accurate computed root.

Throughout, $\mathbb{P}^{\mathrm{MB}}$ denotes multinomial-bootstrap
probability conditional on $\mathcal{D}_{n}$, and $\mathbb{P}_{\mathrm{joint}}$
denotes the joint law of $\mathcal{D}_{n}$ and all auxiliary Monte
Carlo randomness used by the calibration procedure. Conditional $o_{\mathbb{P}^{\mathrm{MB}}}(1)$
and conditional weak convergence statements are understood in $\mathbb{P}$-probability.
Let $\Phi$ denote the standard normal distribution function and write
$z_{\beta}=\Phi^{-1}(\beta)$. We use $d_{\mathrm{TV}}$ for total
variation distance and $\mathcal{L}_{\pi}(X)$ for the law of $X$
under distribution $\pi$. Proofs are given in Section~\ref{sec:proofs}
of the Supplementary Material. 

\subsection{First-order population and bootstrap coverage}

\label{subsec:pop-boot-coverage}

For the theoretical analysis, we regard $\mathcal{D}_{n}$ as random
and write 
\[
R^{\mathrm{MB}}_{n}(\theta;M)=\frac{1}{n}\sum^{n}_{i=1}M_{i}\ell(\theta;Y_{i}).
\]
The corresponding bootstrap posterior is denoted by $\pi^{\mathrm{MB}}_{\omega}(\cdot\mid M,\mathcal{D}_{n})$.
Let 
\[
\psi(y)=\nabla_{\theta}\ell(\theta^{\dagger};y),\qquad J=\mathbb{E}_{P}[\nabla^{2}_{\theta}\ell(\theta^{\dagger};Y)],\qquad V=\mathbb{E}_{P}[\psi(Y)\psi(Y)^{\top}].
\]
For the analysis, let $\widehat{\theta}_{n}\in\arg\min_{\theta}R_{n}(\theta)$
be a measurable empirical-risk minimizer. Let $\widehat{\theta}^{{\rm MB}}_{n}$
be a measurable local minimizer of $R^{{\rm MB}}_{n}(\theta;M)$ in
a neighborhood of $\widehat{\theta}_{n}$, whenever such a local minimizer
exists, and define it arbitrarily otherwise.

\begin{assumption}[M-estimation and bootstrap regularity] \label{as:mest}
The unique population minimizer $\theta^{\dagger}$ is an interior
point of the parameter space $\Theta\subset\mathbb{R}^{d}$. On a
neighborhood $\mathcal{N}$ of $\theta^{\dagger}$, $\ell(\theta;y)$
is twice continuously differentiable in $\theta$ for $P$-almost
every $y$, $\mathbb{E}_{P}[\psi(Y)]=0$ and $\mathbb{E}_{P}[\|\psi(Y)\|^{2+\delta}]<\infty$
for some $\delta>0$, and $J$ and $V$ are positive definite. There
is a measurable $H_{\ell}$ such that 
\begin{equation}
\sup_{\theta\in\mathcal{N}}\|\nabla^{2}_{\theta}\ell(\theta;Y)\|\le H_{\ell}(Y),\qquad\mathbb{E}_{P}[H_{\ell}(Y)^{2}]<\infty.\label{eq:hessian-envelope}
\end{equation}
Moreover, $\widehat{\theta}_{n}-\theta^{\dagger}=O_{\mathbb{P}}(n^{-1/2})$,
and the empirical Hessian converges uniformly to $J$ along the line
segment joining $\theta^{\dagger}$ and $\widehat{\theta}_{n}$. In
addition, 
\[
\mathbb{P}^{{\rm MB}}\!\left\{ \widehat{\theta}^{{\rm MB}}_{n}\text{ is a local minimizer of }R^{{\rm MB}}_{n}(\cdot;M)\right\} \xrightarrow{P}1,\qquad\widehat{\theta}^{{\rm MB}}_{n}-\widehat{\theta}_{n}=o_{\mathbb{P}^{{\rm MB}}}(1),
\]
and the bootstrap-weighted Hessian converges uniformly to $J$ on
the corresponding shrinking neighborhoods. \end{assumption}

Under Assumption~\ref{as:mest}, 
\begin{align}
\sqrt{n}(\widehat{\theta}_{n}-\theta^{\dagger}) & =-J^{-1}\frac{1}{\sqrt{n}}\sum^{n}_{i=1}\psi(Y_{i})+o_{\mathbb{P}}(1),\label{eq:unweighted-M-expansion}\\
\sqrt{n}(\widehat{\theta}^{\mathrm{MB}}_{n}-\widehat{\theta}_{n}) & =-J^{-1}\frac{1}{\sqrt{n}}\sum^{n}_{i=1}(M_{i}-1)\psi(Y_{i})+o_{\mathbb{P}^{\mathrm{MB}}}(1).\label{eq:weighted-M-expansion}
\end{align}
The leading term in the first expansion converges unconditionally
to $N_{d}(0,J^{-1}VJ^{-1})$, while the second converges conditionally
to the same limit, in $\mathbb{P}$-probability.

\begin{assumption}[Uniform posterior normality] \label{as:bvm} Let
$\Omega=[\underline{\omega},\overline{\omega}]\subset(0,\infty)$
be compact. The prior is positive and continuous near $\theta^{\dagger}$.
Let
\[
\mathcal{A}^{\mathrm{MB}}_{n}=\{\pi^{\mathrm{MB}}_{\omega}(\cdot\mid M,\mathcal{D}_{n})\text{ is proper for every }\omega\in\Omega\}.
\]
The original-data posterior is proper for every $\omega\in\Omega$,
and $\mathbb{P}^{\mathrm{MB}}\left\{ \mathcal{A}^{\mathrm{MB}}_{n}\right\} \xrightarrow{\mathbb{P}}1$.
On these proper-posterior events, the following uniform Bernstein--von
Mises approximations hold. For the original data,
\[
\sup_{\omega\in\Omega}d_{\mathrm{TV}}\left(\mathcal{L}_{\pi_{\omega}}\{\sqrt{n}(\theta-\widehat{\theta}_{n})\},N_{d}(0,\omega^{-1}J^{-1})\right)\xrightarrow{\mathbb{P}}0.
\]
For the bootstrap posterior and every $u>0$,
\[
\mathbb{P}^{\mathrm{MB}}\left\{ \mathcal{A}^{\mathrm{MB}}_{n}\cap\left\{ \sup_{\omega\in\Omega}d_{\mathrm{TV}}\left(\mathcal{L}_{\pi^{\mathrm{MB}}_{\omega}}\{\sqrt{n}(\theta-\widehat{\theta}^{\mathrm{MB}}_{n})\},N_{d}(0,\omega^{-1}J^{-1})\right)>u\right\} \right\} \xrightarrow{\mathbb{P}}0.
\]

\end{assumption}

\begin{assumption}[Regular scalar functional] \label{as:functional}
The functional $g$ is continuously differentiable near $\theta^{\dagger}$.
Let $\dot{g}=\nabla g(\theta^{\dagger})$ and 
\[
A_{g}=\dot{g}^{\top}J^{-1}\dot{g},\qquad B_{g}=\dot{g}^{\top}J^{-1}VJ^{-1}\dot{g},
\]
with $A_{g}>0$ and $B_{g}>0$. \end{assumption} 

\begin{lemma}[Uniform posterior quantiles and median]\label{lem:functional-quantiles}
Suppose Assumptions~\ref{as:mest}--\ref{as:functional} hold. For
each fixed $\beta\in(0,1)$, let $q_{\beta,n}(\omega)$ denote the
lower $\beta$-quantile of $g(\theta)$ under the original posterior.
On the bootstrap propriety event in Assumption~\ref{as:bvm}, let
$q^{\mathrm{MB}}_{\beta,n}(\omega)$ denote the corresponding bootstrap
quantile, and define it arbitrarily otherwise. Then 
\begin{equation}
\sup_{\omega\in\Omega}\left|\sqrt{n}\{q_{\beta,n}(\omega)-g(\widehat{\theta}_{n})\}-z_{\beta}\sqrt{\frac{A_{g}}{\omega}}\right|\xrightarrow{\mathbb{P}}0,\label{eq:uniform-original-quantile}
\end{equation}
and 
\begin{equation}
\sup_{\omega\in\Omega}\left|\sqrt{n}\{q^{\mathrm{MB}}_{\beta,n}(\omega)-g(\widehat{\theta}^{\mathrm{MB}}_{n})\}-z_{\beta}\sqrt{\frac{A_{g}}{\omega}}\right|=o_{\mathbb{P}^{\mathrm{MB}}}(1)\label{eq:uniform-bootstrap-quantile}
\end{equation}
in $\mathbb{P}$-probability. In particular, since $\check{g}_{n}(\omega)=q_{1/2,n}(\omega)$,
\begin{equation}
\sup_{\omega\in\Omega}\sqrt{n}\,|\check{g}_{n}(\omega)-g(\widehat{\theta}_{n})|\xrightarrow{\mathbb{P}}0.\label{eq:median-erm-equivalence}
\end{equation}
 \end{lemma}

Equation~\eqref{eq:median-erm-equivalence} links the implemented
median-centered calibration criterion to the first-order theory. We
can therefore establish the common first-order limit of the population
and bootstrap coverage curves.

\begin{proposition}[First-order population and bootstrap coverage]
\label{prp:common-limit} Suppose Assumptions~\ref{as:mest}--\ref{as:functional}
hold and define 
\[
H_{\alpha}(\omega)=\Phi\left(z_{\beta_{U}}\sqrt{\frac{A_{g}}{\omega B_{g}}}\right)-\Phi\left(z_{\beta_{L}}\sqrt{\frac{A_{g}}{\omega B_{g}}}\right).
\]
Then 
\begin{equation}
\sup_{\omega\in\Omega}|c_{n,\alpha}(\omega)-H_{\alpha}(\omega)|\to0,\qquad\sup_{\omega\in\Omega}|c^{\mathrm{MB}}_{n,\alpha}(\omega)-H_{\alpha}(\omega)|=o_{\mathbb{P}}(1).\label{eq:common-H}
\end{equation}
Consequently, 
\begin{equation}
\sup_{\omega\in\Omega}|c^{\mathrm{MB}}_{n,\alpha}(\omega)-c_{n,\alpha}(\omega)|=o_{\mathbb{P}}(1).\label{eq:bootstrap-pop-equivalence}
\end{equation}
\end{proposition}

Since $z_{\beta_{L}}<0<z_{\beta_{U}}$, 
\[
H_{\alpha}'(\omega)=-\frac{t(\omega)}{2\omega}\{z_{\beta_{U}}\phi(z_{\beta_{U}}t(\omega))-z_{\beta_{L}}\phi(z_{\beta_{L}}t(\omega))\}<0,\qquad t(\omega)=\sqrt{\frac{A_{g}}{\omega B_{g}}},
\]
 where $\phi$ is the standard normal density. Moreover, 
\[
H_{\alpha}(A_{g}/B_{g})=\beta_{U}-\beta_{L}=1-\alpha.
\]
Thus the unique first-order target is $\omega^{\circ}_{g}=A_{g}/B_{g}$,
provided it belongs to $\operatorname{int}(\Omega)$. The target $\omega^{\circ}_{g}$
is specific to $g$; simultaneous calibration of multiple directions
generally requires stronger covariance proportionality. Neither the
tail allocation nor median-centering changes its first-order value.

The weighted-Fisher rule of \citet{Onizuka2026} is $\omega_{\mathrm{WS}}(M)=\operatorname{tr}(JM)/\operatorname{tr}(VM)$
(in our notation). Choosing the positive-semidefinite rank-one weight
$M_{g}=J^{-1}\dot{g}\dot{g}^{\top}J^{-1}$ gives $\omega_{\mathrm{WS}}(M_{g})=A_{g}/B_{g}=\omega^{\circ}_{g}$.
This is an equality of first-order targets, not of finite-sample plug-in
and bootstrap-calibrated rates.

Because a finite-sample bootstrap coverage curve need not have an
exact root, let a measurable $\widetilde{\omega}^{\mathrm{MB}}\in\Omega$
satisfy 
\begin{equation}
|c^{\mathrm{MB}}_{n,\alpha}(\widetilde{\omega}^{\mathrm{MB}})-(1-\alpha)|\le\inf_{\omega\in\Omega}|c^{\mathrm{MB}}_{n,\alpha}(\omega)-(1-\alpha)|+\rho_{n},\qquad\rho_{n}\geq0,\quad\rho_{n}=o_{\mathbb{P}}(1).\label{eq:exact-M-approx-root}
\end{equation}
\begin{theorem}[First-order validity of bootstrap calibration]\label{thm:bootstrap-validity}
Suppose Assumptions~\ref{as:mest}--\ref{as:functional} hold and
$\omega^{\circ}_{g}\in\operatorname{int}(\Omega)$. Any $\widetilde{\omega}^{\mathrm{MB}}$
satisfying \eqref{eq:exact-M-approx-root} obeys $\ensuremath{\widetilde{\omega}^{\mathrm{MB}}\xrightarrow{\mathbb{P}}\omega^{\circ}_{g}}$,
and 
\begin{equation}
\mathbb{P}\{g^{\dagger}\in C_{\alpha,\widetilde{\omega}^{\mathrm{MB}}}(\mathcal{D}_{n})\}\longrightarrow1-\alpha.\label{eq:actual-exact-coverage}
\end{equation}
\end{theorem}

These results also extend to independent-cluster sampling. For i.i.d.~clusters,
the original-data posterior can be written as \eqref{eq:cluster-posterior},
so that the preceding analysis applies with the cluster as the independent
sampling unit. Define
\[
\psi^{\mathrm{CL}}(\mathcal{Y}_{c})=\nabla_{\theta}\ell^{\mathrm{CL}}(\theta^{\mathrm{CL},\dagger};\mathcal{Y}_{c}).
\]
\[
J^{\mathrm{CL}}=\mathbb{E}_{P^{\mathrm{CL}}}\left[\nabla^{2}_{\theta}\ell^{\mathrm{CL}}(\theta^{\mathrm{CL},\dagger};\mathcal{Y}_{c})\right],\qquad V^{\mathrm{CL}}=\mathbb{E}_{P^{\mathrm{CL}}}\left[\psi^{\mathrm{CL}}(\mathcal{Y}_{c})\psi^{\mathrm{CL}}(\mathcal{Y}_{c})^{\top}\right].
\]
Thus $V^{\mathrm{CL}}$ incorporates arbitrary dependence among observations
within a cluster. We index the asymptotics by $G$, the number of
independent clusters, with the $\sqrt{G}$ normalization. Let $\check{g}^{\mathrm{CL}}_{G}(\omega)$
denote the median of $g(\theta)$ under the original-data posterior,
and let $C^{\mathrm{CMB}}_{\alpha,\omega}(m^{\mathrm{CMB}})$ denote
the corresponding quantile-based CMB credible interval with levels
$\beta_{L},\beta_{U}$. Define 
\[
c^{\mathrm{CMB}}_{G,\alpha}(\omega)=\mathbb{E}^{\mathrm{CMB}}\left[\mathbb{I}\left\{ \check{g}^{\mathrm{CL}}_{G}(\omega)\in C^{\mathrm{CMB}}_{\alpha,\omega}(M^{\mathrm{CMB}})\right\} \right],
\]
where $\mathbb{E}^{\mathrm{CMB}}$ is conditional on $\mathcal{D}^{\mathrm{CL}}_{G}$.
The population counterpart is 
\[
c^{\mathrm{CL}}_{G,\alpha}(\omega)=\mathbb{P}\left\{ g(\theta^{\mathrm{CL},\dagger})\in C_{\alpha,\omega}(\mathcal{D}^{\mathrm{CL}}_{G})\right\} .
\]

\begin{corollary}[Independent-cluster extension] \label{cor:cluster-extension}
Suppose that the cluster-level analogues of Assumptions~\ref{as:mest}--\ref{as:functional}
hold with $G$, $\mathcal{Y}_{c}$, and $\ell^{\mathrm{CL}}(\theta;\mathcal{Y}_{c})$
replacing $n$, $Y_{i}$, and $\ell(\theta;Y_{i})$, respectively.
Let 
\[
A^{\mathrm{CL}}_{g}=\dot{g}^{\top}(J^{\mathrm{CL}})^{-1}\dot{g},\qquad B^{\mathrm{CL}}_{g}=\dot{g}^{\top}(J^{\mathrm{CL}})^{-1}V^{\mathrm{CL}}(J^{\mathrm{CL}})^{-1}\dot{g},
\]
where $\dot{g}=\nabla g(\theta^{\mathrm{CL},\dagger})$, and define
\[
H^{\mathrm{CL}}_{\alpha}(\omega)=\Phi\left(z_{\beta_{U}}\sqrt{\frac{A^{\mathrm{CL}}_{g}}{\omega B^{\mathrm{CL}}_{g}}}\right)-\Phi\left(z_{\beta_{L}}\sqrt{\frac{A^{\mathrm{CL}}_{g}}{\omega B^{\mathrm{CL}}_{g}}}\right).
\]
Then 
\[
\sup_{\omega\in\Omega}\left|c^{\mathrm{CL}}_{G,\alpha}(\omega)-H^{\mathrm{CL}}_{\alpha}(\omega)\right|\longrightarrow0,
\]
and 
\[
\sup_{\omega\in\Omega}\left|c^{\mathrm{CMB}}_{G,\alpha}(\omega)-H^{\mathrm{CL}}_{\alpha}(\omega)\right|=o_{\mathbb{P}}(1).
\]
Consequently, 
\[
\sup_{\omega\in\Omega}\left|c^{\mathrm{CMB}}_{G,\alpha}(\omega)-c^{\mathrm{CL}}_{G,\alpha}(\omega)\right|=o_{\mathbb{P}}(1).
\]
The unique first-order calibration target is therefore 
\[
\omega^{\mathrm{CL},\circ}_{g}=\frac{A^{\mathrm{CL}}_{g}}{B^{\mathrm{CL}}_{g}},
\]
provided that $\omega^{\mathrm{CL},\circ}_{g}\in\operatorname{int}(\Omega)$.
Moreover, any CMB approximate root satisfying the analogue of \eqref{eq:exact-M-approx-root}
converges in probability to $\omega^{\mathrm{CL},\circ}_{g}$ and
yields 
\[
\mathbb{P}\left\{ g(\theta^{\mathrm{CL},\dagger})\in C_{\alpha,\widetilde{\omega}^{\mathrm{CMB}}}(\mathcal{D}^{\mathrm{CL}}_{G})\right\} \longrightarrow1-\alpha.
\]
\end{corollary}

Within-cluster dependence may be arbitrary provided the cluster-level
moment, smoothness, and Hessian conditions hold. The result assumes
i.i.d.~clusters; extensions to independent, non-identically distributed
clusters would require corresponding triangular-array conditions and
are not pursued here.

\subsection{Consistency of the computed criterion}

\label{subsec:consist-num-app}

Numerical evaluation of the bootstrap calibration criterion requires
approximating both the learning-rate-dependent posterior center and
the bootstrap posterior probabilities evaluated at that center.

\begin{assumption}[Stable numerical evaluation] \label{as:is} With
$\mathbb{P}$-probability tending to one, the following hold conditionally
on the observed data for each pair $(\omega,m)$ in any fixed finite
collection under consideration.
\begin{enumerate}
\item The reweighted original-data posterior CDF consistently estimates
$F^{0}_{n,\omega}$ in a neighborhood of $\check{g}_{n}(\omega)$,
so that 
\begin{equation}
\check{g}_{n,S}(\omega)\longrightarrow\check{g}_{n}(\omega)\label{eq:center-consistency}
\end{equation}
in conditional probability. The posterior CDF is continuous and strictly
increasing in a neighborhood of $\check{g}_{n}(\omega)$.
\item The retained bootstrap CDF estimator, whether obtained by importance
reweighting or a safeguard, is locally uniformly consistent around
$\check{g}_{n}(\omega)$: 
\begin{equation}
\sup_{|x-\check{g}_{n}(\omega)|\le\kappa}|\widehat{F}^{\mathrm{comp}}_{n,\omega,m;S,K}(x)-F_{n,\omega,m}(x)|\longrightarrow0\label{eq:safeguard-consistency}
\end{equation}
for some $\kappa>0$, in conditional probability as $S,K\rightarrow\infty$.
\item There are no ties at the coverage cutoffs:
\[
\mathbb{P}^{\mathrm{MB}}\left\{ F_{n,\omega,M}(\check{g}_{n}(\omega))\in\{\beta_{L},\beta_{U}\}\right\} =0.
\]
\end{enumerate}
\end{assumption} For i.i.d.~bootstrap draws $M^{(1)},M^{(2)},\ldots$,
define 
\begin{equation}
c^{\mathrm{MB}}_{n,\alpha;B}(\omega)=\frac{1}{B}\sum^{B}_{b=1}\mathbb{I}\left\{ \beta_{L}\le F_{n,\omega,M^{(b)}}(\check{g}_{n}(\omega))\le\beta_{U}\right\} .\label{eq:finite-B-exact}
\end{equation}

\begin{proposition}[Consistency of the computed criterion] \label{prp:computation}
Let $\mathcal{E}\subset\Omega$ be a fixed finite set and suppose
Assumption~\ref{as:is} holds. For fixed $B$, on the event that
all posteriors indexed by $\mathcal{E}\times\{M^{(1)},\ldots,M^{(B)}\}$
are proper and none of the corresponding CDF values equals a cutoff,
\begin{equation}
\max_{\omega\in\mathcal{E}}|\widehat{c}^{\mathrm{MB}}_{n,\alpha;S,K,B}(\omega)-c^{\mathrm{MB}}_{n,\alpha;B}(\omega)|\longrightarrow0\label{eq:S-consistency}
\end{equation}
in conditional probability as the numerical budgets diverge. Moreover,
\begin{equation}
\max_{\omega\in\mathcal{E}}|c^{\mathrm{MB}}_{n,\alpha;B}(\omega)-c^{\mathrm{MB}}_{n,\alpha}(\omega)|\longrightarrow0\label{eq:B-consistency}
\end{equation}
almost surely, conditionally on the data, as $B\to\infty$. \end{proposition}

Proposition~\ref{prp:computation} establishes consistency on fixed
finite evaluation sets. For adaptively selected learning rates, Corollary~\ref{cor:computed}
instead imposes the numerical accuracy conditions in \eqref{eq:computed-root}
and \eqref{eq:computed-error}, without requiring a separate convergence
theory for the search recursion.

\begin{corollary}[Validity under vanishing numerical error]\label{cor:computed}
Suppose the assumptions of Theorem~\ref{thm:bootstrap-validity}
hold. Let $\widehat{\omega}^{\mathrm{MB}}\in\Omega$ be the output
of a numerical calibration procedure, and suppose
\begin{equation}
|\widehat{c}^{\mathrm{MB}}_{n,\alpha;S,K,B}(\widehat{\omega}^{\mathrm{MB}})-(1-\alpha)|=o_{\mathbb{P}_{\mathrm{joint}}}(1),\label{eq:computed-root}
\end{equation}
and 
\begin{equation}
|\widehat{c}^{\mathrm{MB}}_{n,\alpha;S,K,B}(\widehat{\omega}^{\mathrm{MB}})-c^{\mathrm{MB}}_{n,\alpha}(\widehat{\omega}^{\mathrm{MB}})|=o_{\mathbb{P}_{\mathrm{joint}}}(1).\label{eq:computed-error}
\end{equation}
Then $\widehat{\omega}^{\mathrm{MB}}\xrightarrow{\mathbb{P}_{\mathrm{joint}}}\omega^{\circ}_{g}$,
and 
\begin{equation}
\mathbb{P}_{\mathrm{joint}}\{g^{\dagger}\in C_{\alpha,\widehat{\omega}^{\mathrm{MB}}}(\mathcal{D}_{n})\}\longrightarrow1-\alpha.\label{eq:actual-computed-coverage}
\end{equation}
\end{corollary}

Here $C_{\alpha,\widehat{\omega}^{\mathrm{MB}}}(\mathcal{D}_{n})$
is the exact posterior interval at the selected rate; numerical endpoint
estimates must also be consistent. The Monte Carlo budgets and tolerance
may depend on $n$ to satisfy \eqref{eq:computed-root}--\eqref{eq:computed-error}.
Although the importance-weight ESS safeguard improves reliability,
it does not guarantee these conditions. Corollary~\ref{cor:computed}
establishes validity under vanishing numerical error, not convergence
of Algorithm~\ref{alg:gpcirw} itself.

\section{Simulation}

\label{sec:simulation}

We evaluate the statistical and computational performance of GPC-IRW
for independent observations and independent clusters, followed by
a stress test with a non-smooth loss.

\subsection{Independent sampling}

\label{subsec:sim-independet}

We adapt the heteroskedastic linear-regression design of \citet{Wu2023}.
We consider two data-generating processes (DGPs): DGP-1 has homoskedastic
Gaussian errors, whereas DGP-2 has heteroskedastic Gaussian errors.
Inference is based on the quadratic loss
\[
\ell(\theta;(X_{i},Y_{i}))=\frac{1}{2}(Y_{i}-X^{\top}_{i}\theta)^{2}
\]
and the prior $\theta\sim N_{4}(0,10^{2}I_{4})$. The true coefficient
is $\theta_{0}=(1,1,2,-1)^{\top}$. We calibrate the nominal $95\%$
quantile-based credible interval for $g(\theta)=\theta_{1}$, with
$(\beta_{L},\beta_{U})=(0.025,0.975)$. The DGPs are specified in
Section~\ref{sec:add-sim-result} of the Supplementary Material.
We consider sample sizes $n\in\{100,200,400,800\}$. For reference,
the first-order calibration target $\omega^{\circ}_{g}=A_{g}/B_{g}$
derived in Section~\ref{sec:theory} equals $1$ under DGP-1 and
approximately $0.270$ under DGP-2.

We conduct two experiments for each DGP and sample size. The first
assesses empirical coverage and interval length over $1{,}000$ independently
generated datasets. Whenever direct posterior sampling is required,
we generate independent draws from the known Gaussian posterior rather
than using MCMC.

The second experiment compares computational efficiency on $50$ datasets
generated using the first $50$ random seeds from the statistical
experiment. Here we use the No-U-Turn Sampler (NUTS) \citep{Hoffman2014}
for the direct posterior simulations required by GPC-MCMC, for the
initial posterior simulations required by GPC-SMC, and whenever a
direct posterior simulation is triggered by GPC-IRW. For each NUTS
fit, we run a single chain with $1{,}000$ warmup iterations followed
by $5{,}000$ retained draws, so that the posterior samples used by
GPC-IRW have $S=K=5{,}000$. We use the default \texttt{cmdstanpy}
tuning settings, with adaptation performed separately for each posterior
fit. The initial original-data fit starts at the ordinary least-squares
estimator. When posterior draws from an original-data fit are available,
subsequent posterior fits are initialized at the coordinatewise median
of those draws. 

Unless otherwise stated, the procedures use $B=500$, $\Omega=[0.01,10]$,
$\gamma_{l}=l^{-0.51}$, $\omega^{(0)}=0.3$, $\epsilon_{\mathrm{tol}}=0.01$,
and at most $T=50$ coverage evaluations. For the proposed GPC-IRW
procedure, we use $\zeta_{0}=0.5$, $E_{\min}=100$, and $\kappa_{\mathrm{sg}}=2$.
When a procedure requires replicate-specific calculations, we parallelize
them across bootstrap replicates; for GPC-IRW, this includes safeguard-triggered
bootstrap-posterior simulations.

For GPC-SMC, we use $5{,}000$ particles, matching the nominal Monte
Carlo sample size used for the other procedures. \citet{Tanaka2025}
used a target importance-weight ESS parameter of $0.999$ for selecting
intermediate learning rates and noted that this value is near the
upper end of those used in the adaptive SMC literature. We use the
less conservative value $0.95$ as the primary computational setting.
The remaining implementation details of GPC-SMC follow \citet{Tanaka2025}.

We report empirical coverage and interval length from the $1{,}000$
statistical replications, and wall-clock time and calibration iterations
from the $50$ NUTS-based replications. Timing excludes data generation
and preprocessing common to all procedures. For GPC-IRW, we also report
the mean number of safeguard-triggered direct bootstrap-posterior
fits per bootstrap replicate across the $50$ computational runs.
All programs are written in Python and run on an Ubuntu workstation
with an AMD Ryzen Threadripper 9980X processor, using at most $63$
parallel workers.

\begin{figure}
\caption{Empirical coverage of nominal $95\%$ credible intervals: Independent
sampling}

\label{fig:sim-ind-coverage}
\centering{}\subfloat[DGP-1]{
\centering{}\includegraphics[width=0.4\columnwidth]{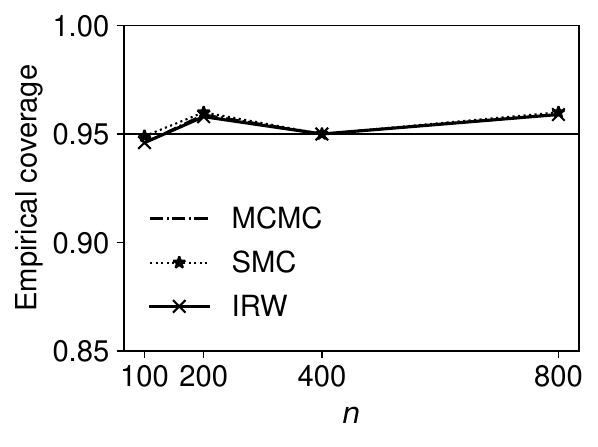}} \subfloat[DGP-2]{
\begin{centering}
\includegraphics[width=0.4\columnwidth]{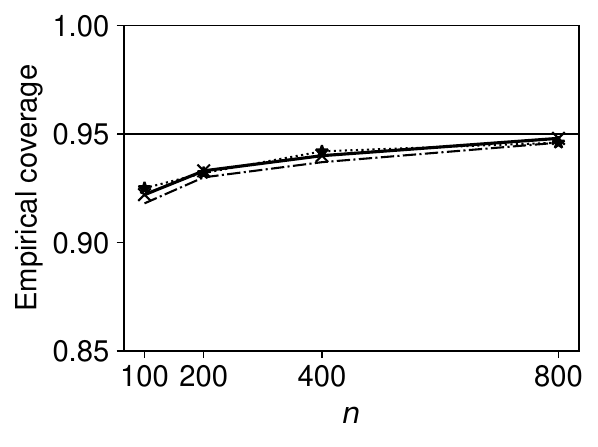}
\par\end{centering}
}
\end{figure}

\begin{figure}
\caption{Length of nominal $95\%$ credible intervals: Independent sampling}

\label{fig:sim-ind-length}
\centering{}\subfloat[DGP-1, $n=100$]{
\centering{}\includegraphics[width=0.2\columnwidth]{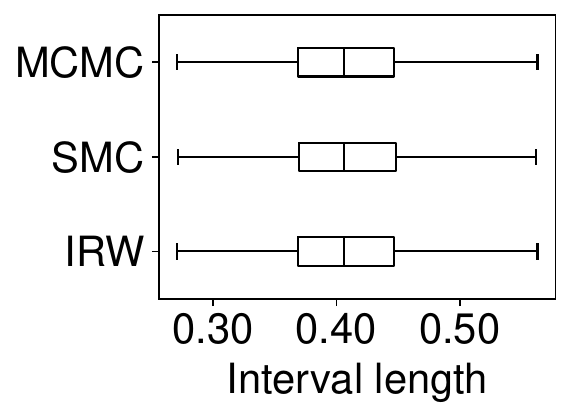}} \subfloat[DGP-1, $n=200$]{
\centering{}\includegraphics[width=0.2\columnwidth]{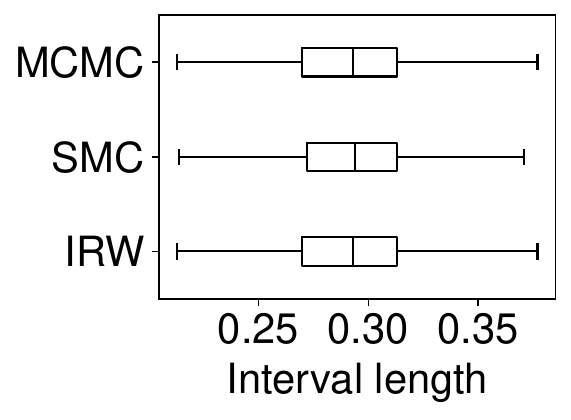}} \subfloat[DGP-1, $n=400$]{
\centering{}\includegraphics[width=0.2\columnwidth]{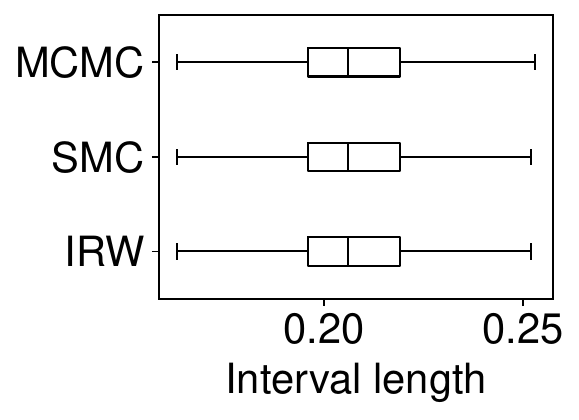}} \subfloat[DGP-1, $n=800$]{
\centering{}\includegraphics[width=0.2\columnwidth]{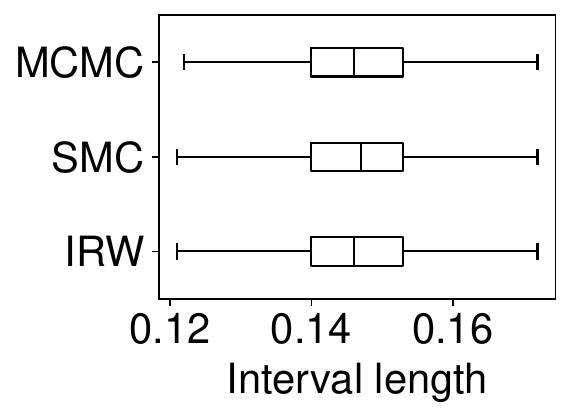}}\\
\subfloat[DGP-2, $n=100$]{
\centering{}\includegraphics[width=0.2\columnwidth]{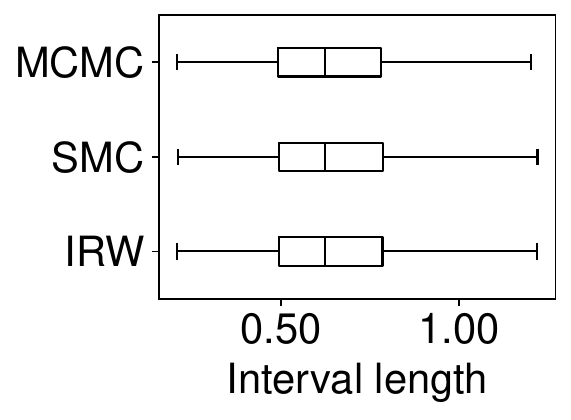}} \subfloat[DGP-2, $n=200$]{
\centering{}\includegraphics[width=0.2\columnwidth]{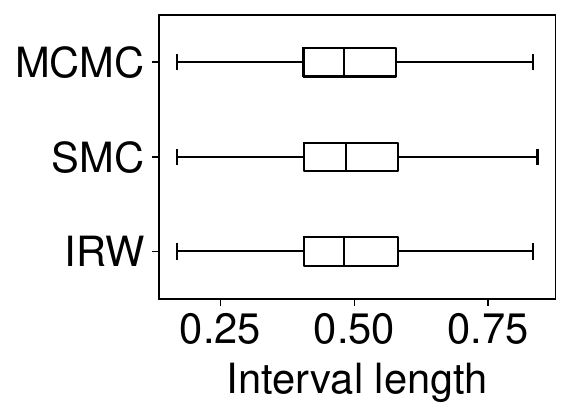}} \subfloat[DGP-2, $n=400$]{
\centering{}\includegraphics[width=0.2\columnwidth]{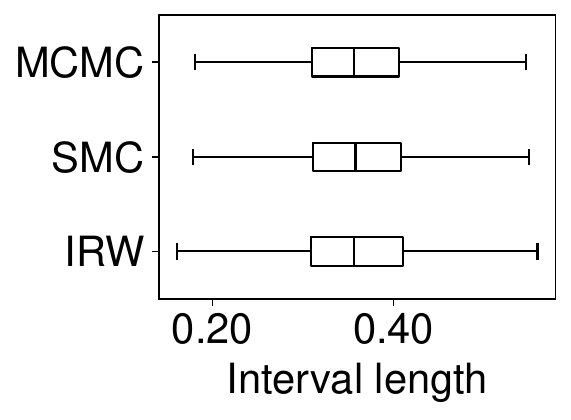}} \subfloat[DGP-2, $n=800$]{
\centering{}\includegraphics[width=0.2\columnwidth]{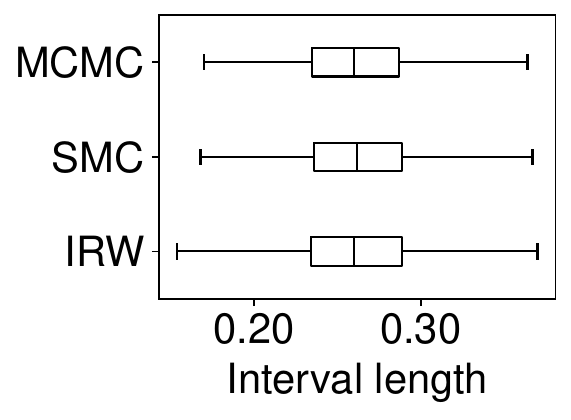}}
\end{figure}

Figures~\ref{fig:sim-ind-coverage} and~\ref{fig:sim-ind-length}
show similar coverage and interval-length distributions across the
three procedures. Under DGP-1, coverage is close to nominal across
sample sizes; under DGP-2, coverage is lower in smaller samples but
approaches the nominal level as $n$ increases. Intervals shorten
with increasing sample size under both DGPs.

\begin{figure}
\caption{Wall-clock time in the $50$ NUTS-based runs: Independent sampling}

\label{fig:sim-ind-time}
\centering{}\subfloat[DGP-1, $n=100$]{
\centering{}\includegraphics[width=0.2\columnwidth]{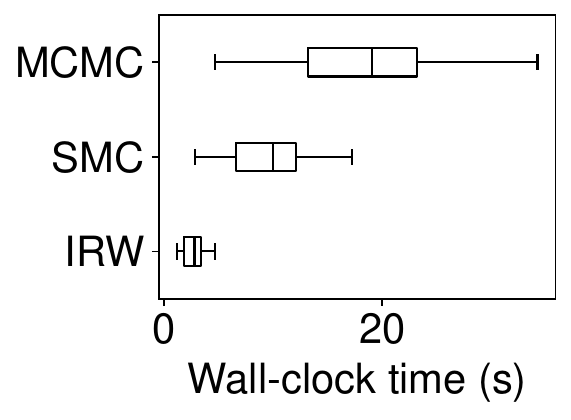}} \subfloat[DGP-1, $n=200$]{
\centering{}\includegraphics[width=0.2\columnwidth]{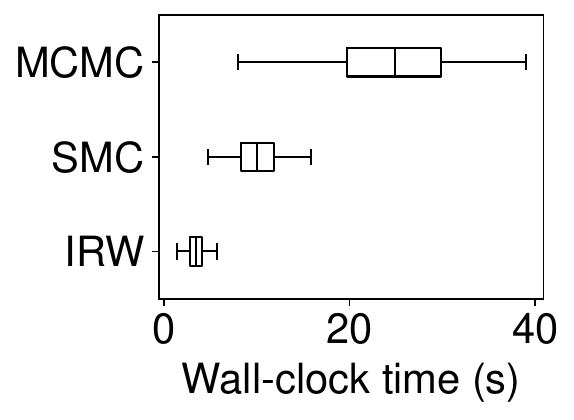}} \subfloat[DGP-1, $n=400$]{
\centering{}\includegraphics[width=0.2\columnwidth]{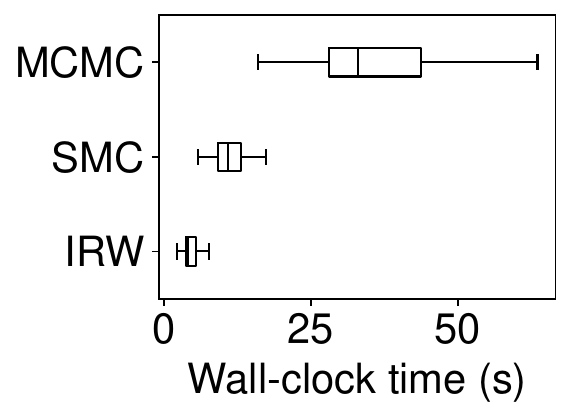}} \subfloat[DGP-1, $n=800$]{
\centering{}\includegraphics[width=0.2\columnwidth]{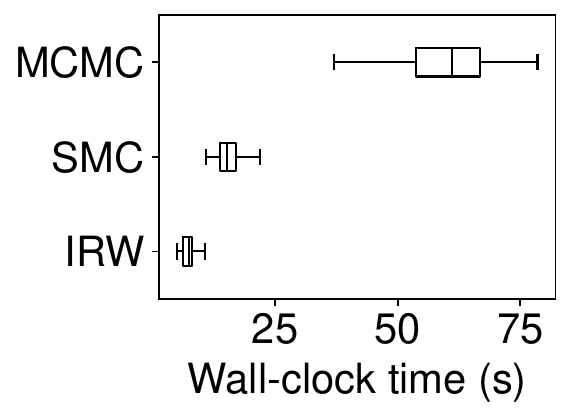}}\\
\subfloat[DGP-2, $n=100$]{
\centering{}\includegraphics[width=0.2\columnwidth]{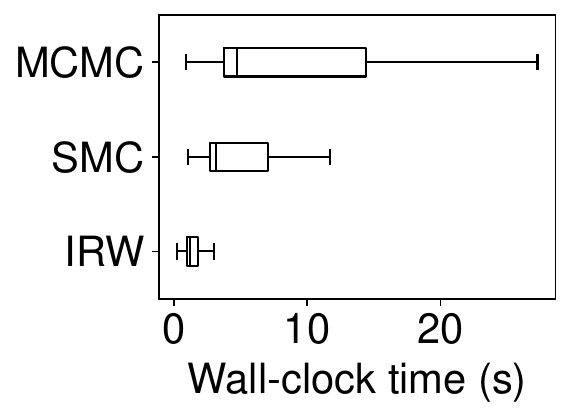}} \subfloat[DGP-2, $n=200$]{
\centering{}\includegraphics[width=0.2\columnwidth]{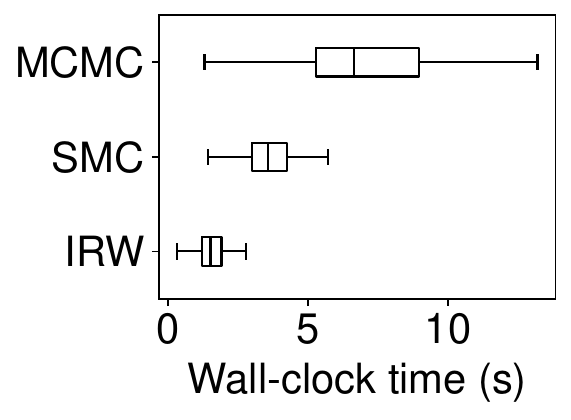}} \subfloat[DGP-2, $n=400$]{
\centering{}\includegraphics[width=0.2\columnwidth]{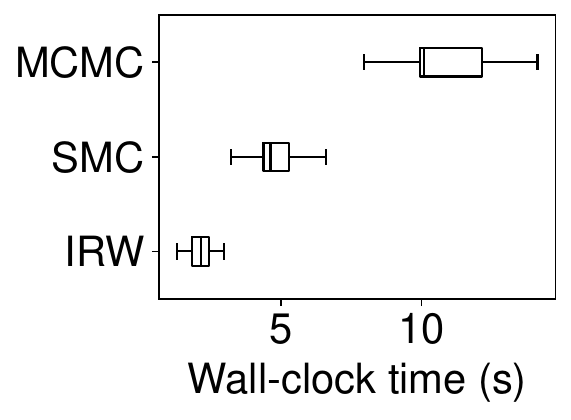}} \subfloat[DGP-2, $n=800$]{
\centering{}\includegraphics[width=0.2\columnwidth]{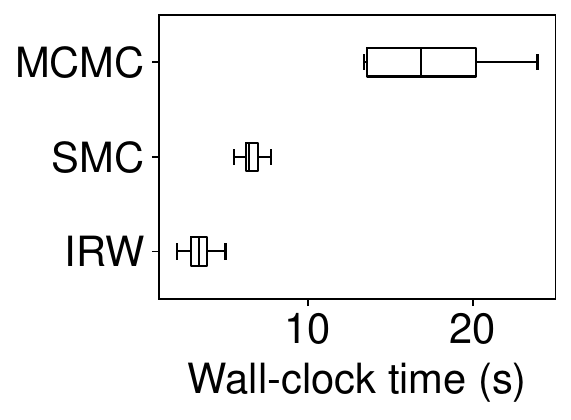}}
\end{figure}

Figure~\ref{fig:sim-ind-time} shows that GPC-IRW is consistently
faster than GPC-MCMC and GPC-SMC across both DGPs and all sample sizes,
with larger gains at larger $n$. Together with its comparable statistical
performance (Figures~\ref{fig:sim-ind-coverage} and~\ref{fig:sim-ind-length}),
this indicates that GPC-IRW reduces computational cost without materially
changing inference. Because calibration iterations are broadly comparable
across procedures (Figure~\ref{fig:sim-ind-iter} in the Supplementary
Material), the speedups arise primarily from less costly coverage
evaluations rather than fewer iterations.

Supplementary Table~B.1 provides further insight into these computational
gains: the average number of safeguard-triggered direct posterior
fits per bootstrap replicate is only $0.20$--$0.39$ across the
settings considered, indicating that most bootstrap coverage evaluations
can be resolved by importance reweighting without additional posterior
simulation.

Supplementary Figures~\ref{fig:sim-ind-omega-dgp1} and~\ref{fig:sim-ind-omega-dgp2}
compare the calibrated learning rates obtained using independent Gaussian
posterior draws with those obtained using NUTS on the same datasets.
The points lie close to the $45$-degree line across all three procedures
and both DGPs, indicating that the use of NUTS has little effect on
the resulting calibration.

\subsection{Independent-cluster sampling}

\label{subsec:sim-cluster}

We next assess the cluster extension of GPC-IRW, using multinomial
bootstrap counts at the cluster level as in Section~\ref{subsec:clustered}.
We consider only GPC-IRW rather than repeating the comparison among
calibration procedures.

We use a clustered analogue of the linear-regression design in the
preceding subsection. We retain the quadratic loss, true parameters,
Gaussian prior, and target functional from the preceding subsection:
\[
\ell^{\mathrm{CL}}(\theta;\mathcal{Y}_{c})=\frac{1}{2}\sum^{m}_{j=1}(Y_{cj}-X^{\top}_{cj}\theta)^{2},\qquad\theta\sim N_{4}(0,10^{2}I_{4}),
\]
with $\mathcal{Y}_{c}=\{(X_{cj},Y_{cj}):j=1,\ldots,m\}$. We calibrate
the nominal $95\%$ quantile-based credible interval for $g(\theta)=\theta_{1}$,
with $(\beta_{L},\beta_{U})=(0.025,0.975)$. The details of the DGP
are given in Section~\ref{sec:add-sim-result} of the Supplementary
Material. The first-order calibration target is $\omega^{\mathrm{CL},\circ}_{g}=0.5$.
We consider $G\in\{20,40,80,160\}$, corresponding to total sample
sizes $n_{G}\in\{100,200,400,800\}$. These values match the total
sample sizes considered in the preceding subsection, although the
number of independent sampling units is now $G$. Unless otherwise
stated, we use the same calibration and importance-reweighting settings
as in the independent-sampling study. 

For each $G$, we assess empirical coverage over $1{,}000$ independently
generated datasets, using independent Gaussian posterior draws as
above. We also conduct a NUTS-based computational experiment using
the first $50$ replication seeds and the settings above.

\begin{figure}
\caption{Empirical coverage of nominal $95\%$ credible intervals: Independent-cluster
sampling}

\label{fig:sim-clust-coverage}
\centering{}\includegraphics[width=0.4\columnwidth]{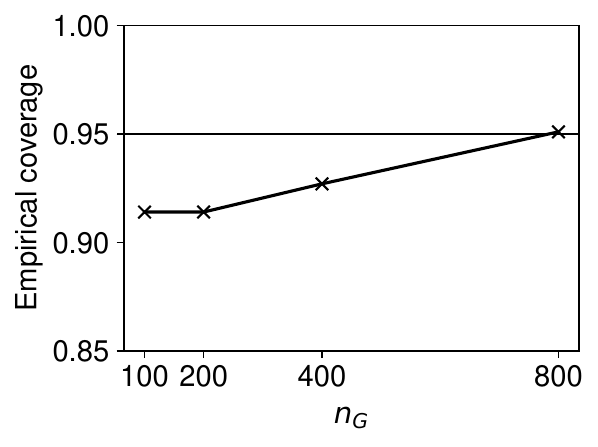}
\end{figure}

Figure~\ref{fig:sim-clust-coverage} shows that coverage approaches
the nominal level as $G$ increases, with some undercoverage at smaller
$G$, consistent with Corollary~\ref{cor:cluster-extension}. In
the $50$ computational runs, Gaussian-draw and NUTS calibrations
agree closely (Supplementary Figure~\ref{fig:sim-clust-omega}).
Median wall-clock time increases from $3.25$ to $13.98$ seconds
as $n_{G}$ increases from $100$ to $800$, while median calibration
iterations remain $11$, suggesting higher per-iteration cost rather
than more iterations. Safeguard-triggered direct fits average $0.181$--$0.385$
per replicate (Supplementary Table~\ref{tab:sim-clust-safeguard}),
supporting the effectiveness of importance reweighting under independent-cluster
sampling.

\subsection{Stress test with a non-smooth loss: support vector machines}

\label{subsec:sim-svm}

Although the first-order theory in Section~\ref{sec:theory} assumes
smooth losses, the GPC-IRW algorithm itself does not require loss
differentiability. We therefore consider a binary-classification example
using a support vector machine with hinge loss, adapting the design
of \citet{Martin2022}. We calibrate nominal $95\%$ quantile-based
credible intervals for the slope coefficient over $n\in\{100,200,400,800\}$
and compare GPC-IRW with GPC-MCMC. The DGP and implementation details
are given in Section~\ref{sec:stress-test} of the Supplementary
Material.

Supplementary Figures~\ref{fig:svm-coverage} and~\ref{fig:svm-length}
show nearly identical coverage and similar interval lengths, with
coverage generally near nominal and shorter intervals as $n$ increases.
GPC-IRW is faster than GPC-MCMC (Supplementary Figure~\ref{fig:svm-time}),
with comparable calibration iterations (Supplementary Figure~\ref{fig:svm-iter})
and an average of $0.117$--$0.138$ safeguard-triggered direct fits
per replicate (Supplementary Table~\ref{tab:svm-safeguard}). These
findings suggest that GPC-IRW may also be applicable to non-smooth
losses in settings such as this one, although its validity in such
cases is not established by the present theory.

\section{Application}

\label{sec:application}

\subsection{General empirical design}

The exoplanet mass--radius relationship informs planetary bulk composition
and structure, but its substantial intrinsic scatter motivates flexible
modeling \citep{Wolfgang2016,ChenKipping2017,Ning2018}. Reported
masses may also be subject to method-dependent selection effects \citep{Steffen2016,Mills2017},
while planets within the same host may have correlated properties
\citep{Millholland2017,Weiss2018V}. We therefore fit a flexible mass--radius
trend with a robust loss and compare observation-level and host-level
bootstrap calibration to assess GPC-IRW's computational performance
and the sensitivity of posterior uncertainty to within-host dependence.

We obtain exoplanet mass and radius measurements from the NASA Exoplanet
Archive's composite-parameter table \citep{Akeson2013,Christiansen2025,NEA2026PSCompPars}.
We select transiting planets orbiting single stars, requiring each
selected planet to have a positive reported mass measurement (rather
than a limit or minimum-mass estimate) and a reported radius between
$1$ and $4$ Earth radii. We also exclude archive-calculated mass
or radius values. Selection details are given in Section~\ref{sec:exoplanet-detail}
of the Supplementary Material.

Let $r_{i}=R^{\mathrm{obs}}_{i}/R_{\oplus}$ and $y_{i}=\log(M^{\mathrm{obs}}_{i}/M_{\oplus})$.
We fit the robust log-mass trend $f_{\theta}(r)=x(r)^{\top}\theta$,
where $x(r)$ contains an intercept and a natural cubic spline basis
in $\log r$. The boundary knots are at $\log1$ and $\log4$, and
the interior knots are at $\log1.5$, $\log2$, and $\log3$. We assign
independent $N(0,10^{2})$ priors to the coefficients and use the
pseudo-Huber loss \citep{Huber1973,Charbonnier1997}
\[
\ell^{\mathrm{pH}}_{\eta}(\theta;y_{i},r_{i})=\eta^{2}\left[\sqrt{1+\{y_{i}-f_{\theta}(r_{i})\}^{2}/\eta^{2}}-1\right],\qquad\eta=\log2.
\]
We calibrate the quantile-based $95\%$ credible interval for the
scalar functional 
\begin{equation}
g(\theta)=f_{\theta}(3)-f_{\theta}(1.5),\label{eq:target-functional}
\end{equation}
with $(\beta_{L},\beta_{U})=(0.025,0.975)$. We then report the posterior
median and corresponding interval for $\exp\{g(\theta)\}$, the ratio
of fitted masses at radii of $3$ and $1.5$ Earth radii. We use $B=500$,
$S=K=5{,}000$, and NUTS for direct posterior simulation; further
settings are in Section~\ref{subsec:exoplanet-model} of the Supplementary
Material. GPC-MCMC and GPC-IRW use identical bootstrap replicates
when comparing the same calibration criterion.

\subsection{Single-planet systems: independent sampling}

\label{subsec:app-independent}

We first consider $113$ planets from $113$ hosts, each with exactly
one confirmed planet. Treating the hosts as independent, we use the
observation-level multinomial bootstrap (MB) for calibration. The
initial learning rate is $\omega^{(0)}=6$, selected by a preliminary
grid search whose cost is excluded from the reported wall-clock times.

\begin{figure}
\caption{Fitted mass--radius curves}

\label{fig:single-fit}
\centering{}\includegraphics[width=0.6\columnwidth]{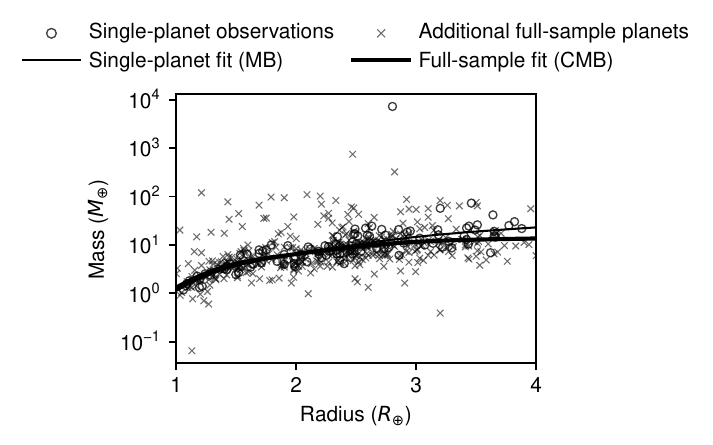}
\end{figure}

Figure~\ref{fig:single-fit} shows the fitted curves, and Table~\ref{tab:exo-summary}
summarizes the mass-ratio estimates. GPC-MCMC-MB and GPC-IRW-MB select
learning rates of $7.204$ and $7.840$, respectively, and yield identical
mass-ratio medians of $3.954$, with similar $95\%$ intervals that
are considerably narrower than the uncalibrated interval. Despite
requiring more calibration iterations ($101$ versus $51$; Supplementary
Table~\ref{tab:exo-add-table}), GPC-IRW-MB reduces computation time
from $2{,}653.7$ to $396.0$ seconds, a $6.7$-fold speedup (an $85\%$
reduction). Over the entire calibration run, the safeguards trigger
an average of $0.462$ direct bootstrap-posterior fits per replicate
(Supplementary Table~\ref{tab:exo-add-table}).

\begin{table}
\caption{Summary of exoplanet applications}
\label{tab:exo-summary}
\centering{}%
\begin{tabular}{lrrcr}
\hline 
Calibration & Learning rate & \multicolumn{2}{c}{$\exp\{g(\theta)\}$} & Wall-clock time (s)\tabularnewline
\cline{3-4}
 &  & Median & $95\%$ interval  & \tabularnewline
\hline 
(a) Single-planet systems &  &  &  & \tabularnewline
Uncalibrated & 1.000  & 4.074 & {[}2.106, 7.694{]} & 3.1\tabularnewline
GPC-MCMC-MB & 7.204 & 3.954 & {[}3.138, 5.032{]} & 2653.7\tabularnewline
GPC-IRW-MB & 7.840 & 3.954 & {[}3.151, 4.973{]} & 396.0\tabularnewline
\multicolumn{5}{l}{(b) All eligible systems}\tabularnewline
Uncalibrated & 1.000 & 2.845 & {[}2.100, 3.847{]} & 14.82\tabularnewline
GPC-MCMC-CMB & 3.564 & 2.809 & {[}2.402, 3.311{]} & 7252.8\tabularnewline
GPC-IRW-MB & 3.856 & 2.809 & {[}2.412, 3.279{]} & 913.7\tabularnewline
GPC-IRW-CMB & 3.612 & 2.825 & {[}2.389, 3.325{]} & 758.2\tabularnewline
\hline 
\end{tabular}
\end{table}

\subsection{All eligible systems: independent-cluster sampling}

\label{subsec:app-cluster}

We next analyze all $521$ eligible planets from $347$ hosts. Because
planets within the same host may be dependent \citep{Weiss2018V,Millholland2017},
we calibrate by resampling entire hosts using the cluster multinomial
bootstrap (CMB). The original-data posterior still sums planet-level
losses, retaining an observation-weighted target; only the bootstrap
unit changes. We set the initial learning rate to $\omega^{(0)}=3$.

For comparison, we also apply observation-level MB to the full sample.
GPC-IRW-MB selects a higher learning rate than GPC-IRW-CMB ($3.856$
versus $3.612$) and gives a narrower $95\%$ mass-ratio interval
($[2.412,3.279]$ versus $[2.389,3.325]$). At any fixed learning
rate, the two schemes have the same original-data posterior; their
difference arises from calibration with different resampling units.

Under host-level calibration, GPC-MCMC-CMB and GPC-IRW-CMB yield similar
learning rates ($3.564$ and $3.612$), medians ($2.809$ and $2.825$),
and intervals (Table~\ref{tab:exo-summary}). GPC-IRW-CMB reduces
computation time from $7{,}252.8$ to $758.2$ seconds, a $9.6$-fold
speedup (a $90\%$ reduction), with $35$ versus $32$ calibration
iterations. Its safeguards trigger an average of $0.486$ direct bootstrap-posterior
fits per replicate (Supplementary Table~\ref{tab:exo-add-table}).
Differences between the single-planet and full-sample fitted curves
in Figure~\ref{fig:single-fit} should also be interpreted in light
of their different sample compositions.

\section{Conclusion}

\label{sec:conclusion}

We developed GPC-IRW to reduce the computational cost of generalized
posterior calibration by reusing posterior draws across bootstrap
resamples and learning rates. A CDF-based coverage criterion avoids
computing bootstrap-specific credible intervals, while diagnostic
safeguards invoke additional posterior simulation when importance
reweighting is unreliable. We established that median-centered calibration
preserves the first-order learning-rate target and that sufficiently
accurate numerical solutions achieve asymptotically nominal coverage
under regularity conditions. The framework also accommodates independent-cluster
sampling through cluster-level resampling.

Simulations demonstrated coverage and interval lengths comparable
to those of existing calibration methods at substantially lower computational
cost. In the exoplanet application, GPC-IRW achieved $6.7$- and $9.6$-fold
speedups over GPC-MCMC in the single-planet and full-sample analyses,
respectively, while highlighting the importance of accounting for
within-host dependence. Although computational gains depend on adequate
overlap between reference and target posteriors, diagnostic safeguards
help identify cases requiring additional simulation. Extending the
theoretical guarantees to non-smooth losses and more general credible
regions remains an important direction for future research.

\section*{Acknowledgement}

The author is grateful to Shonosuke Sugasawa for suggesting the use
of importance reweighting to accelerate generalized posterior calibration
and for valuable discussions on this topic.

\section*{Funding}

This work was supported by JSPS KAKENHI Grant Number 25K21168. The
funder had no involvement in the study design, analysis, interpretation,
or manuscript preparation.

\section*{Disclosure Statement}

The author reports there are no competing interests to declare.

\section*{Declaration of Generative AI and AI-assisted Technologies in the
Manuscript Preparation Process}

During the preparation of this work, the author used ChatGPT 5.6 for
English-language editing. After using this tool, the author reviewed
and edited the content as needed and takes full responsibility for
the content of the manuscript.

\bibliographystyle{apalike2}
\bibliography{reference}

%\begin{comment}


\section*{Supplementary material}

# Supplementary Material for "Efficient Bootstrap Calibration of Generalized Posteriors with Importance Reweighting"

Masahiro Tanaka *

October 27, 2026

## A Proofs for Section 4

### A.1 Preliminary $M$-estimation expansions

We verify the ordinary and bootstrap expansions used in the proof of Proposition 1. Write

$$\dot{\ell}_i(\theta) = \nabla_\theta \ell(\theta; Y_i), \qquad \ddot{\ell}_i(\theta) = \nabla_\theta^2 \ell(\theta; Y_i), \qquad \psi_i = \dot{\ell}_i(\theta^\dagger),$$

and let

$$J_n(\theta) = \frac{1}{n}\sum_{i=1}^n \ddot{\ell}_i(\theta).$$

For the ordinary empirical-risk minimizer,

$$0 = \frac{1}{n}\sum_{i=1}^n \dot{\ell}_i(\widehat{\theta}_n).$$

With $\mathbb{P}$-probability tending to one, $\widehat{\theta}_n$ is interior; conditionally on the data, $\widehat{\theta}_n^{\mathrm{MB}}$ is interior with $\mathbb{P}^{\mathrm{MB}}$-probability tending to one. Applying the fundamental theorem of calculus componentwise gives

$$0 = \frac{1}{n}\sum_{i=1}^n \psi_i + \overline{J}_n(\widehat{\theta}_n - \theta^\dagger), \qquad \overline{J}_n = \int_0^1 J_n\{\theta^\dagger + t(\widehat{\theta}_n - \theta^\dagger)\}\, dt. \tag{A.1}$$

By the empirical-Hessian condition in Assumption 1,

$$\overline{J}_n = J + o_{\mathbb{P}}(1).$$

*Faculty of Economics, Fukuoka University, Fukuoka, Japan. Address: 8-19-1, Nanakuma, Jonan, Fukuoka, Japan 814-0180. E-mail: m.tanaka.tt@fukuoka-u.ac.jp.

Since $J$ is nonsingular,

$$\sqrt{n}(\widehat{\theta}_n - \theta^\dagger) = -J^{-1}\frac{1}{\sqrt{n}}\sum_{i=1}^n \psi_i + o_{\mathbb{P}}(1),$$

which is (10). The multivariate central limit theorem and Slutsky's theorem give

$$\sqrt{n}(\widehat{\theta}_n - \theta^\dagger) \rightsquigarrow N_d(0, J^{-1}VJ^{-1}).$$

We next consider the multinomial-bootstrap minimizer. Define

$$J_{n,M}(\theta) = \frac{1}{n}\sum_{i=1}^n M_i \ddot{\ell}_i(\theta).$$

The bootstrap first-order condition and the fundamental theorem of calculus give

$$0 = \frac{1}{n}\sum_{i=1}^n M_i \dot{\ell}_i(\widehat{\theta}_n) + \overline{J}_{n,M}(\widehat{\theta}_n^{\mathrm{MB}} - \widehat{\theta}_n), \tag{A.2}$$

where

$$\overline{J}_{n,M} = \int_0^1 J_{n,M}\{\widehat{\theta}_n + t(\widehat{\theta}_n^{\mathrm{MB}} - \widehat{\theta}_n)\}dt.$$

By the bootstrap Hessian condition in Assumption 1,

$$\overline{J}_{n,M} = J + o_{\mathbb{P}^{\mathrm{MB}}}(1), \tag{A.3}$$

in $\mathbb{P}$-probability.

Since $n^{-1}\sum_i \dot{\ell}_i(\widehat{\theta}_n) = 0$,

$$\frac{1}{\sqrt{n}}\sum_{i=1}^n M_i \dot{\ell}_i(\widehat{\theta}_n) = \frac{1}{\sqrt{n}}\sum_{i=1}^n (M_i - 1)\dot{\ell}_i(\widehat{\theta}_n).$$

It remains to replace $\dot{\ell}_i(\widehat{\theta}_n)$ by $\psi_i$. Put

$$u_{ni} = \dot{\ell}_i(\widehat{\theta}_n) - \psi_i.$$

For multinomial counts,

$$\mathbb{E}^{\mathrm{MB}}\left[\left\|\frac{1}{\sqrt{n}}\sum_{i=1}^n (M_i - 1)u_{ni}\right\|^2\right] \le \frac{1}{n}\sum_{i=1}^n \|u_{ni}\|^2.$$

By the fundamental theorem of calculus and the envelope condition (9), with probability tending to one,

$$\|u_{ni}\| \le H_\ell(Y_i)\|\widehat{\theta}_n - \theta^\dagger\|.$$

Hence

$$\frac{1}{n}\sum_{i=1}^n \|u_{ni}\|^2 \le \|\widehat{\theta}_n - \theta^\dagger\|^2 \frac{1}{n}\sum_{i=1}^n H_\ell(Y_i)^2 = O_{\mathbb{P}}(n^{-1}),$$

where the last step follows from $\mathbb{E}_P[H_\ell(Y)^2] < \infty$. Conditional Markov's inequality then yields

$$\frac{1}{\sqrt{n}}\sum_{i=1}^{n}(M_i-1)\{\dot{\ell}_i(\widehat{\theta}_n)-\psi_i\} = o_{\mathbb{P}^{\mathrm{MB}}}(1), \tag{A.4}$$

in $\mathbb{P}$-probability. Combining (A.2), (A.3), and (A.4) gives

$$\sqrt{n}(\widehat{\theta}_n^{\mathrm{MB}}-\widehat{\theta}_n) = -J^{-1}\frac{1}{\sqrt{n}}\sum_{i=1}^{n}(M_i-1)\psi_i + o_{\mathbb{P}^{\mathrm{MB}}}(1),$$

which is (11).

For completeness, let $\iota_1,\dots,\iota_n$ be conditionally i.i.d. uniform on $\{1,\dots,n\}$. Then

$$\frac{1}{\sqrt{n}}\sum_{i=1}^{n}(M_i-1)\psi_i = \frac{1}{\sqrt{n}}\sum_{j=1}^{n}(\psi_{\iota_j}-\bar{\psi}_n), \qquad \bar{\psi}_n = \frac{1}{n}\sum_{i=1}^{n}\psi_i.$$

Conditional on the data, the summands on the right are i.i.d. with mean zero and covariance

$$V_n = \frac{1}{n}\sum_{i=1}^{n}(\psi_i-\bar{\psi}_n)(\psi_i-\bar{\psi}_n)^\top,$$

where $V_n \xrightarrow{\mathbb{P}} V$. The moment condition $\mathbb{E}_P[\|\psi(Y)\|^{2+\delta}] < \infty$ implies the conditional Lindeberg condition, so

$$\frac{1}{\sqrt{n}}\sum_{i=1}^{n}(M_i-1)\psi_i \rightsquigarrow N_d(0,V).$$

Therefore

$$\sqrt{n}(\widehat{\theta}_n^{\mathrm{MB}}-\widehat{\theta}_n) \rightsquigarrow N_d(0,J^{-1}VJ^{-1}).$$

Finally, the delta method gives

$$T_n := \sqrt{n}\{g(\widehat{\theta}_n)-g^\dagger\} \rightsquigarrow N(0,B_g), \tag{A.5}$$

and

$$T_n^{\mathrm{MB}} := \sqrt{n}\{g(\widehat{\theta}_n^{\mathrm{MB}})-g(\widehat{\theta}_n)\} \rightsquigarrow N(0,B_g), \tag{A.6}$$

in $\mathbb{P}$-probability, where

$$B_g = \dot{g}^\top J^{-1}VJ^{-1}\dot{g}.$$

## A.2 Proof of Lemma 1

For the original posterior, write $X_{n,\omega} = \sqrt{n}(\theta-\widehat{\theta}_n)$ and let $Z_\omega \sim N_d(0,\omega^{-1}J^{-1})$. Define

$$h_n(x) = \sqrt{n}\left\{g\left(\widehat{\theta}_n+\frac{x}{\sqrt{n}}\right)-g(\widehat{\theta}_n)\right\}.$$

By the contraction property of total variation distance and Assumption 2, the law of $h_n(X_{n,\omega})$ is uniformly close in total variation to that of $h_n(Z_\omega)$. Since $\widehat{\theta}_n \to_{\mathbb{P}} \theta^\dagger$ and $\{Z_\omega : \omega \in \Omega\}$ is uniformly tight, continuity of $\nabla g$ at $\theta^\dagger$ implies, for every $u > 0$,

$$\sup_{\omega\in\Omega} \mathbb{P}_{Z_\omega}\{|h_n(Z_\omega) - \dot{g}^\top Z_\omega| > u \mid \mathcal{D}_n\} \xrightarrow{\mathbb{P}} 0,$$

where $\mathbb{P}_{Z_\omega}$ denotes probability with respect to $Z_\omega$, conditionally on $\mathcal{D}_n$. Since $\dot{g}^\top Z_\omega \sim N(0, A_g/\omega)$ and $\Omega$ is compact in $(0, \infty)$, this normal family has uniformly bounded densities, which are also uniformly bounded away from zero near the $\beta$-quantiles $z_\beta\sqrt{A_g/\omega}$. Hence the preceding approximation gives uniform CDF convergence, and quantile inversion yields the stated expansion.

For the bootstrap posterior, Assumption 1 gives

$$\widehat{\theta}_n^{\mathrm{MB}} - \widehat{\theta}_n = o_{\mathbb{P}^{\mathrm{MB}}}(1),$$

with $\widehat{\theta}_n \xrightarrow{\mathbb{P}} \theta^\dagger$. Uniform tightness of the Gaussian family therefore permits the same conditional delta-method and quantile-inversion argument, yielding (13). Taking $\beta = 1/2$ gives (14). □

## A.3 Proof of Proposition 1

This proof adapts the first-order coverage calculation underlying exact bootstrap calibration in Tanaka (2026) to the moving posterior-median center. Lemma 1 justifies the median-centering step specific to the present criterion.

For $\beta \in \{\beta_L, \beta_U\}$, write

$$q_{\beta,n}(\omega) = g(\widehat{\theta}_n) + n^{-1/2} z_\beta \sqrt{\frac{A_g}{\omega}} + n^{-1/2}\Delta_{\beta,n}(\omega), \tag{A.7}$$

where

$$\max_{\beta} \sup_{\omega\in\Omega} |\Delta_{\beta,n}(\omega)| = o_{\mathbb{P}}(1).$$

Let $T_n = \sqrt{n}\{g(\widehat{\theta}_n) - g^\dagger\}$. The quantile expansion therefore gives, uniformly in $\omega \in \Omega$,

$$-z_{\beta_U}\sqrt{\frac{A_g}{\omega}} + o_{\mathbb{P}}(1) \le T_n \le -z_{\beta_L}\sqrt{\frac{A_g}{\omega}} + o_{\mathbb{P}}(1).$$

Since $\Phi(-x) = 1 - \Phi(x)$, the normal probability between the two limiting endpoints is $H_\alpha(\omega)$. By (A.5) and Pólya's theorem, the endpoint perturbations change the coverage probability by $o(1)$, uniformly in $\omega \in \Omega$. Together with the uniform remainder above, this proves the first assertion in (15) uniformly over $\omega \in \Omega$.

By Assumption 2, $\mathbb{P}^{\mathrm{MB}}\{(\mathcal{A}_n^{\mathrm{MB}})^c\} = o_{\mathbb{P}}(1)$; hence it suffices to establish the bootstrap approximation on $\mathcal{A}_n^{\mathrm{MB}}$, since its complement contributes at most $o_{\mathbb{P}}(1)$ to the coverage criterion.

For the bootstrap posterior, similarly write

$$q_{\beta,n}^{\mathrm{MB}}(\omega) = g(\widehat{\theta}_n^{\mathrm{MB}}) + n^{-1/2} z_\beta \sqrt{\frac{A_g}{\omega}} + n^{-1/2}\Delta_{\beta,n}^{\mathrm{MB}}(\omega),$$

with the remainder uniformly $o_{\mathbb{P}^{\mathrm{MB}}}(1)$ in $\mathbb{P}$-probability. Put

$$D_n(\omega) = \sqrt{n}\{\check{g}_n(\omega) - g(\widehat{\theta}_n)\}.$$

Lemma 1 gives $\sup_{\omega\in\Omega} |D_n(\omega)| = o_{\mathbb{P}}(1)$. Hence $\check{g}_n(\omega) \in C^{\mathrm{MB}}_{\alpha,\omega}(M)$ is equivalent, uniformly over $\omega$, to

$$-z_{\beta_U}\sqrt{\frac{A_g}{\omega}} + o_{\mathbb{P}^{\mathrm{MB}}}(1) \le T_n^{\mathrm{MB}} \le -z_{\beta_L}\sqrt{\frac{A_g}{\omega}} + o_{\mathbb{P}^{\mathrm{MB}}}(1),$$

where $T_n^{\mathrm{MB}}$ is defined in (A.6). Conditional weak convergence to $N(0, B_g)$ and the conditional version of Pólya's theorem give uniform convergence of the corresponding conditional distribution functions, in $\mathbb{P}$-probability. Together with the uniform remainder above, this proves the second assertion in (15) uniformly over $\omega \in \Omega$. The triangle inequality gives (16). □

## A.4 Proof of Theorem 1

Given the uniform first-order approximation in Proposition 1, the root-stability argument is a first-order version of the identification argument in Tanaka (2026). We include it to link that argument to the explicit limiting coverage curve $H_\alpha$ and the approximate-root definition used here.

Let

$$D_n^{\mathrm{MB}} = \sup_{\omega\in\Omega} |c^{\mathrm{MB}}_{n,\alpha}(\omega) - H_\alpha(\omega)| = o_{\mathbb{P}}(1).$$

Since $\omega_g^\circ$ is the unique root of $H_\alpha(\omega) = 1 - \alpha$, (17) implies

$$|H_\alpha(\widetilde{\omega}^{\mathrm{MB}}) - (1-\alpha)| \le 2D_n^{\mathrm{MB}} + \rho_n = o_{\mathbb{P}}(1).$$

Strict monotonicity of $H_\alpha$ on the compact $\Omega$ therefore yields $\widetilde{\omega}^{\mathrm{MB}} \to \omega_g^\circ$ in probability.

Evaluate (A.7) at $\widetilde{\omega}^{\mathrm{MB}}$. Its uniform remainder remains $o_{\mathbb{P}}(1)$, while

$$z_\beta\sqrt{\frac{A_g}{\widetilde{\omega}^{\mathrm{MB}}}} \xrightarrow{\mathbb{P}} z_\beta\sqrt{B_g}, \qquad \beta \in \{\beta_L, \beta_U\}.$$

Together with $T_n \rightsquigarrow N(0, B_g)$, Slutsky's theorem gives (18). □

## A.5 Proof of Corollary 1

Regard each cluster $\mathcal{Y}_c$ as a single observation with loss $\ell^{\mathrm{CL}}(\theta; \mathcal{Y}_c)$. Then

$$R_G^{\mathrm{CL}}(\theta) = \frac{1}{G}\sum_{c=1}^{G} \ell^{\mathrm{CL}}(\theta; \mathcal{Y}_c),$$

and the generalized posterior can be written as

$$\pi_\omega(\theta \mid \mathcal{D}_G^{\mathrm{CL}}) \propto \exp\{-G\omega R_G^{\mathrm{CL}}(\theta)\}p(\theta).$$

Thus it has exactly the form of (1), with $G$, $\mathcal{Y}_c$, and $\ell^{\mathrm{CL}}$ replacing $n$, $Y_i$, and $\ell$, respectively.

Likewise, if

$$M^{\mathrm{CMB}} = (M_1^{\mathrm{CMB}}, \ldots, M_G^{\mathrm{CMB}}) \sim \text{Multinomial}\left(G; \frac{1}{G}, \ldots, \frac{1}{G}\right),$$

then the CMB empirical risk is

$$R_G^{\mathrm{CMB}}(\theta; M^{\mathrm{CMB}}) = \frac{1}{G}\sum_{c=1}^{G} M_c^{\mathrm{CMB}} \ell^{\mathrm{CL}}(\theta; \mathcal{Y}_c),$$

so the cluster multinomial bootstrap is precisely the ordinary multinomial bootstrap applied to the i.i.d. cluster-level observations $\mathcal{Y}_1, \ldots, \mathcal{Y}_G$.

By assumption, the cluster-level analogues of Assumptions 1–3 hold. In particular, under the above identification, the matrices $J$ and $V$ in the i.i.d. analysis become

$$J^{\mathrm{CL}} = \mathbb{E}_{P^{\mathrm{CL}}}\left[\nabla_\theta^2 \ell^{\mathrm{CL}}(\theta^{\mathrm{CL},\dagger}; \mathcal{Y}_c)\right],$$

and

$$V^{\mathrm{CL}} = \mathbb{E}_{P^{\mathrm{CL}}}\left[\psi^{\mathrm{CL}}(\mathcal{Y}_c)\psi^{\mathrm{CL}}(\mathcal{Y}_c)^\top\right].$$

Consequently, Proposition 1, applied at the cluster level, gives

$$\sup_{\omega\in\Omega}\left|c_{G,\alpha}^{\mathrm{CL}}(\omega) - H_\alpha^{\mathrm{CL}}(\omega)\right| \longrightarrow 0$$

and

$$\sup_{\omega\in\Omega}\left|c_{G,\alpha}^{\mathrm{CMB}}(\omega) - H_\alpha^{\mathrm{CL}}(\omega)\right| = o_{\mathbb{P}}(1).$$

The triangle inequality therefore yields

$$\sup_{\omega\in\Omega}\left|c_{G,\alpha}^{\mathrm{CMB}}(\omega) - c_{G,\alpha}^{\mathrm{CL}}(\omega)\right| = o_{\mathbb{P}}(1).$$

Since

$$H_\alpha^{\mathrm{CL}}(\omega) = \Phi\left(z_{\beta_U}\sqrt{\frac{A_g^{\mathrm{CL}}}{\omega B_g^{\mathrm{CL}}}}\right) - \Phi\left(z_{\beta_L}\sqrt{\frac{A_g^{\mathrm{CL}}}{\omega B_g^{\mathrm{CL}}}}\right)$$

is strictly decreasing in $\omega$, its unique solution to $H_\alpha^{\mathrm{CL}}(\omega) = 1-\alpha$ is

$$\omega_g^{\mathrm{CL},\circ} = \frac{A_g^{\mathrm{CL}}}{B_g^{\mathrm{CL}}}.$$

Finally, Theorem 1, again applied to the i.i.d. cluster-level observations, implies that any CMB approximate root satisfying the cluster analogue of (17) obeys

$$\widetilde{\omega}^{\mathrm{CMB}} \xrightarrow{\mathbb{P}} \omega_g^{\mathrm{CL},\circ},$$

and that

$$\mathbb{P}\left\{g(\theta^{\mathrm{CL},\dagger}) \in C_{\alpha,\widetilde{\omega}^{\mathrm{CMB}}}(\mathcal{D}_G^{\mathrm{CL}})\right\} \longrightarrow 1-\alpha.$$

This proves the result. □

## A.6 Proof of Proposition 2

Fix $\omega \in \mathcal{E}$ and a bootstrap vector $m$. By (19), $\check{g}_{n,S}(\omega) \to \check{g}_n(\omega)$ in conditional probability. Local uniform consistency in (20) and continuity of $F_{n,\omega,m}$ at $\check{g}_n(\omega)$ imply

$$\widehat{F}^{\text{comp}}_{n,\omega,m;S,K}(\check{g}_{n,S}(\omega)) \longrightarrow F_{n,\omega,m}(\check{g}_n(\omega))$$

in conditional probability. Since $\mathcal{E}$ is finite (and $B$ is fixed in the first part), convergence holds simultaneously over all evaluated pairs $(\omega, M^{(b)})$. The no-tie condition makes the coverage indicator locally constant at each limiting value, proving (22).

For (23), conditional on $\mathcal{D}_n$ and for fixed $\omega$, the summands in (21) are i.i.d. Bernoulli with mean $c^{\text{MB}}_{n,\alpha}(\omega)$. The strong law of large numbers and finiteness of $\mathcal{E}$ complete the proof. □

## A.7 Proof of Corollary 2

By (24)–(25),

$$|c^{\text{MB}}_{n,\alpha}(\widehat{\omega}^{\text{MB}}) - (1-\alpha)| = o_{\mathbb{P}_{\text{joint}}}(1).$$

Proposition 1 therefore implies

$$|H_\alpha(\widehat{\omega}^{\text{MB}}) - (1-\alpha)| = o_{\mathbb{P}_{\text{joint}}}(1).$$

Uniqueness of the root and strict monotonicity of $H_\alpha$ give $\widehat{\omega}^{\text{MB}} \to \omega^\circ_g$ under the joint law. Evaluating (A.7) at $\widehat{\omega}^{\text{MB}}$ and repeating the final step in the proof of Theorem 1 yields (26). □

# B Additional Information for Section 5

## B.1 Data-generating processes

### B.1.1 Independent sampling

Independent observations are generated according to

$$Y_i = X_i^\top \theta_0 + \varepsilon_i, \qquad i = 1, \dots, n,$$

where

$$X_i \overset{\text{i.i.d.}}{\sim} N_4(0, \Sigma), \qquad \Sigma_{jk} = \varrho^{|j-k|}, \qquad \varrho = 0.2,$$

and

$$\theta_0 = (1, 1, 2, -1)^\top.$$

Under DGP-1, the errors are homoskedastic Gaussian:

$$\varepsilon_i \mid X_i \sim N(0, 1).$$

Under DGP-2, the errors are heteroskedastic Gaussian:

$$\varepsilon_i \mid X_i \sim N(0, \sigma_i^2),$$

where

$$\sigma_i = \begin{cases} 0.01/\varsigma, & X_{i1} < \Phi^{-1}(0.05), \\ 0.1/\varsigma, & \Phi^{-1}(0.05) \le X_{i1} \le \Phi^{-1}(0.95), \\ 1/\varsigma, & X_{i1} > \Phi^{-1}(0.95), \end{cases}$$

and

$$\varsigma = \left\{0.05(0.01)^2 + 0.90(0.1)^2 + 0.05(1)^2\right\}^{1/2}.$$

Because $X_{i1} \sim N(0,1)$, this scaling ensures that $\mathbb{E}_P[\varepsilon_i^2] = 1$ under both DGPs. Consequently, $g^\dagger = g(\theta^\dagger) = 1$. For $g(\theta) = \theta_1$, $J = \Sigma$ and $V = \mathbb{E}_P[\varepsilon_i^2 X_i X_i^\top]$. Thus DGP-1 has $\omega_g^\circ = 1$, whereas DGP-2 gives

$$\omega_g^\circ = \left\{(1-\varrho^2)\mathbb{E}_P(\sigma_i^2 X_{i1}^2) + \varrho^2\right\}^{-1} \simeq 0.270.$$

### B.1.2 Independent-cluster sampling

We generate $G$ i.i.d. clusters, each containing $m = 5$ observations, so that $N_c \equiv m$ and the total sample size is $n_G = mG = 5G$. To induce within-cluster dependence in the covariates while preserving the marginal distribution used in the independent-sampling design, generate

$$Z_c, Z_{c1}, \ldots, Z_{cm} \overset{\text{i.i.d.}}{\sim} N_4(0, \Sigma), \qquad c = 1, \ldots, G,$$

using the same $\Sigma$ as above. For $\lambda_X \in [0,1]$, define

$$X_{cj} = \sqrt{\lambda_X}\, Z_c + \sqrt{1-\lambda_X}\, Z_{cj}, \qquad j = 1, \ldots, m.$$

Thus, marginally, $X_{cj} \sim N_4(0, \Sigma)$, whereas observations within the same cluster satisfy

$$\mathrm{Cov}(X_{cj}, X_{ck}) = \lambda_X \Sigma, \qquad j \neq k.$$

The responses are generated according to

$$Y_{cj} = X_{cj}^\top \theta_0 + \varepsilon_{cj}, \qquad \theta_0 = (1, 1, 2, -1)^\top.$$

Independently of the covariates, generate

$$U_c, E_{c1}, \ldots, E_{cm} \overset{\text{i.i.d.}}{\sim} N(0,1)$$

and, for $\lambda_\varepsilon \in [0,1]$, define

$$\varepsilon_{cj} = \sqrt{\lambda_\varepsilon}\, U_c + \sqrt{1-\lambda_\varepsilon}\, E_{cj}.$$

Hence $\varepsilon_{cj} \sim N(0,1)$ marginally, with

$$\mathrm{Corr}(\varepsilon_{cj}, \varepsilon_{ck}) = \lambda_\varepsilon, \qquad j \neq k.$$

All latent variables are independent across clusters, and the covariate-related latent variables are independent of the error-related latent variables. We set $\lambda_X = \lambda_\varepsilon = 0.5$.

Because $\mathbb{E}_{P^{\mathrm{CL}}}(\varepsilon_{cj} \mid X_{c1}, \ldots, X_{cm}) = 0$, the cluster-level population minimizer is $\theta^{\mathrm{CL},\dagger} = \theta_0$, and hence the repeated-sampling coverage target is $g(\theta^{\mathrm{CL},\dagger}) = 1$. Moreover, direct calculation gives

$$J^{\mathrm{CL}} = m\Sigma, \qquad V^{\mathrm{CL}} = m\{1 + (m-1)\lambda_X\lambda_\varepsilon\}\Sigma = \{1 + (m-1)\lambda_X\lambda_\varepsilon\}J^{\mathrm{CL}}.$$

Therefore, the first-order calibration target in Corollary 1 is

$$\omega_g^{\mathrm{CL},\circ} = \frac{1}{1 + (m-1)\lambda_X \lambda_\varepsilon}.$$

For $m = 5$ and $\lambda_X = \lambda_\varepsilon = 0.5$, this yields $\omega_g^{\mathrm{CL},\circ} = 0.5$. Thus, although each observation has the same marginal covariate and error distributions as under DGP-1 in the independent-sampling design, the generalized posterior with $\omega = 1$ is too concentrated relative to the repeated-sampling distribution because its curvature does not account for the additional variability induced by within-cluster dependence.

## B.2 Additional results

This section reports additional diagnostics and computational results for the simulations in Section 5.

Figure B.1: Calibration iterations in the 50 NUTS-based runs: Independent sampling

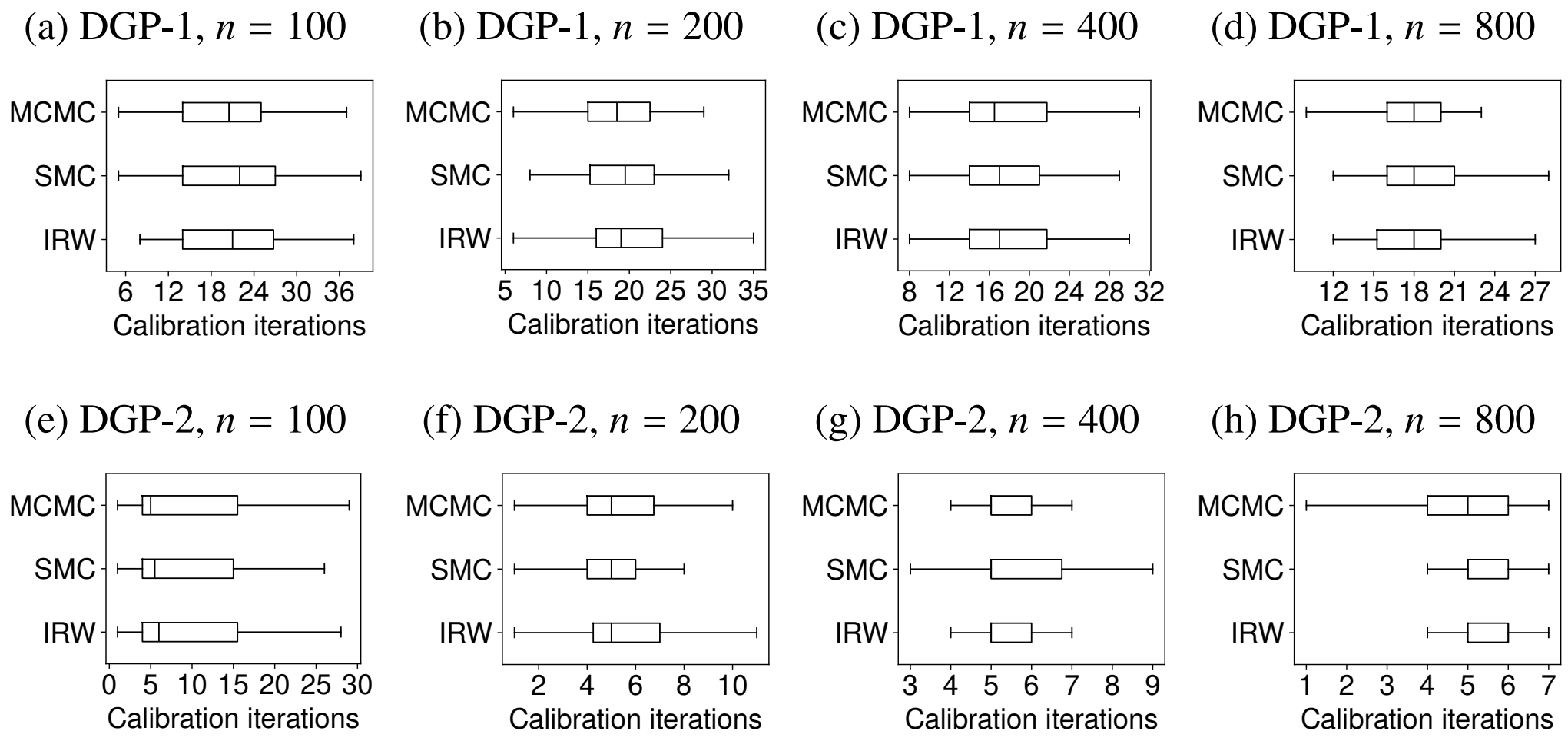


Table B.1: Mean safeguard-triggered direct bootstrap-posterior fits per bootstrap replicate (50 NUTS-based runs): Independent sampling

| DGP | $n$ | | | |
|---|---|---|---|---|
| | 100 | 200 | 400 | 800 |
| DGP-1 | 0.312 | 0.253 | 0.234 | 0.200 |
| DGP-2 | 0.328 | 0.313 | 0.377 | 0.386 |

Figure B.2: Calibrated learning rate: Independent sampling, DGP-1

(a) GPC-MCMC

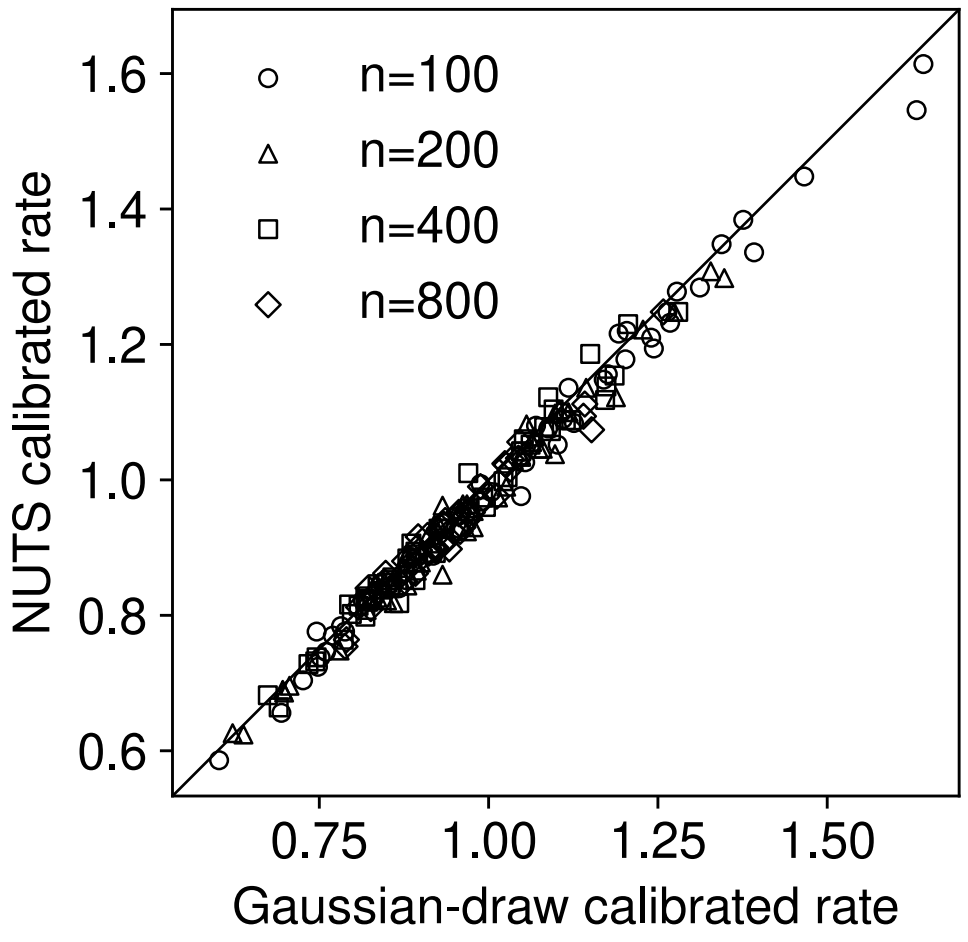


(b) GPC-SMC

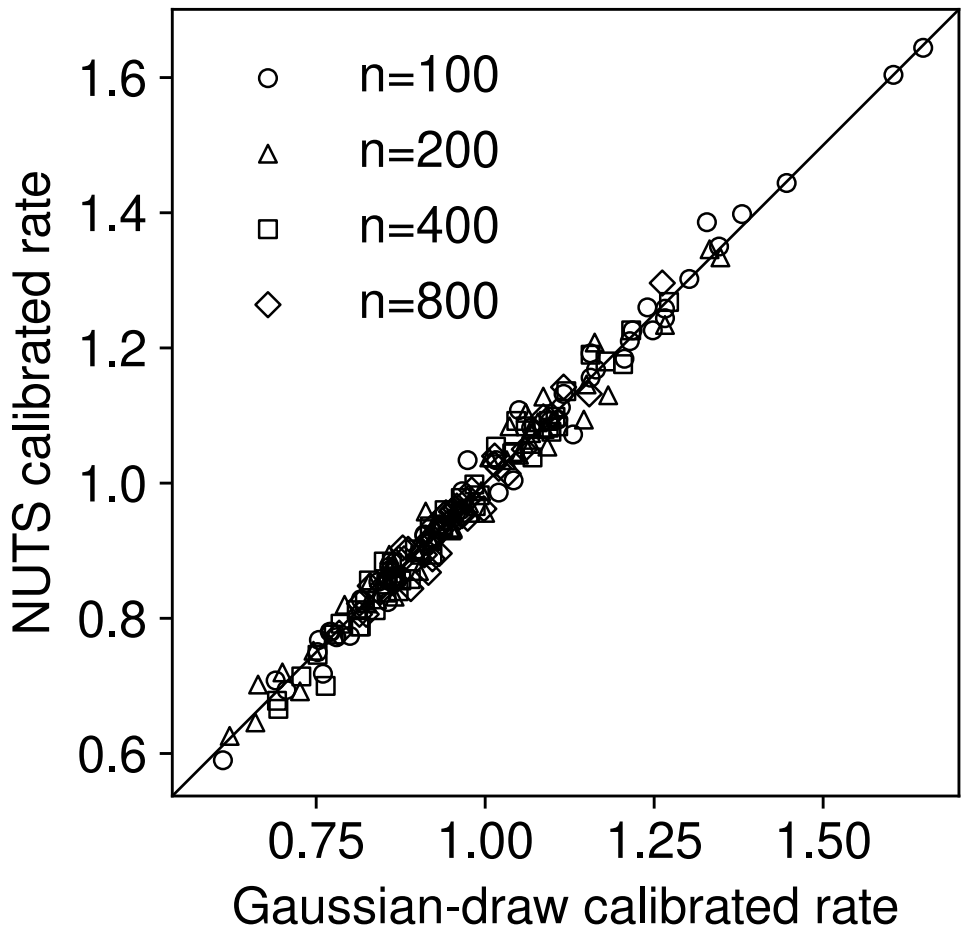


(c) GPC-IRW

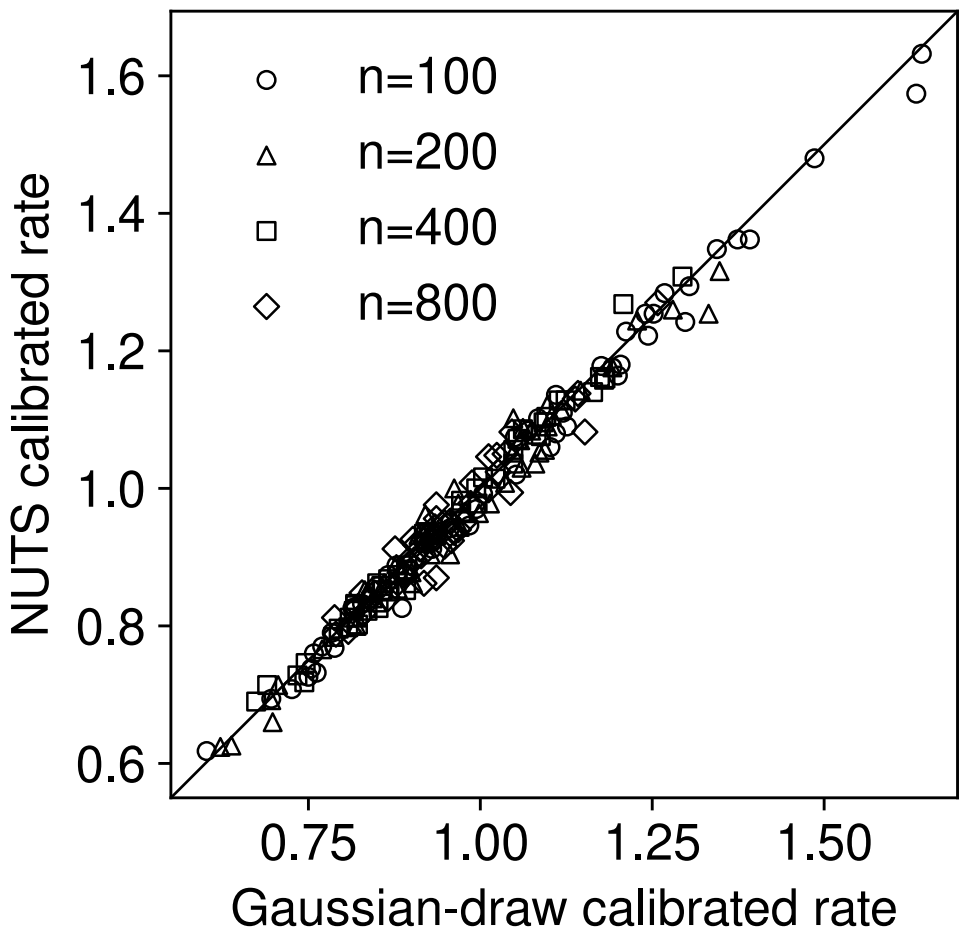

Figure B.3: Calibrated learning rate: Independent sampling, DGP-2

(a) GPC-MCMC

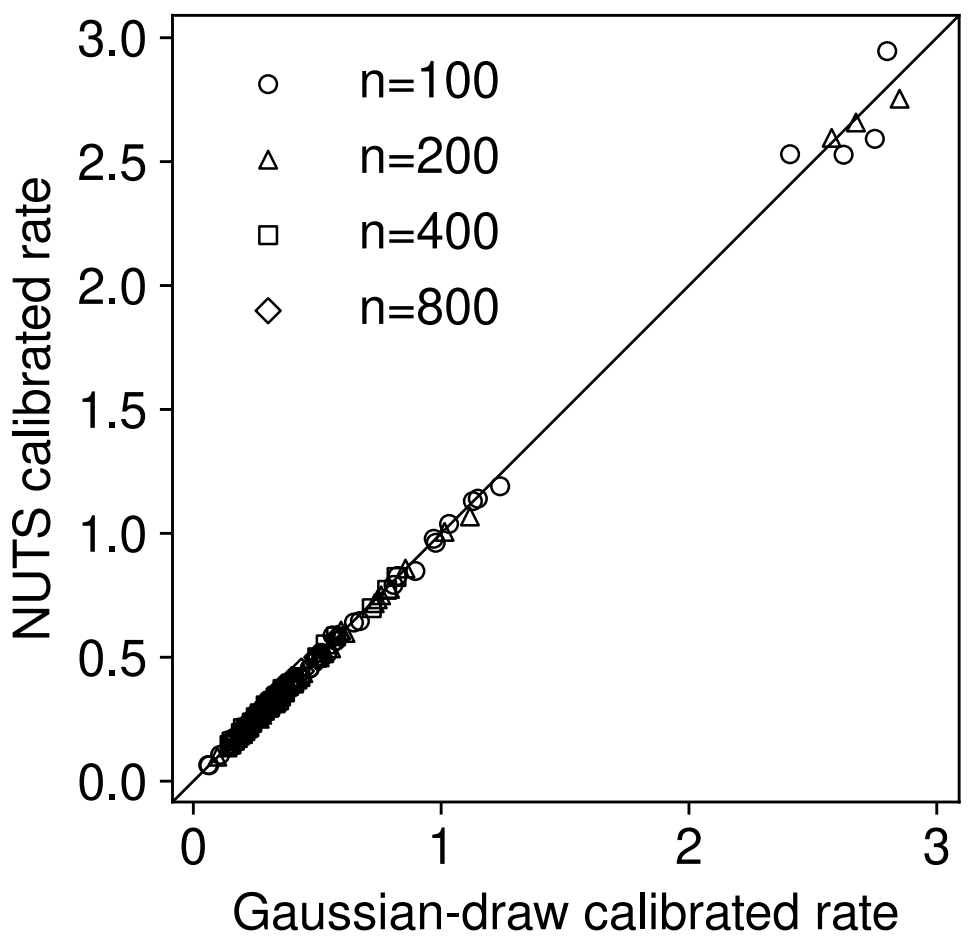


(b) GPC-SMC

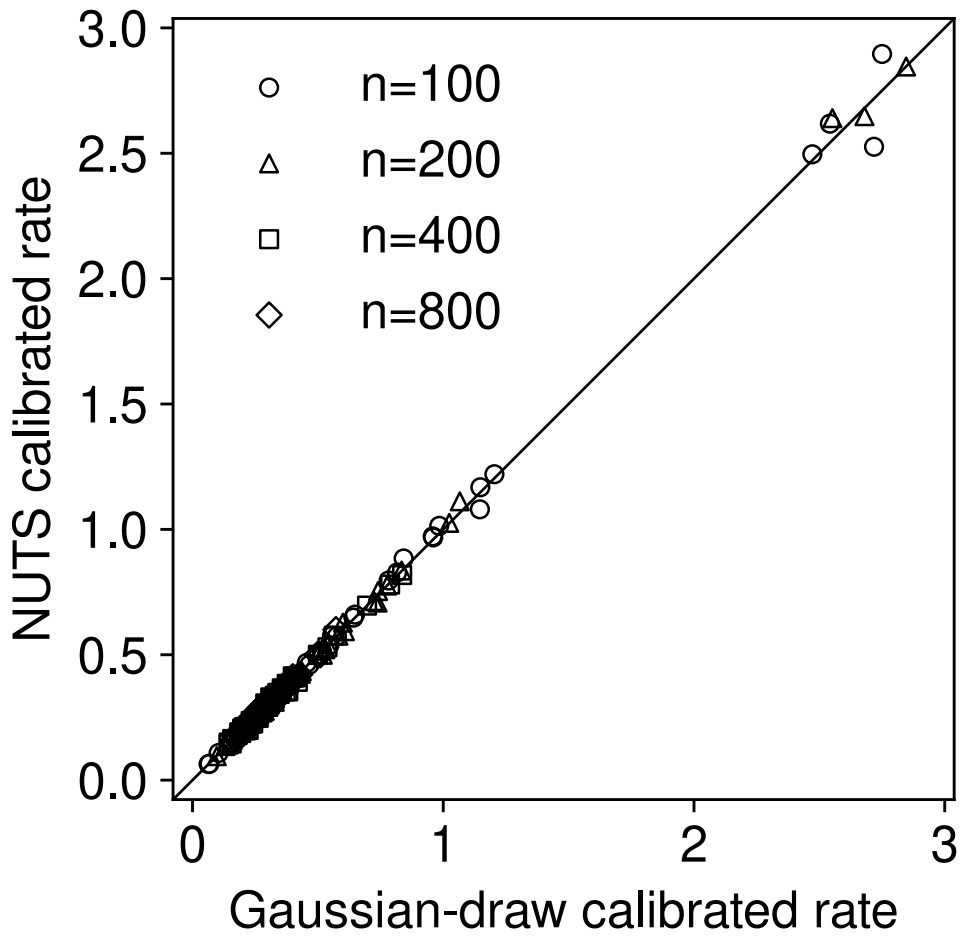


(c) GPC-IRW

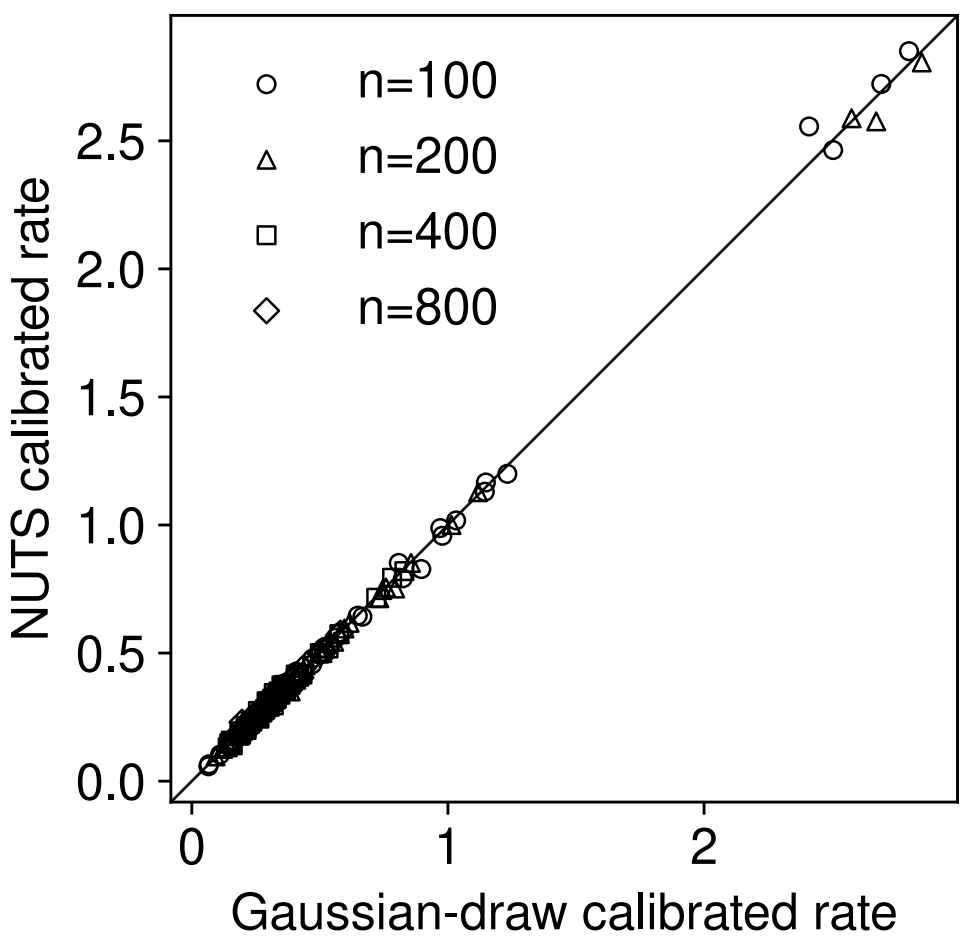


Table B.2: Mean safeguard-triggered direct bootstrap-posterior fits per bootstrap replicate (50 NUTS-based runs): Independent-cluster sampling

| $n_G$ | | | |
|---|---|---|---|
| 100 | 200 | 400 | 800 |
| 0.385 | 0.254 | 0.245 | 0.181 |

Figure B.4: Calibrated learning rate: GPC-IRW, Independent-cluster sampling

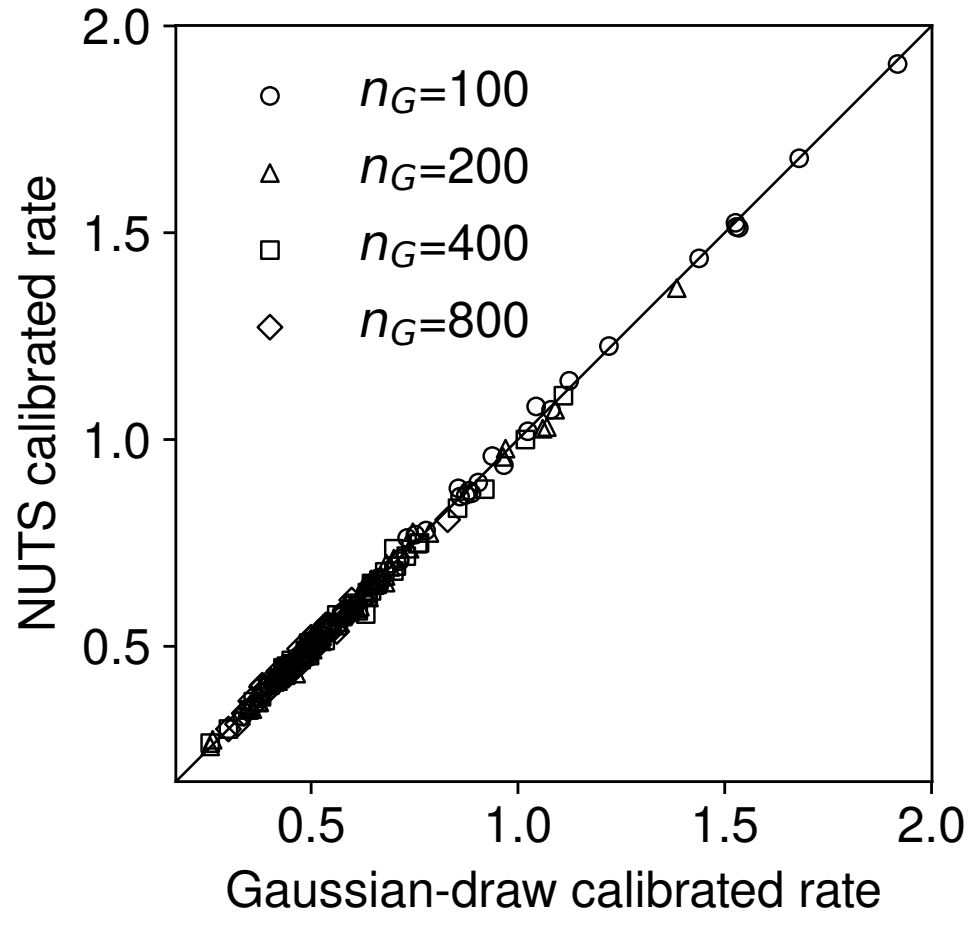


# C Details of the Non-smooth-loss Stress Test

## C.1 Support vector machines

We consider a support vector machine (SVM) for binary outcomes $y_i \in \{-1, 1\}$, with feature vector $x_i = (1, x_{i1}, \ldots, x_{i,p-1})^\top$. Let $\theta$ denote a parameter vector. The hinge loss is

$$\ell(\theta; y_i, x_i) = 2\max(1 - y_i x_i^\top \theta, 0),$$

giving the empirical loss

$$R_n(\theta) = \frac{2}{n}\sum_{i=1}^{n} \max(1 - y_i x_i^\top \theta, 0).$$

We use a prior corresponding to an $L^1$-penalty

$$p(\theta) \propto \exp\left(-\frac{1}{\nu}\sum_{j=1}^{p} \frac{|\theta_j|}{\sigma_j}\right),$$

where $\nu > 0$ is a fixed hyperparameter and $\sigma_j$ denotes the standard deviation of design column $j$ for $j \geq 2$, with $\sigma_1 = 1$ for the intercept. The hinge loss provides a non-smooth test case for GPC-IRW.

We formulate the sampler for an arbitrary nonnegative observation-weight vector $v = (v_1, \ldots, v_n)$. The original-data posterior corresponds to $v_i = 1$ for every $i$, whereas a multinomial-bootstrap posterior with realized count vector $m$ corresponds to $v_i = m_i$. Under the hinge loss above, the resulting weighted generalized posterior is

$$\pi_\omega(\theta \mid v, \mathcal{D}_n) \propto \exp\left\{-2\omega \sum_{i=1}^{n} v_i \max(1 - y_i x_i^\top \theta, 0)\right\} \prod_{j=1}^{p} \exp\left(-\frac{|\theta_j|}{\nu\sigma_j}\right). \tag{C.1}$$

As in the main text, the learning rate scales the empirical loss but not the prior penalty. We adapt the data-augmentation representation of Polson & Scott (2011) to (C.1). We denote the latent variables for the hinge terms by $\phi_i$ and those for the $L^1$ prior by $\chi_j$.

For $q > 0$ and any real $r$, the Polson–Scott mixture identity extends to

$$\exp\{-2q \max(r, 0)\} = \int_0^\infty \left(\frac{q}{2\pi\phi}\right)^{1/2} \exp\left\{-\frac{q(\phi + r)^2}{2\phi}\right\} d\phi. \tag{C.2}$$

To see this, expand the quadratic in the exponent. Apart from the factor $\exp(-qr)$, the resulting integral is

$$\left(\frac{q}{2\pi}\right)^{1/2} \int_0^\infty \phi^{-1/2} \exp\left\{-\frac{q}{2}\left(\phi + \frac{r^2}{\phi}\right)\right\} d\phi = \exp(-q|r|),$$

so that the right-hand side of (C.2) equals $\exp\{-q(r+|r|)\} = \exp\{-2q \max(r, 0)\}$. Taking $q = \omega v_i$ and $r = 1 - y_i x_i^\top \theta$ therefore gives an exact augmentation of each positively weighted hinge-loss contribution. Observations with $v_i = 0$ contribute a factor of one to (C.1) and are omitted from the augmentation.

The $L^1$-penalty prior also admits the normal scale-mixture representation used by Polson & Scott (2011),

$$\frac{1}{2\nu\sigma_j} \exp\left(-\frac{|\theta_j|}{\nu\sigma_j}\right) = \int_0^\infty \frac{1}{\sqrt{2\pi\nu^2\sigma_j^2\chi_j}} \exp\left(-\frac{\theta_j^2}{2\nu^2\sigma_j^2\chi_j}\right) \frac{1}{2} \exp\left(-\frac{\chi_j}{2}\right) d\chi_j. \tag{C.3}$$

Conditional on the latent variables, the full conditional distribution of $\theta$ is consequently multivariate normal. Let

$$\mathcal{I}(v) = \{i : v_i > 0\}, \qquad d_i = \frac{\omega v_i}{\phi_i}, \qquad c_j = \frac{1}{\nu^2\sigma_j^2\chi_j},$$

and define

$$Q = \sum_{i\in\mathcal{I}(v)} d_i x_i x_i^\top + \operatorname{diag}(c_1, \ldots, c_p)$$

and

$$b = \sum_{i\in\mathcal{I}(v)} d_i y_i (1 + \phi_i) x_i.$$

Completing the square gives

$$\theta \mid \phi, \chi, v, \mathcal{D}_n, \omega \sim N_p(Q^{-1}b, Q^{-1}). \tag{C.4}$$

The mixing variables have the following full conditional distributions. We parameterize $\mathrm{IG}(\mu, a)$ as the inverse-Gaussian distribution with mean $\mu$ and shape parameter $a$. For every $i \in \mathcal{I}(v)$,

$$\phi_i^{-1} \mid \theta, v, \mathcal{D}_n, \omega \sim \mathrm{IG}\left(\left|1 - y_i x_i^\top \theta\right|^{-1}, \omega v_i\right), \tag{C.5}$$

whereas, for $j = 1, \ldots, p$,

$$\chi_j^{-1} \mid \theta_j \sim \mathrm{IG}\left(\frac{\nu\sigma_j}{|\theta_j|}, 1\right). \tag{C.6}$$

Equations (C.4)–(C.6) therefore yield a Gibbs sampler for both the original-data and bootstrap-weighted generalized posteriors. Relative to the sampler of Polson & Scott (2011), the only modification to the hinge-loss augmentation is that the unit contribution of observation $i$ is replaced by the product $\omega v_i$.

The observation weights $v_i$ in this Gibbs sampler should be distinguished from the importance weights used by GPC-IRW. In particular, if $\theta^{(s)}$ is drawn from the original-data reference posterior at learning rate $\omega_{\text{ref}}$, the unnormalized importance ratio for a target with observation weights $v$ and learning rate $\omega$ is

$$r\{\theta^{(s)}; \omega, v\} = \exp\left[-2\sum_{i=1}^{n}(\omega v_i - \omega_{\text{ref}}) \max\{1 - y_i x_i^\top \theta^{(s)}, 0\}\right].$$

These importance ratios reweight retained Gibbs draws and do not enter the Gibbs full conditional distributions. When importance reweighting is inadequate and a direct target-posterior simulation is required, the Gibbs sampler above is run with $v_i = m_i$ for the corresponding bootstrap count vector.

## C.2 Simulation design

We adapt the classification design of Martin & Syring (2022). For each observation, let

$$Z_i \overset{\text{i.i.d.}}{\sim} N(1, 1), \qquad x_i = (1, Z_i)^\top,$$

and generate $y_i \in \{-1, 1\}$ according to

$$\Pr(y_i = 1 \mid Z_i = z) = F_3(1 - z),$$

where $F_3$ is the Student-$t$ distribution function with three degrees of freedom. Under this DGP, numerical integration of the population hinge risk gives

$$\theta^\dagger = c^\dagger(1, -1)^\top, \qquad c^\dagger = 1.260499\ldots.$$

Thus the decision boundary is $z = 1$, and we target

$$g(\theta) = \theta_2, \qquad g^\dagger = -1.260499\ldots.$$

Our individual loss is twice that used by Martin & Syring (2022); this rescaling leaves $\theta^\dagger$ unchanged and only rescales the learning rate. The hinge loss is not twice continuously differentiable, so this example falls outside Assumption 1; nevertheless, its population risk is smooth under the continuous-predictor design, making it a useful stress test beyond the scope of the present theory.

We use the same basic simulation design and calibration settings as in the independent-sampling experiment of Section 5.1, while replacing the quadratic-loss posterior computation by the Gibbs sampler described above. In particular, we consider $n \in \{100, 200, 400, 800\}$ and calibrate nominal 95% quantile-based credible intervals for $g(\theta) = \theta_2$, with $(\beta_L, \beta_U) = (0.025, 0.975)$. We compare GPC-MCMC and the proposed GPC-IRW procedure using the same bootstrap size, learning-rate search interval, stochastic-approximation sequence, stopping criterion, and safeguard settings as

in Section 5.1. These sample sizes differ from those considered by Martin & Syring (2022), who used $n = 400$ and 1000, but match those in our main independent-sampling experiment.

For the $L^1$-penalty prior, we set $\nu = 10$ and fix the feature-scale parameters at their population values rather than re-estimating them in each simulated dataset. Since the sole non-intercept predictor satisfies $\mathrm{Var}(Z_i) = 1$, this gives $\sigma_1 = \sigma_2 = 1$. Thus the prior is held fixed across simulation replications and bootstrap perturbations. Direct simulations from the original-data and bootstrap posteriors are carried out using the data-augmented Gibbs sampler in equations (C.4)–(C.6). For GPC-IRW, posterior draws from a reference Gibbs fit are reused across learning-rate values and bootstrap perturbations through the importance ratios described above, with direct Gibbs simulation invoked when the safeguard rejects importance reweighting.

We generate 1,000 independent datasets at each sample size and report empirical coverage and average length of the calibrated 95% credible intervals for $\theta_2^\dagger$. Wall-clock times, calibration iteration counts, and safeguard summaries are computed from these same 1,000 paired replications, with no separate computational simulation or smaller timing subset. For each required posterior simulation, we run 1,000 burn-in iterations and retain 5,000 draws. Unless otherwise stated, all remaining calibration and computational settings are identical to those used in the independent-sampling experiment of Section 5.1. The programs are written in Python, with the Gibbs sampler implemented in C++.

## C.3 Simulation results

Figures C.1–C.4 report empirical coverage, credible-interval lengths, wall-clock times, and numbers of calibration iterations for the SVM experiment. Table C.1 summarizes the average number of safeguard-triggered direct bootstrap-posterior fits per bootstrap replicate. The main findings are discussed in Section 5.3 of the main text.

Figure C.1: Empirical coverage of nominal 95% credible intervals: SVM

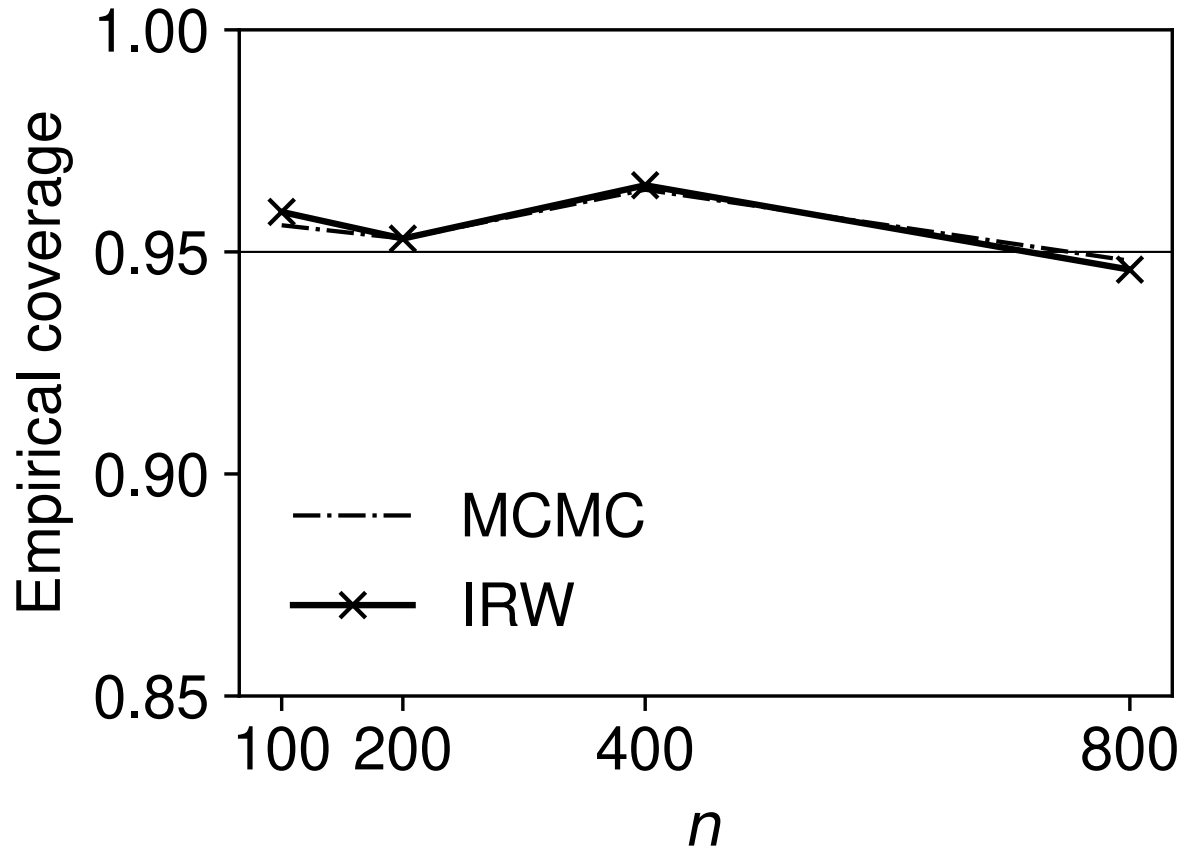

Figure C.2: Length of nominal 95% credible intervals: SVM

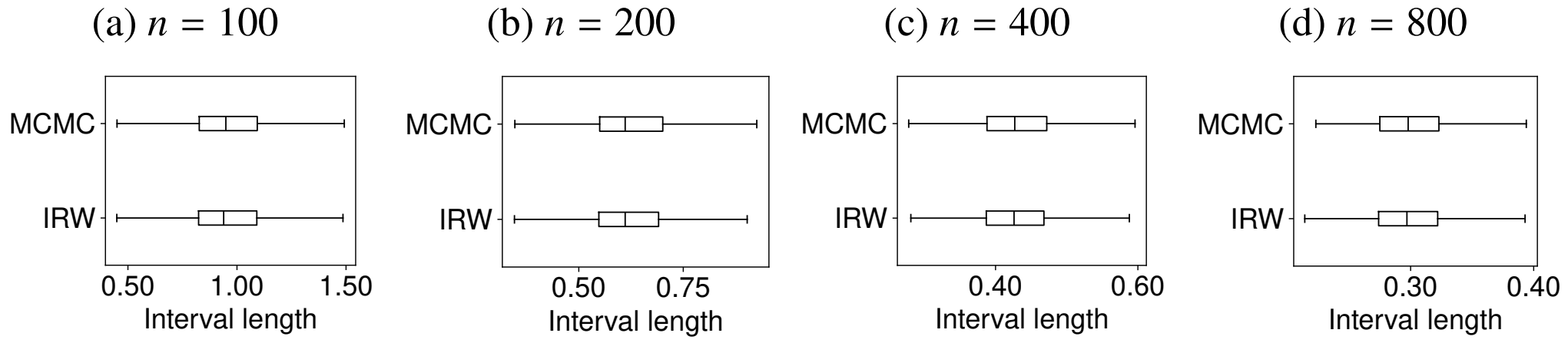


Figure C.3: Wall-clock time: SVM

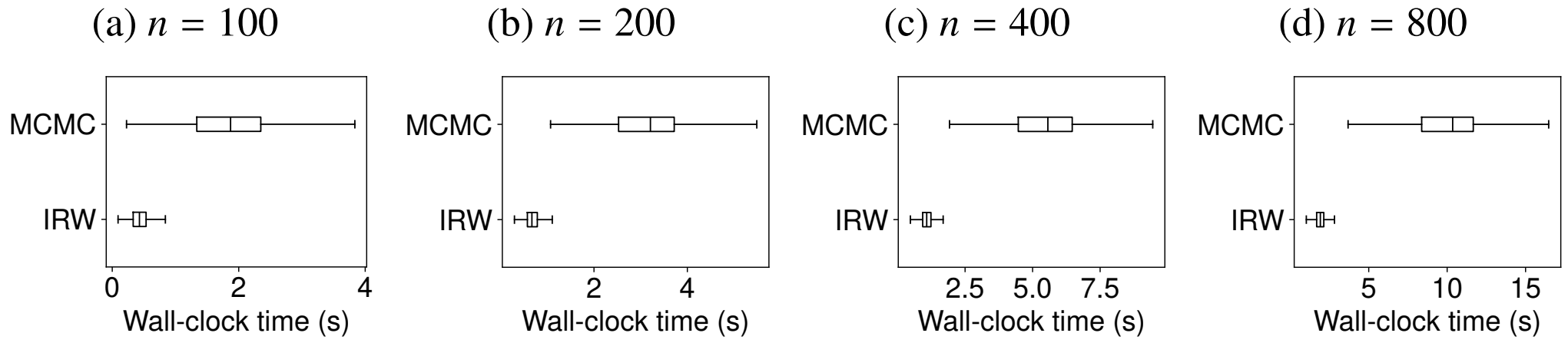


Figure C.4: Number of calibration iterations: SVM

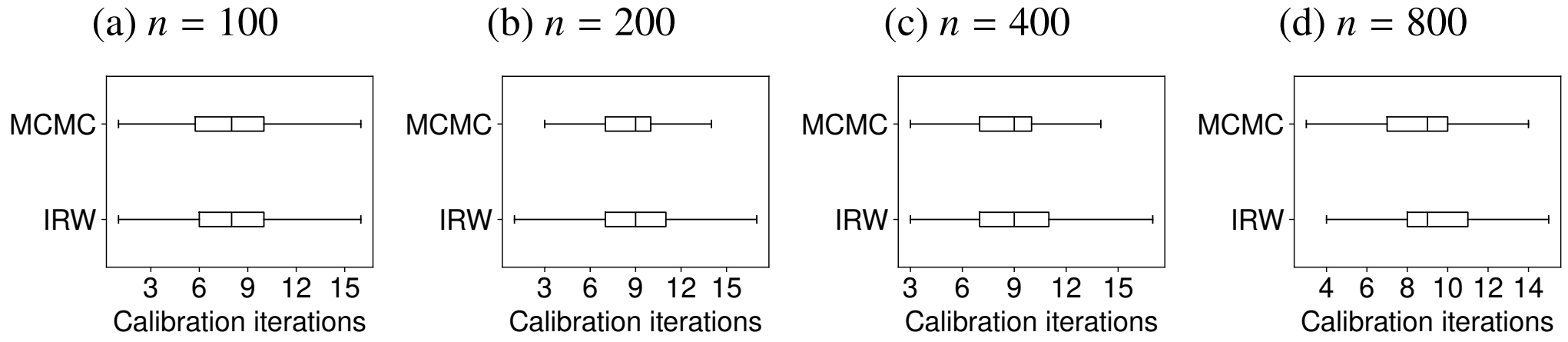


Table C.1: Average number of safeguard-triggered direct bootstrap-posterior fits per bootstrap replicate: SVM

| $n$ | | | |
|---|---|---|---|
| 100 | 200 | 400 | 800 |
| 0.138 | 0.125 | 0.120 | 0.117 |

# D Real Data Application Details

## D.1 Data extraction and sample construction

On October 10, 2026, at 08:34:28 UTC, we retrieved the `pscomppars` table (Planetary Systems Composite Parameters) from the NASA Exoplanet Archive using the Table Access Protocol (TAP)

(Akeson et al., 2013; Christiansen et al., 2025; NASA Exoplanet Archive, 2026). Because this table is subject to updates, we archived the returned CSV, the query, and the selected planet identifiers. The fields `pl_bmasse` and `pl_rade` give masses and radii in Earth masses and Earth radii, respectively; `hostname` and `pl_name` identify the host and planet. We used the following initial ADQL query:

```
SELECT pl_name, hostname, sy_pnum, sy_snum, tran_flag,
       pl_controv_flag, pl_bmasse, pl_rade, pl_bmassprov,
       pl_bmasselim, pl_radelim, pl_bmasse_reflink,
       pl_rade_reflink
FROM pscomppars
WHERE tran_flag = 1
  AND sy_snum = 1
  AND pl_controv_flag = 0
  AND pl_bmasse > 0
  AND pl_rade BETWEEN 1 AND 4
  AND pl_bmassprov = 'Mass'
  AND pl_bmasselim = 0
  AND pl_radelim = 0
ORDER BY hostname, pl_name
```

The limit flags exclude upper and lower limits, and requiring `pl_bmassprov = 'Mass'` excludes minimum-mass estimates. Because the composite table can substitute calculated values for missing measurements (NASA Exoplanet Archive, 2026), we screened the mass reference field (`pl_bmasse_reflink`) and the radius reference field (`pl_rade_reflink`) for "Calculated Value" using a case-insensitive substring match and excluded rows with a match in either field. We did not filter on mass-measurement significance. All analyses used this archived extract rather than a new live query.

For the independent-sampling analysis, we further restricted the sample to hosts with `sy_pnum=1`, where `sy_pnum` denotes the number of confirmed planets associated with a host. This yielded 113 planets from 113 hosts. The full sample contained $n_G = 521$ eligible planets from $G = 347$ hosts; 124 hosts contributed at least two eligible planets, and eligible cluster sizes ranged from one to six. A host with several confirmed planets could still contribute only one eligible planet. Thus, $N_c$, the number of eligible planets for host $c$, need not equal `sy_pnum`. We ordered rows by `hostname` and `pl_name` before forming observation- and host-level bootstrap units.

We treat hosts as independent observational units within the selected archive population. The composite catalog can combine measurements from different publications; neither measurement uncertainty nor catalog selection is modeled explicitly. Inference therefore concerns the robust association between *reported* masses and radii in this population, not a selection-corrected relation for all exoplanets.

## D.2 Model and computational implementation

We defined the response as $y = \log(\texttt{pl_bmasse})$ and the predictor as $t = \log(\texttt{pl_rade})$. The natural cubic spline had boundary knots at $\log 1$ and $\log 4$ and interior knots at $\log 1.5$, $\log 2$, and $\log 3$. We assigned independent $N(0, 10^2)$ priors to the coefficients. The archived analysis code

specifies the spline design matrix and the design vectors at radii 1.5 and 3. The knots and basis transformation were fixed before calibration and were not re-estimated in bootstrap samples. We used the unstandardized pseudo-Huber loss with $\eta = \log 2$; the prior and planet-level loss were identical for observation- and host-level calibration.

We used $B = 500$, $\Omega = [0.01, 10]$, $\gamma_l = l^{-0.51}$, $\epsilon_{\text{tol}} = 0.01$, and at most $T = 200$ coverage evaluations. For GPC-IRW, $S = K = 5{,}000$, $\zeta_0 = 0.5$, $E_{\min} = 100$, and $\kappa_{\text{sg}} = 2$. Each direct NUTS fit used one chain, 1,000 warmup iterations, and 5,000 retained draws. In the single-planet analysis, the initial learning rate $\omega^{(0)} = 6$ was selected by a preliminary grid search, whose cost was excluded from the reported wall-clock times. In the full-sample analysis, we set $\omega^{(0)} = 3$.

For each sample, we generated the $B$ multinomial bootstrap replicates once and retained them across calibration evaluations. GPC-MCMC and GPC-IRW used identical replicate weights when targeting the same criterion. In the full sample, observation-level MB resampled planets, whereas CMB resampled `hostname` groups, assigning each host's count to all its eligible planets. Both used the same sum of planet-level losses in the original-data posterior; only the bootstrap unit differed.

The uncalibrated posterior used $\omega = 1$. We computed posterior medians and quantile-based 95% credible intervals for $\exp\{f_\theta(3) - f_\theta(1.5)\}$ from the original-data posterior at each indicated learning rate. All methods were run on the same computational platform.

## D.3 Additional results

Table D.1 supplements the learning-rate, interval, and timing comparisons in Section 6 of the main text by reporting calibration iteration counts and safeguard diagnostics.

Table D.1: Additional results for exoplanet applications

| Method | Calibration iterations | Direct fits per bootstrap replicate |
|---|---:|---:|
| (a) Single-planet systems | | |
| GPC-MCMC-MB | 51 | – |
| GPC-IRW-MB | 101 | 0.462 |
| (b) All eligible systems | | |
| GPC-MCMC-CMB | 32 | – |
| GPC-IRW-MB | 47 | 0.430 |
| GPC-IRW-CMB | 35 | 0.486 |

*Note:* Direct fits are safeguard-triggered bootstrap-posterior simulations, averaged per replicate over the $B = 500$ bootstrap replicates for the entire calibration run; "–" indicates that the safeguard does not apply.

# E Dirichlet Weighting

## E.1 Dirichlet-weight posteriors

Although the multinomial bootstrap naturally approximates repeated sampling, each resample typically assigns zero weight to many observations. Scaled Dirichlet weights, closely related to the

Bayesian bootstrap (Rubin, 1981), instead assign positive weight to every observation while introducing perturbations that mimic sampling variability.

Dirichlet-weighted generalized posterior calibration combined with importance reweighting was proposed by Sugasawa (2026). Here we consider a Dirichlet-weight version of GPC-IRW that retains its median-centered calibration criterion and bootstrap-replicate-specific diagnostic safeguards, allowing the multinomial and Dirichlet weighting schemes to be compared within the same algorithmic framework.

Specifically, let

$$W = (W_1, \dots, W_n), \qquad \left(\frac{W_1}{n}, \dots, \frac{W_n}{n}\right) \sim \text{Dirichlet}(1, \dots, 1),$$

so that $\sum_{i=1}^n W_i = n$ almost surely. For a realization $w = (w_1, \dots, w_n)$, define the Dirichlet-weighted posterior

$$\pi_\omega^{\text{DW}}(\theta \mid w, \mathcal{D}_n) \propto \exp\left\{-\omega \sum_{i=1}^n w_i \ell(\theta; y_i)\right\} p(\theta).$$

A superscript DW denotes a quantity associated with Dirichlet weighting. The GPC-IRW-DW procedure applies Algorithm 1 with the multinomial count vectors $M^{(b)}$ replaced by independent scaled Dirichlet weight vectors $W^{(b)}$. Section E.2 shows that its calibration criterion is first-order equivalent to the multinomial-bootstrap criterion, with the same asymptotic calibration target.

The cluster-level Dirichlet-weight variant is obtained analogously to the CMB extension in Section 3.4. Let

$$W^{\text{CDW}} = (W_1^{\text{CDW}}, \dots, W_G^{\text{CDW}}), \qquad \left(\frac{W_1^{\text{CDW}}}{G}, \dots, \frac{W_G^{\text{CDW}}}{G}\right) \sim \text{Dirichlet}(1, \dots, 1),$$

so that $\sum_{c=1}^G W_c^{\text{CDW}} = G$ almost surely. Replacing the cluster multinomial counts by these scaled Dirichlet weights gives

$$\pi_\omega^{\text{CDW}}(\theta \mid w^{\text{CDW}}, \mathcal{D}_G^{\text{CL}}) \propto \exp\left\{-\omega \sum_{c=1}^G w_c^{\text{CDW}} \ell^{\text{CL}}(\theta; \mathcal{Y}_c)\right\} p(\theta).$$

Accordingly, both the multinomial-bootstrap and Dirichlet-weight versions of GPC-IRW extend to independent-cluster sampling simply by generating weights at the cluster level and applying the resulting common weight to all observations within each cluster.

## E.2 First-order equivalence of multinomial-bootstrap and Dirichlet-weight calibration

We establish the first-order equivalence of multinomial-bootstrap and Dirichlet-weight calibration. As with the multinomial-bootstrap posterior, we set the coverage indicator to zero when the weighted posterior is improper. Let $\mathbb{P}^{\text{DW}}$ and $\mathbb{E}^{\text{DW}}$ denote probability and expectation with respect to $W$, conditional on $\mathcal{D}_n$, and interpret $o_{\mathbb{P}^{\text{DW}}}(1)$, $O_{\mathbb{P}^{\text{DW}}}(1)$, and conditional weak convergence statements in $\mathbb{P}$-probability.

Let $C^{\mathrm{DW}}_{\alpha,\omega}(w)$ denote the quantile-based $(1-\alpha)$ credible interval for $g(\theta)$ under the Dirichlet-weighted posterior, using levels $\beta_L, \beta_U$, and define

$$c^{\mathrm{DW}}_{n,\alpha}(\omega) = \mathbb{E}^{\mathrm{DW}}\left[\mathbb{I}\{\breve{g}_n(\omega) \in C^{\mathrm{DW}}_{\alpha,\omega}(W)\}\right].$$

Thus multinomial-bootstrap and Dirichlet-weight calibration use the same original-data posterior center.

For the theoretical argument, write

$$R^{\mathrm{DW}}_n(\theta; w) = \frac{1}{n}\sum_{i=1}^{n} w_i \ell(\theta; Y_i), \qquad R^{\mathrm{DW}}_n(\theta) := R^{\mathrm{DW}}_n(\theta; W),$$

and let $\widehat{\theta}^{\mathrm{DW}}_n$ denote the measurable local minimizer specified in Assumption E.1 below. The scaled Dirichlet weights admit the normalized-exponential representation

$$W_i = \frac{\tau_i}{\bar{\tau}_n}, \qquad \tau_1, \dots, \tau_n \overset{\text{i.i.d.}}{\sim} \text{Exponential}(1), \qquad \bar{\tau}_n = \frac{1}{n}\sum_{i=1}^{n} \tau_i.$$

Let $\mathbb{P}_\tau$ and $\mathbb{E}_\tau$ denote probability and expectation with respect to $\tau_1, \dots, \tau_n$, conditionally on $\mathcal{D}_n$.

Since the normalized exponential representation induces the Dirichlet law of $W$, for every measurable function $h$ of $W$,

$$\mathbb{E}^{\mathrm{DW}}[h(W)] = \mathbb{E}_\tau\left[h\left(\frac{\tau_1}{\bar{\tau}_n}, \dots, \frac{\tau_n}{\bar{\tau}_n}\right)\right].$$

Consequently,

$$\mathbb{E}^{\mathrm{DW}}[W_i] = 1, \qquad \mathrm{Var}^{\mathrm{DW}}(W_i) = \frac{n-1}{n+1},$$

and, for $i \neq j$,

$$\mathrm{Cov}^{\mathrm{DW}}(W_i, W_j) = -\frac{1}{n+1}.$$

In particular,

$$\frac{1}{n}\sum_{i=1}^{n}(W_i - 1)^2 = \frac{n^{-1}\sum_{i=1}^{n}\tau_i^2}{\bar{\tau}_n^2} - 1 \xrightarrow{\mathbb{P}^{\mathrm{DW}}} 1.$$

Thus the Dirichlet weights have the same asymptotic variance normalization as the multinomial-bootstrap weights. For the result below, rather than invoke a general theorem for exchangeably weighted bootstraps (e.g., Mason & Newton, 1992; Præstgaard & Wellner, 1993), we establish the required conditional central limit theorem directly from the exponential representation.

We impose the following Dirichlet-weight counterparts of the local $M$-estimation and posterior-quantile conditions used for the multinomial bootstrap.

**Assumption E.1** (Dirichlet-weighted regularity)**.** Conditionally on $\mathcal{D}_n$, let $\widehat{\theta}^{\mathrm{DW}}_n$ be a measurable local minimizer of $R^{\mathrm{DW}}_n(\theta; W)$ whenever such a local minimizer exists, and define it arbitrarily otherwise. Assume that

$$\mathbb{P}^{\mathrm{DW}}\left\{\widehat{\theta}^{\mathrm{DW}}_n \text{ is a local minimizer of } R^{\mathrm{DW}}_n(\cdot; W)\right\} \xrightarrow{\mathbb{P}} 1$$

and

$$\widehat{\theta}_n^{\mathrm{DW}} - \widehat{\theta}_n = o_{\mathbb{P}^{\mathrm{DW}}}(1).$$

Moreover, the Dirichlet-weighted empirical Hessian converges to $J$, in conditional probability, uniformly along the line segment joining $\widehat{\theta}_n$ and $\widehat{\theta}_n^{\mathrm{DW}}$. In addition,

$$\mathbb{P}^{\mathrm{DW}} \left\{ \pi_\omega^{\mathrm{DW}}(\cdot \mid W, \mathcal{D}_n) \text{ is proper for every } \omega \in \Omega \right\} \xrightarrow{\mathbb{P}} 1.$$

For $\beta \in \{\beta_L, \beta_U\}$, on the propriety event above, let $q_{\beta,n}^{\mathrm{DW}}(\omega)$ denote the $\beta$-quantile of $g(\theta)$ under $\pi_\omega^{\mathrm{DW}}(\theta \mid W, \mathcal{D}_n)$, and define it arbitrarily otherwise. Uniformly over $\omega \in \Omega$,

$$\sup_{\omega \in \Omega} \left| \sqrt{n}\{q_{\beta,n}^{\mathrm{DW}}(\omega) - g(\widehat{\theta}_n^{\mathrm{DW}})\} - z_\beta \sqrt{\frac{A_g}{\omega}} \right| = o_{\mathbb{P}^{\mathrm{DW}}}(1).$$

The posterior-quantile condition is the Dirichlet counterpart of Lemma 1. For example, it can be obtained from an appropriate uniform Bernstein–von Mises approximation for the Dirichlet-weighted posterior. We state the quantile expansion directly because it is the approximation needed for the coverage argument.

**Proposition E.1** (First-order equivalence and validity of Dirichlet-weight calibration)**.** *Suppose Assumptions 1–3 of the main paper and the Dirichlet-weighted regularity assumption above hold. Then*

$$\sqrt{n}(\widehat{\theta}_n^{\mathrm{DW}} - \widehat{\theta}_n) \rightsquigarrow N_d(0, J^{-1} V J^{-1})$$

*and*

$$\sup_{\omega \in \Omega} \left| c_{n,\alpha}^{\mathrm{DW}}(\omega) - H_\alpha(\omega) \right| = o_{\mathbb{P}}(1).$$

*Consequently,*

$$\sup_{\omega \in \Omega} \left| c_{n,\alpha}^{\mathrm{DW}}(\omega) - c_{n,\alpha}^{\mathrm{MB}}(\omega) \right| = o_{\mathbb{P}}(1).$$

*If*

$$\omega_g^\circ = \frac{A_g}{B_g} \in \operatorname{int}(\Omega)$$

*and a measurable $\widetilde{\omega}^{\mathrm{DW}} \in \Omega$ satisfies*

$$\left| c_{n,\alpha}^{\mathrm{DW}}(\widetilde{\omega}^{\mathrm{DW}}) - (1-\alpha) \right| \le \inf_{\omega \in \Omega} \left| c_{n,\alpha}^{\mathrm{DW}}(\omega) - (1-\alpha) \right| + \rho_n, \qquad \rho_n \ge 0, \quad \rho_n = o_{\mathbb{P}}(1), \tag{E.1}$$

*then $\widetilde{\omega}^{\mathrm{DW}} \xrightarrow{\mathbb{P}} \omega_g^\circ$ and*

$$\mathbb{P}\{g^\dagger \in C_{\alpha, \widetilde{\omega}^{\mathrm{DW}}}(\mathcal{D}_n)\} \longrightarrow 1 - \alpha.$$

*Proof.* Write

$$\psi_i = \psi(Y_i), \qquad \bar{\psi}_n = \frac{1}{n}\sum_{i=1}^n \psi_i, \qquad V_n = \frac{1}{n}\sum_{i=1}^n (\psi_i - \bar{\psi}_n)(\psi_i - \bar{\psi}_n)^\top.$$

Under Assumption 1, $V_n \xrightarrow{\mathbb{P}} V$. Since $\sum_i (W_i - 1) = 0$, the normalized-exponential representation gives

$$\frac{1}{\sqrt{n}} \sum_{i=1}^{n} (W_i - 1)\psi_i = \frac{1}{\bar{\tau}_n} \frac{1}{\sqrt{n}} \sum_{i=1}^{n} (\tau_i - 1)(\psi_i - \bar{\psi}_n).$$

Conditionally on $\mathcal{D}_n$, the variables $n^{-1/2}(\tau_i - 1)(\psi_i - \bar{\psi}_n)$ are independent with mean zero, and their covariance matrices sum to $V_n$. Moreover, the moment condition in Assumption 1 implies the conditional Lyapunov condition:

$$\frac{1}{n^{1+\delta/2}} \sum_{i=1}^{n} \mathbb{E}_\tau \left[ \|(\tau_i - 1)(\psi_i - \bar{\psi}_n)\|^{2+\delta} \,\middle|\, \mathcal{D}_n \right] = C_\delta n^{-\delta/2} \frac{1}{n} \sum_i \|\psi_i - \bar{\psi}_n\|^{2+\delta} = o_{\mathbb{P}}(1),$$

where

$$C_\delta = \mathbb{E}_\tau |\tau_1 - 1|^{2+\delta} < \infty.$$

Since $\bar{\tau}_n \to 1$ in $\mathbb{P}_\tau$-probability, the conditional multivariate central limit theorem and Slutsky's theorem yield

$$\frac{1}{\sqrt{n}} \sum_{i=1}^{n} (W_i - 1)\psi_i \rightsquigarrow N_d(0, V). \tag{E.2}$$

Let

$$\overline{J}_{n,W} = \int_0^1 \left[ \frac{1}{n} \sum_{i=1}^{n} W_i \ddot{\ell}_i \{\widehat{\theta}_n + t(\widehat{\theta}_n^{\mathrm{DW}} - \widehat{\theta}_n)\} \right] dt.$$

On the event that $\widehat{\theta}_n^{\mathrm{DW}}$ is a local minimizer—an event whose $\mathbb{P}^{\mathrm{DW}}$-probability tends to one in $\mathbb{P}$-probability by Assumption E.1—the weighted and ordinary first-order conditions give

$$\sqrt{n}(\widehat{\theta}_n^{\mathrm{DW}} - \widehat{\theta}_n) = -\overline{J}_{n,W}^{-1} \frac{1}{\sqrt{n}} \sum_{i=1}^{n} (W_i - 1)\dot{\ell}_i(\widehat{\theta}_n).$$

Assumption E.1 gives

$$\overline{J}_{n,W} = J + o_{\mathbb{P}^{\mathrm{DW}}}(1).$$

To replace $\dot{\ell}_i(\widehat{\theta}_n)$ by $\psi_i$, put

$$d_{ni} = \dot{\ell}_i(\widehat{\theta}_n) - \psi_i.$$

The covariance formula for the Dirichlet weights gives

$$\mathbb{E}^{\mathrm{DW}} \left[ \left\| \frac{1}{\sqrt{n}} \sum_{i=1}^{n} (W_i - 1) d_{ni} \right\|^2 \right] \leq \frac{1}{n+1} \sum_{i=1}^{n} \|d_{ni}\|^2.$$

By (9) and

$$\widehat{\theta}_n - \theta^\dagger = O_{\mathbb{P}}(n^{-1/2}),$$

it follows that

$$\frac{1}{n} \sum_{i=1}^{n} \|d_{ni}\|^2 = O_{\mathbb{P}}(n^{-1}),$$

so conditional Markov's inequality yields

$$\frac{1}{\sqrt{n}}\sum_{i=1}^{n}(W_i-1)\{\dot{\ell}_i(\widehat{\theta}_n)-\psi_i\}=o_{\mathbb{P}^{\mathrm{DW}}}(1).$$

Together with (E.2), this also shows that the weighted score at $\widehat{\theta}_n$ is $O_{\mathbb{P}^{\mathrm{DW}}}(1)$. Consequently,

$$\sqrt{n}(\widehat{\theta}_n^{\mathrm{DW}}-\widehat{\theta}_n)=-J^{-1}\frac{1}{\sqrt{n}}\sum_{i=1}^{n}(W_i-1)\psi_i+o_{\mathbb{P}^{\mathrm{DW}}}(1),$$

and (E.2) gives

$$\sqrt{n}(\widehat{\theta}_n^{\mathrm{DW}}-\widehat{\theta}_n)\rightsquigarrow N_d(0,J^{-1}VJ^{-1}).$$

Hence, by the delta method,

$$T_n^{\mathrm{DW}}:=\sqrt{n}\{g(\widehat{\theta}_n^{\mathrm{DW}})-g(\widehat{\theta}_n)\}\rightsquigarrow N(0,B_g).$$

The complement of the all-$\omega$ propriety event in Assumption E.1 has conditional probability $o_{\mathbb{P}}(1)$. Assumption E.1 and (14) now imply, uniformly over $\omega\in\Omega$, that $\check{g}_n(\omega)\in C_{\alpha,\omega}^{\mathrm{DW}}(W)$ is equivalent to

$$-z_{\beta_U}\sqrt{\frac{A_g}{\omega}}+o_{\mathbb{P}^{\mathrm{DW}}}(1)\le T_n^{\mathrm{DW}}\le -z_{\beta_L}\sqrt{\frac{A_g}{\omega}}+o_{\mathbb{P}^{\mathrm{DW}}}(1).$$

The preceding conditional weak convergence, continuity of the Gaussian limit, and compactness of $\Omega$ therefore give

$$\sup_{\omega\in\Omega}|c_{n,\alpha}^{\mathrm{DW}}(\omega)-H_\alpha(\omega)|=o_{\mathbb{P}}(1).$$

Together with Proposition 1, this implies

$$\sup_{\omega\in\Omega}|c_{n,\alpha}^{\mathrm{DW}}(\omega)-c_{n,\alpha}^{\mathrm{MB}}(\omega)|=o_{\mathbb{P}}(1).$$

Finally, let

$$D_n^{\mathrm{DW}}=\sup_{\omega\in\Omega}|c_{n,\alpha}^{\mathrm{DW}}(\omega)-H_\alpha(\omega)|=o_{\mathbb{P}}(1).$$

By (E.1),

$$|H_\alpha(\widetilde{\omega}^{\mathrm{DW}})-(1-\alpha)|\le 2D_n^{\mathrm{DW}}+\rho_n=o_{\mathbb{P}}(1).$$

Since $H_\alpha$ has the unique root $\omega_g^\circ=A_g/B_g$ and is strictly monotone on $\Omega$,

$$\widetilde{\omega}^{\mathrm{DW}}\xrightarrow{\mathbb{P}}\omega_g^\circ.$$

The final coverage conclusion follows by evaluating (A.7) at $\widetilde{\omega}^{\mathrm{DW}}$ and applying the same argument as in the proof of Theorem 1. □

Proposition E.1 establishes only first-order equivalence. For example, for a multinomial-bootstrap count $M_i$,

$$\mathbb{E}^{\mathrm{MB}}[(M_i-1)^3]=\left(1-\frac{1}{n}\right)\left(1-\frac{2}{n}\right)\longrightarrow 1,$$

whereas for a scaled Dirichlet weight

$$\mathbb{E}^{\mathrm{DW}}[(W_i - 1)^3] = \frac{2(n-1)(n-2)}{(n+1)(n+2)} \longrightarrow 2.$$

Thus the two weighting schemes have the same first-order Gaussian limit but different higher-order weight moments, and no second-order equivalence is claimed here.

## E.3 Simulation

We compare GPC-IRW based on multinomial bootstrap counts (IRW-MB) with its scaled Dirichlet-weight counterpart (IRW-DW), using the same simulation settings as in Section 5. Let IRW-CMB denote GPC-IRW using cluster multinomial bootstrap counts, whereas IRW-CDW denotes GPC-IRW using scaled cluster-level Dirichlet weights. In both cluster-level methods, all observations within a cluster receive the same weight.

Figures E.1 and E.2 show that the two methods yield similar empirical coverage and nearly identical interval-length distributions under independent sampling. Under DGP-1, coverage is close to the nominal level across sample sizes, whereas under DGP-2, coverage approaches the nominal level as the sample size increases. Similar patterns are observed under independent-cluster sampling (Figures E.6 and E.7), with both methods exhibiting some undercoverage in smaller samples.

The computational results also reveal no systematic advantage for either weighting scheme. Figures E.3–E.4 and E.8–E.9 show broadly comparable wall-clock times and numbers of calibration iterations. Tables E.1 and E.2 further indicate similar frequencies of safeguard-triggered direct posterior simulations. Moreover, Figures E.5 and E.10 show close agreement between calibrated learning rates obtained using independent Gaussian posterior draws and NUTS for IRW-DW and IRW-CDW, respectively. Overall, these findings are consistent with the first-order equivalence established in Section E.2, but they do not suggest a clear practical advantage of Dirichlet weighting in the settings considered.

Figure E.1: Empirical coverage of nominal 95% credible intervals: Independent sampling

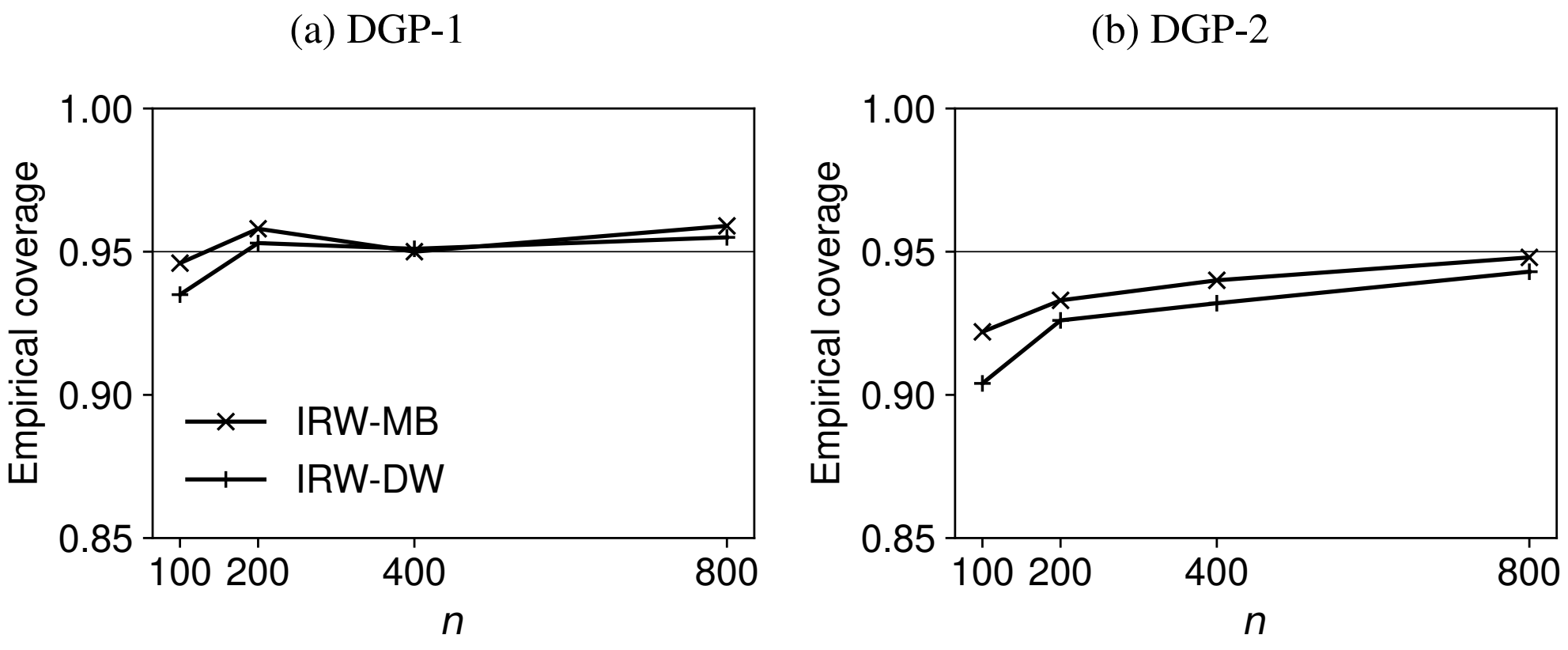

Figure E.2: Length of nominal 95% credible intervals: Independent sampling

(a) DGP-1, $n = 100$ (b) DGP-1, $n = 200$ (c) DGP-1, $n = 400$ (d) DGP-1, $n = 800$

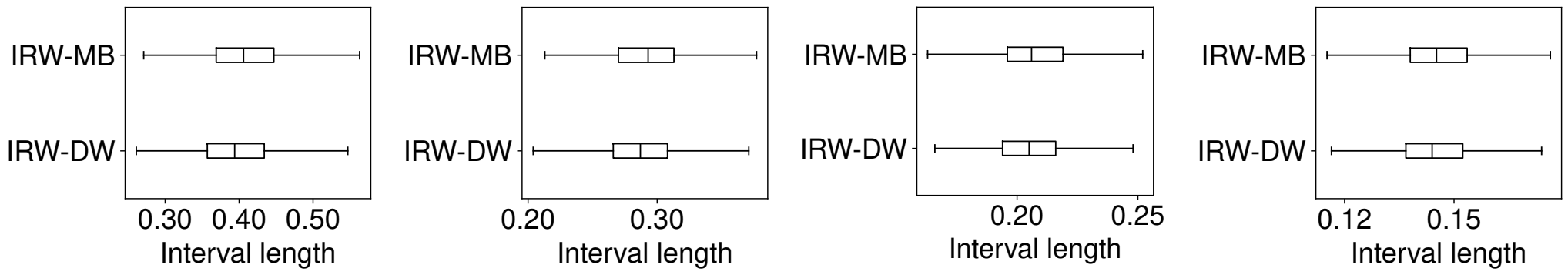


(e) DGP-2, $n = 100$ (f) DGP-2, $n = 200$ (g) DGP-2, $n = 400$ (h) DGP-2, $n = 800$

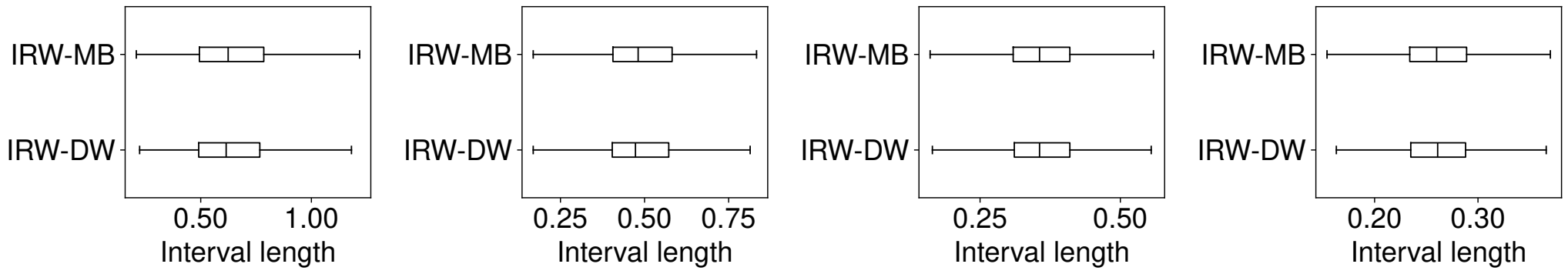


Figure E.3: Wall-clock time: Independent sampling

(a) DGP-1, $n = 100$ (b) DGP-1, $n = 200$ (c) DGP-1, $n = 400$ (d) DGP-1, $n = 800$

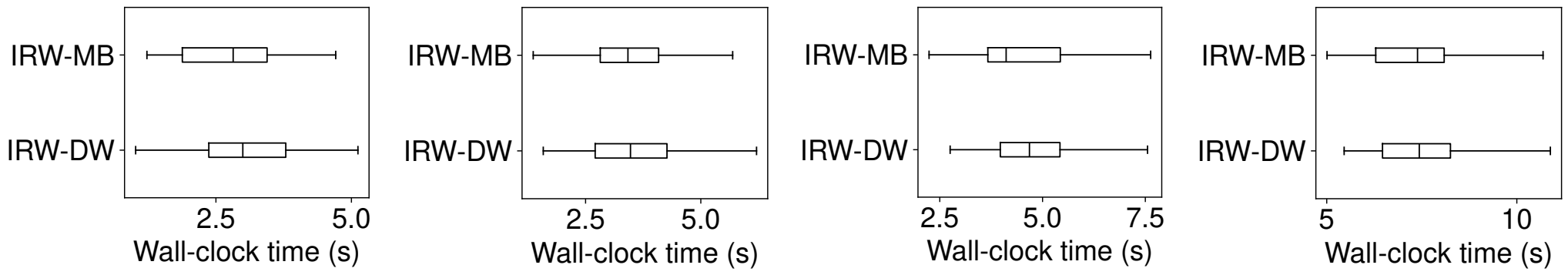


(e) DGP-2, $n = 100$ (f) DGP-2, $n = 200$ (g) DGP-2, $n = 400$ (h) DGP-2, $n = 800$

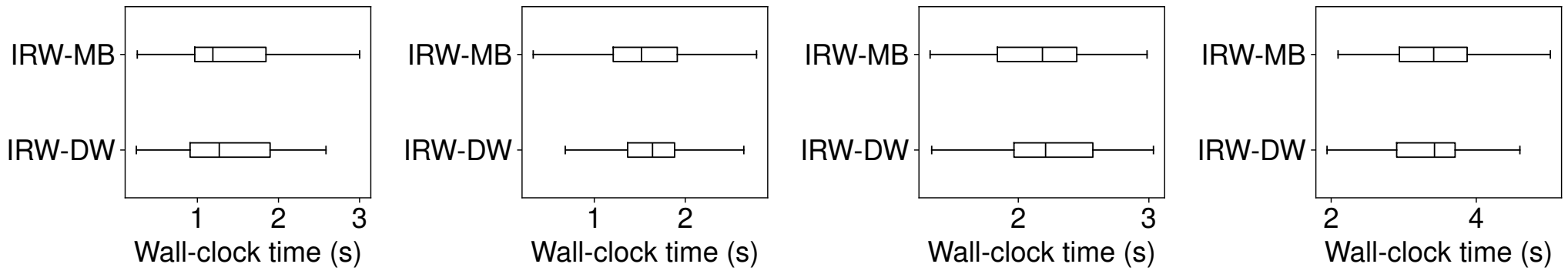

Figure E.4: Number of calibration iterations: Independent sampling

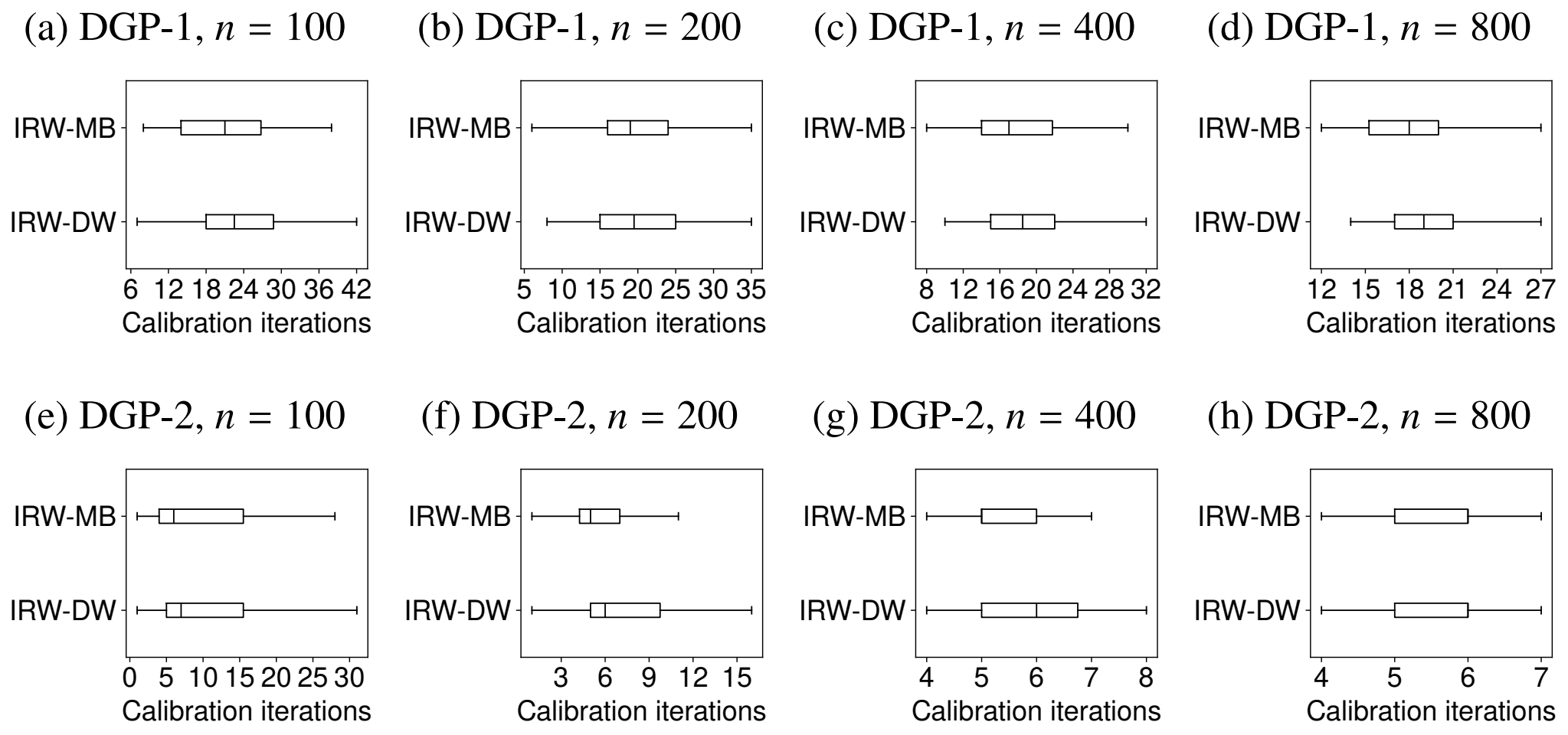


Table E.1: Average number of safeguard-triggered direct bootstrap-posterior fits per bootstrap replicate: Independent sampling

| DGP | Method | $n$ | | | |
|---|---|---|---|---|---|
| | | 100 | 200 | 400 | 800 |
| DGP-1 | IRW-MB | 0.312 | 0.253 | 0.234 | 0.200 |
| | IRW-DW | 0.324 | 0.254 | 0.233 | 0.209 |
| DGP-2 | IRW-MB | 0.328 | 0.313 | 0.377 | 0.386 |
| | IRW-DW | 0.272 | 0.272 | 0.375 | 0.361 |

Figure E.5: Calibrated learning rate: GPC-IRW-DW, Independent sampling

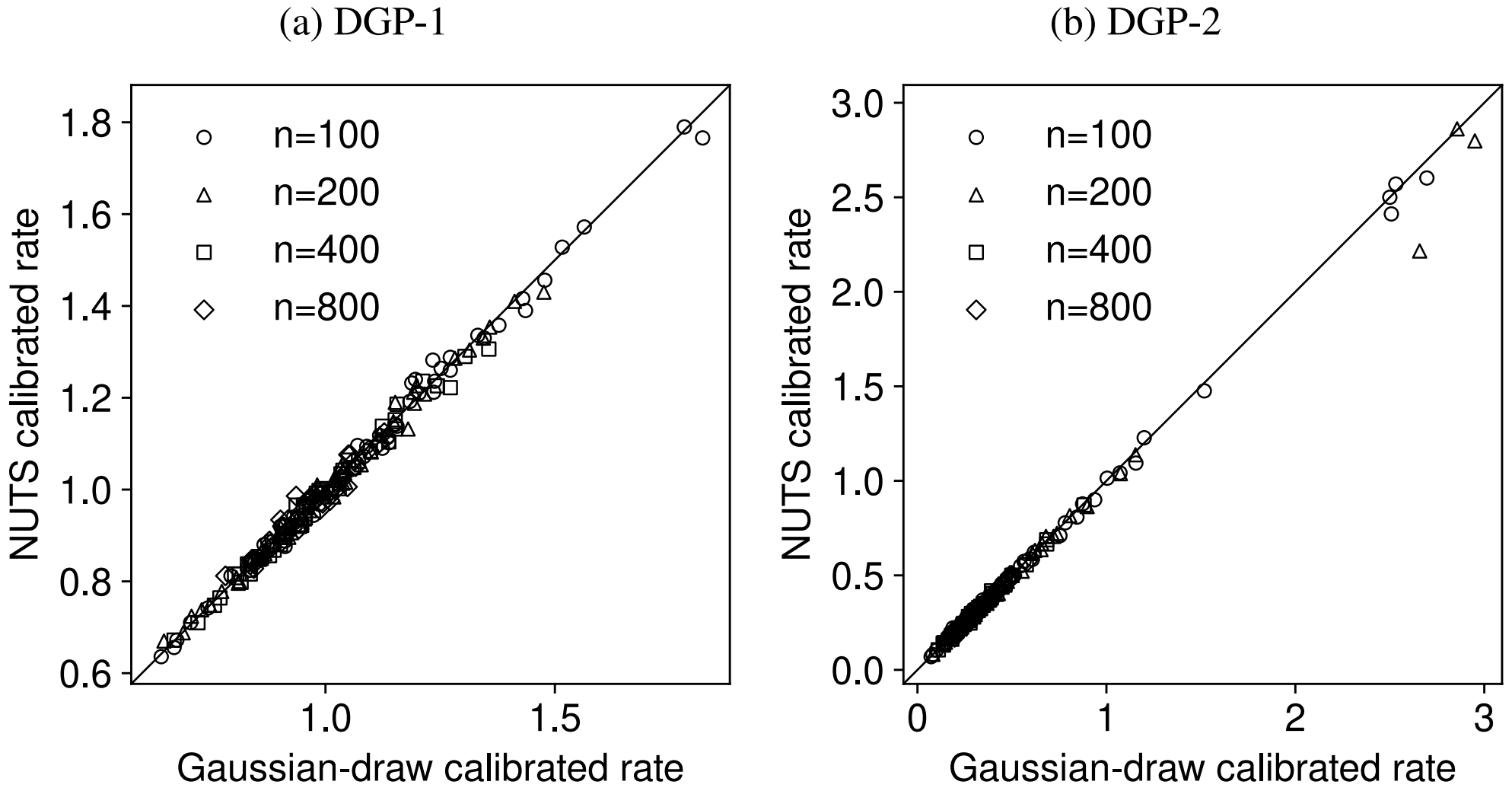

Figure E.6: Empirical coverage of nominal 95% credible intervals: Independent-cluster sampling

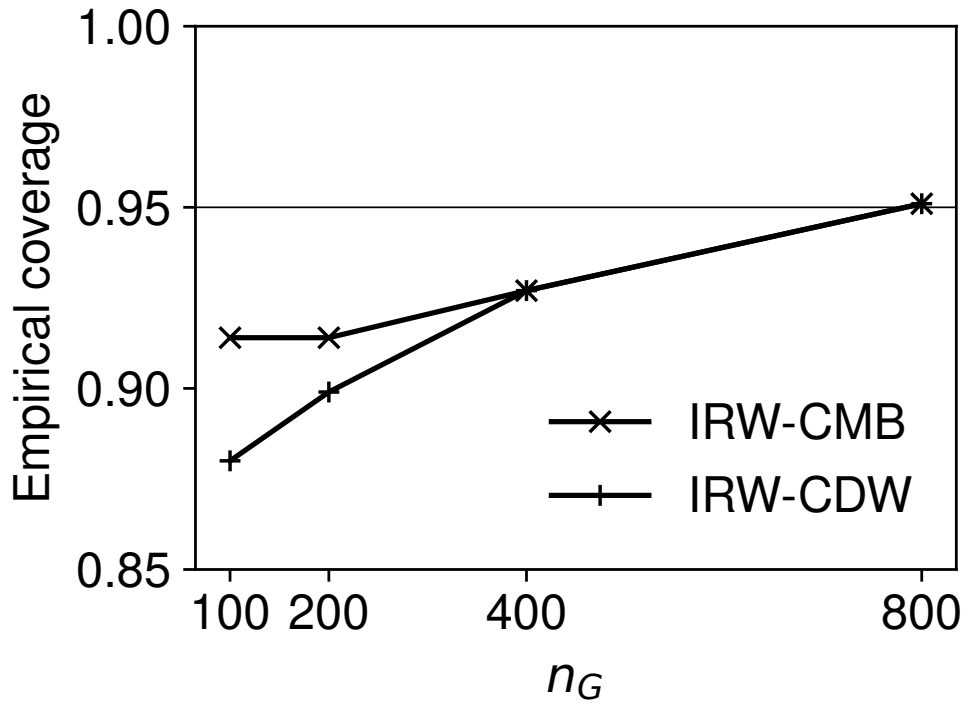


Figure E.7: Length of nominal 95% credible intervals: Independent-cluster sampling

(a) $n_G = 100$ (b) $n_G = 200$ (c) $n_G = 400$ (d) $n_G = 800$

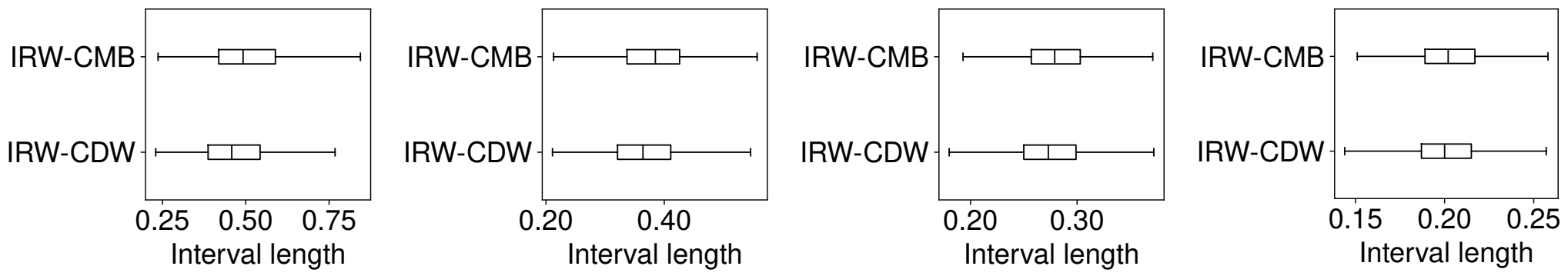


Figure E.8: Wall-clock time: Independent-cluster sampling

(a) $n_G = 100$ (b) $n_G = 200$ (c) $n_G = 400$ (d) $n_G = 800$

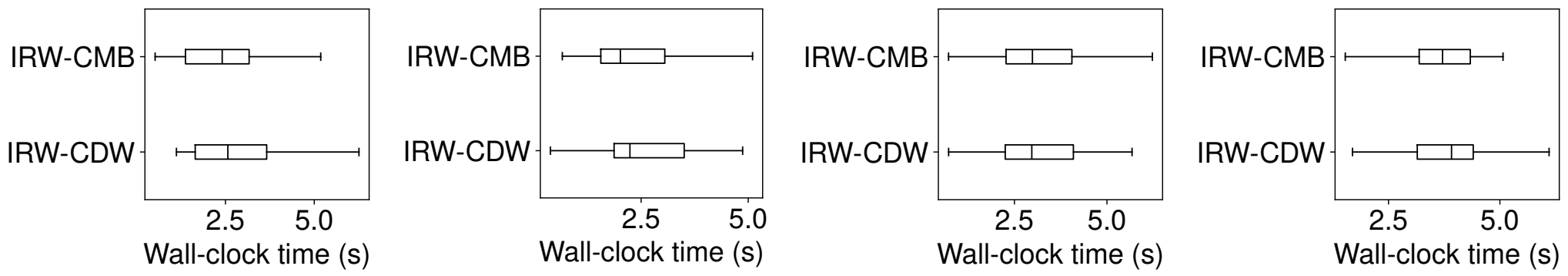


Figure E.9: Number of calibration iterations: Independent-cluster sampling

(a) $n_G = 100$ (b) $n_G = 200$ (c) $n_G = 400$ (d) $n_G = 800$

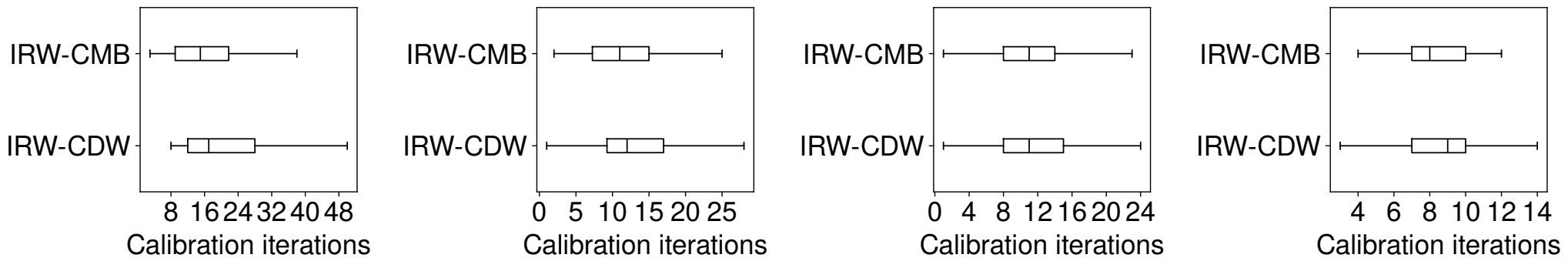

Table E.2: Average number of safeguard-triggered direct bootstrap-posterior fits per bootstrap replicate: Independent-cluster sampling

| Method | $n_G$ | | | |
|---|---|---|---|---|
| | 100 | 200 | 400 | 800 |
| IRW-CMB | 0.385 | 0.254 | 0.245 | 0.181 |
| IRW-CDW | 0.365 | 0.247 | 0.223 | 0.192 |

Figure E.10: Calibrated learning rate: GPC-IRW-CDW, independent-cluster sampling

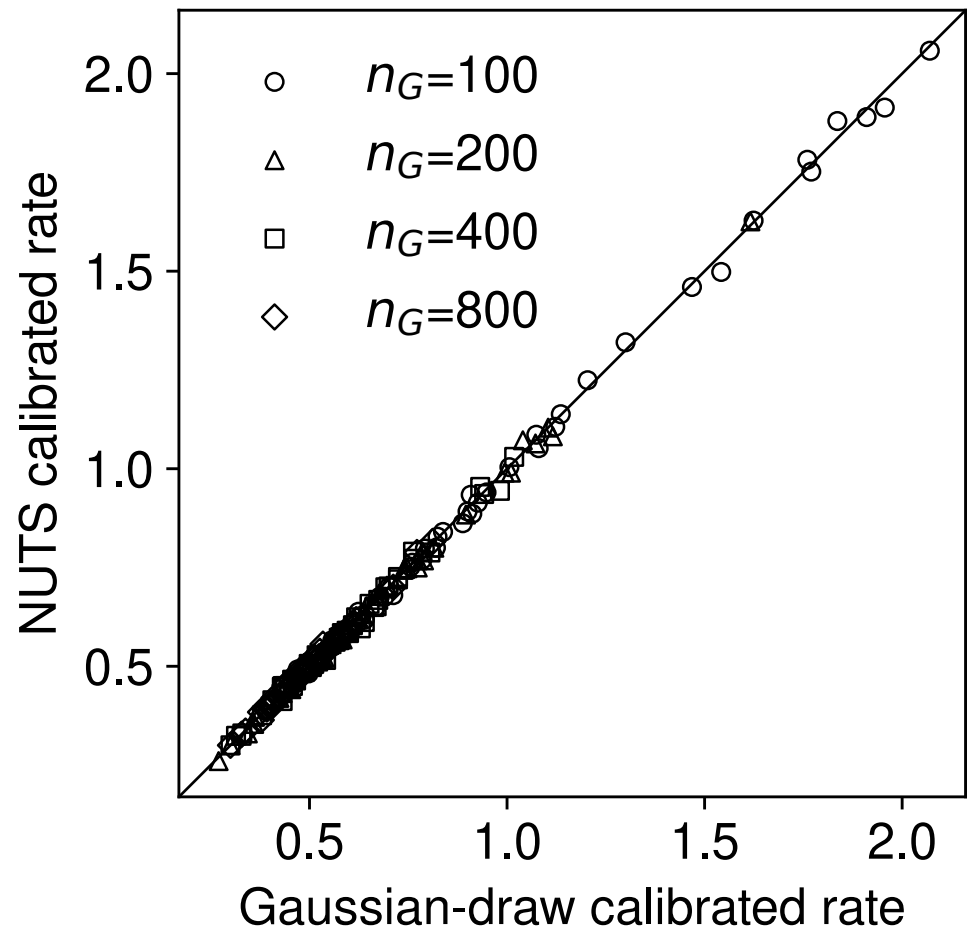


%\end{comment}

\end{document}